\documentclass{article}
\usepackage[utf8]{inputenc}
\usepackage{color}

\usepackage[ruled,vlined]{algorithm2e}
\usepackage{amsmath}
\usepackage{amssymb}
\usepackage{caption}
\usepackage{graphicx}
\usepackage{natbib}
\usepackage{setspace}
\usepackage{subcaption}
\usepackage{tabularx}
\usepackage{verbatimbox} 
\usepackage[dvipsnames]{xcolor}

\DeclareMathOperator*{\argmax}{argmax}

\title{Fast Bayesian estimation of brain activation with cortical surface fMRI data using EM}
\author{Daniel A. Spencer, David Bolin, Amanda F. Mejia}

\date{\today}
\providecommand{\keywords}[1]
{
  \small	
  \textbf{\textit{Keywords---}} #1
}

\newcommand\Includegraphics[2][]{\addvbuffer[3pt 0pt]{\includegraphics[#1]{#2}}} 

\begin{document}
	
\maketitle

\begin{abstract}
	Task functional magnetic resonance imaging (fMRI) is a type of neuroimaging data used to identify areas of the brain that activate during specific tasks or stimuli. These data are conventionally modeled using a massive univariate approach across all data locations, which ignores spatial dependence at the cost of model power. We previously developed and validated a spatial Bayesian model leveraging dependencies along the cortical surface of the brain in order to improve accuracy and power. This model utilizes stochastic partial differential equation spatial priors with sparse precision matrices to allow for appropriate modeling of spatially-dependent activations seen in the neuroimaging literature, resulting in substantial increases in model power. Our original implementation relies on the computational efficiencies of the integrated nested Laplace approximation (INLA) to overcome the computational challenges of analyzing high-dimensional fMRI data while avoiding issues associated with variational Bayes implementations. However, this requires significant memory resources, extra software, and software licenses to run. In this article, we develop an exact Bayesian analysis method for the general linear model, employing an efficient expectation-maximization algorithm to find maximum \textit{a posteriori} estimates of task-based regressors on cortical surface fMRI data. Through an extensive simulation study of cortical surface-based fMRI data, we compare our proposed method to the existing INLA implementation, as well as a conventional massive univariate approach employing ad-hoc spatial smoothing. We also apply the method to task fMRI data from the Human Connectome Project and show that our proposed implementation produces similar results to the validated INLA implementation. Both the INLA and EM-based implementations are available through our open-source \texttt{BayesfMRI} R package.
\end{abstract}


\section{Introduction}

Task-based functional magnetic resonance imaging (fMRI) data measure the blood-oxygen level-dependent (BOLD) response, an indirect measure of neural activity, while a subject performs tasks \citep{lindquist2008statistical,poldrack2011handbook}. These relatively high spatial and temporal resolution data are widely-used, non-invasive experimental measures that allow researchers to study the way that the human brain works \textit{in vivo}. fMRI data exhibit a low signal-to-noise ratio, requiring analyses to be as statistically efficient and powerful as possible to produce accurate results that reflect the full extent of activation rather than detecting only the extreme values. Traditional massive univariate analysis methods treat locations within the brain as independent before any clustering analysis methods are applied to the results, which diminishes the ability for models to make reliable conclusions about which areas of subjects’ brains are associated with a task \citep{elliott2020test}. Sophisticated models that leverage the spatiotemporal dependence structure of the data would improve inferential power, but estimating such models is difficult due to the size and complexity of the data. Recent advancements in computer processing and storage technology have begun to enable more complex analyses of fMRI data, improving the ability of scientists to detect brain signals amidst the noise of measurement and physiological mechanics. Recent models attempt to account for either spatial or temporal dependence, though often at the cost of computational efficiency.

The BOLD response is measured as an image array comprised of volumetric pixels, or voxels, each typically having a volume of about 8$mm^3$. A commonly-used modeling technique is known to the neuroscience community as the \textit{general linear model} (GLM)\footnote{The meaning of GLM in the neuroimaging community is not to be confused with generalized linear models in the statistical sense. The term GLM has expanded beyond task fMRI to basically refer to any massive univariate regression model approach.}. The classical GLM regresses the time series of measurements for each voxel individually against task covariates to determine the effect of the task on neural activity around the brain \citep{chatfield2018introduction,friston1994statistical}. These regressions can be used to test whether a specific voxel has a statistically significant association with the task covariates and is thus ``activated" by a task.  The classical GLM requires multiple comparisons corrections due to the massive number of voxels being tested. Cluster-based methods help to incorporate spatial information in determining activation \citep{poline1993analysis,poline1997combining,smith2009threshold} but are post-hoc and do not benefit the statistical efficiency of the estimates, which will limit power. Parametric assumptions about spatial dependence used in these methods have been shown to be unrealistic in volumetric fMRI data \citep{eklund2016cluster,eklund2019cluster}. 

An alternative is to avoid massive univariate analysis and instead perform analysis using spatial Bayesian methods, examining the posterior probability distribution of all voxels together to determine where activations occur. Such techniques explicitly state the assumptions about spatial dependence made in the model through the prior distributions, avoiding the pitfalls of null hypothesis testing. Bayesian models can leverage spatial information through a prior that allows for nonzero covariance between adjacent data locations in space and time. Several spatial Bayesian methods for volumetric fMRI have been proposed, typically relying on variational Bayes (VB) methods to decrease computational load \citep{penny2005bayesian} at the cost of underestimating posterior variance and poorly estimating the posterior mode \citep{wang2005inadequacy,bishop2006pattern,rue2009approximate,siden2017fast}. \cite{zhang2016spatiotemporal} utilized a Bayesian nonparametric framework to detect activations using both Markov chain Monte Carlo (MCMC) and variational Bayes (VB) methods for both single and multiple subjects, though even the VB method is rather expensive and requires dimension reduction. \cite{siden2017fast} proposed a fast method using a spatial prior on volumetric single-subject data using both MCMC and VB methods that scale well, but does not allow for group analyses. \cite{guhaniyogi2021bayesian} and \cite{spencer2020joint} used shrinkage priors and tensor decompositions to model volumetric task-based fMRI for both single and multiple-subject studies that rely on MCMC methods, which are time-consuming. However, all of these studies apply spatial priors on volumetric data using Euclidean distance to determine covariance, which does not represent the folded nature of the cerebral cortex. This results in signal being averaged across volumetric pixels that are close together in space though they may be neurologically distal due to their location on different folds of the cortex, often as part of functionally or anatomically distinct cortical regions. In addition, spatial methods used with volumetric data implicitly blur activation signal with noise from white matter and cerebrospinal fluid.

Recently, \cite{mejia2020bayesian} proposed a surface-based spatial Bayesian (SBSB) GLM for cortical surface fMRI (cs-fMRI) data. Such data use a triangular mesh to represent the spatial configuration of the cortical surface. This mesh can be used to find the geodesic distances between points on the surface. These distances appropriately account for the cortex's folded structure, unlike Euclidean distances in the volume. The SBSB GLM uses a stochastic partial differential equation (SPDE) prior, which is built on a triangular mesh and does not require a regular lattice structure. The SPDE prior is advantageous because it approximates a continuous Markov random field, it is invariant to finite resamplings, which allows for different resolutions in fMRI data, and it allows for dependence in neighboring data locations while keeping computational cost low through sparse precision matrices. In addition, results from single-subject analyses can be combined in a principled manner for group-level inference. \cite{spencer2022spatial} performed a comprehensive validation study of the SBSB GLM based on test-retest task fMRI data from the Human Connectome Project, showing gains in accuracy and power over the common practice of spatially smoothing data before model-fitting. 

For Bayesian computation, \cite{mejia2020bayesian} fit the SBSB GLM model using an empirical Bayes version of the integrated nested Laplace approximation (INLA) through the R software package \texttt{R-INLA}. The package is a powerful computational resource for spatial modeling, offering efficient backend code written in C++ that utilizes fast matrix operations and allows for parallelization via the PARDISO sparse linear algebra library \citep{pardiso-7.2a,pardiso-7.2b,pardiso-7.2c}. These features of R-INLA drastically reduce model fitting times. However, there are some limitations of INLA in this context. First, the numerical approach to integrating out model hyperparameters used in INLA requires approximating their joint posterior using a multivariate Gaussian, which is memory intensive for more than 5 to 10 hyperparameters, effectively restricting its application to relatively simple task fMRI experimental designs \citep{opitz2017latent}. Previous work has reported memory requirements of 30 - 50 GB for simple single-subject task analysis \citep{mejia2020bayesian}, and in our experience, it is not uncommon to exceed 75GB of memory usage in many practical settings. Second, there are some challenges in developing and maintaining software that depends on \texttt{R-INLA}, which is not available on the Comprehensive R Archive Network (CRAN) and does not install easily on many systems. Therefore, carrying out model estimation can be especially difficult for researchers that do not administer their own computing. Third, in order to achieve reasonable computing speeds, a license for the  PARDISO library is required, which is not always freely available. Finally, at the data scale of functional MRI, INLA requires significant system memory to run, which is prohibitive for some research teams.

In order to address these issues, in this paper we develop an efficient expectation-maximization (EM) algorithm \citep{dempster1977maximum,gelman2013bayesian} to produce posterior estimates and inference of activation for subject-level task in the SBSB GLM. We also develop results from these analyses that are combined across subjects in a principled way to identify group-level effects in multi-subject analysis. We compare this EM algorithm to the INLA-based implementation of the SBSB GLM from \cite{mejia2020bayesian} using both simulated data and motor task fMRI data from the Human Connectome Project (HCP) \citep{barch2013function}. Using simulated data, we find that the EM algorithm achieves similar accuracy while cutting computation time dramatically. All model fitting and preprocessing steps have been implemented within the open source R package \texttt{BayesfMRI}\footnote{https://github.com/mandymejia/BayesfMRI/tree/1.8.EM}.

This article will proceed with a description of the SBSB GLM and details of our EM algorithm in Section \ref{sec:methodology}. Next, we compare our proposed EM algorithm with INLA in a study of simulated cortical surface fMRI data in Section \ref{sec:simulated}. We show results from a study of task fMRI data from the HCP in Section \ref{sec:real}. We end with conclusions and discussion of future work in Section \ref{sec:conclusion}. 

\section{Methodology}
\label{sec:methodology}

We will use the following matrix notation throughout this section. Fixed matrix-valued quantities will be represented using upper-case bold font, $\mathbf{A}$, while vectors are in lower-case bold font $\mathbf{a}$, and scalars are non-bold font in both upper- and lower-case, $a$. Parameters in the model follow the same convention using Greek letters, i.e. $\boldsymbol\Sigma$ represents a matrix, $\boldsymbol\sigma$ represents a vector, and $\sigma$ represents a scalar. $|\mathbf{A}|$ and $\text{Tr}(\mathbf{A})$ represent the determinant and the trace of matrix $\mathbf{A}$, respectively. Following this convention, we now introduce the modeling methodology.

\subsection{The surface-based spatial Bayesian general linear model} \label{sec:single_subjectSBSBGLM}

Consider cortical surface fMRI (cs-fMRI) data from a scan, a time series of length $T$ represented as $\mathbf{y}_v \in \mathbb{R}^T$ for locations $v = 1,\ldots,N$ for $N$ locations in order to model data from a single subject. These data are gathered while a subject completes $K$ different tasks at specific time intervals. The expected BOLD response to task $k$, $\mathbf{x}_k \in \mathbb{R}^T$, is generated by convolving the binary stimulus timing function (on/off) with a canonical haemodynamic response function (HRF) representing the time-lagged BOLD response that follows neuronal firing. After preprocessing the data to reduce noise, eliminating the baseline signal, and inducing residual independence as outlined in \ref{sec:preproc}, the general linear model can be written in the form
\begin{align}
    \mathbf{y}_v = \sum_{k=1}^K \beta_{v,k}\mathbf{x}_{v,k} + \mathbf{e}_v, \quad \mathbf{e}_v \sim \text{Normal}(\mathbf{0},\sigma^2\mathbf{I}), \quad v = 1,\ldots,N, \label{eq:preprocessed_model}
\end{align}
where $\beta_{v,k}$ is the activation effect in terms of percent signal change at location $v$ for task $k$. Note that the task-based regressors $x_{k,v}$ vary across locations $v$. This is because our preprocessing includes location-specific (or spatially variable) prewhitening to account for differences in residual autocorrelation across the cortex \citep{parlak2022sources}. It is typical to fit this model in a massive univariate manner which we refer to as the classical GLM, which is able to be fitted quickly, even to high-resolution data. However, this advantage comes at the cost of ignoring the spatial dependence in brain activation, which reduces model power significantly when compared to the Bayesian GLM, as shown in \cite{spencer2022spatial}.

In order to facilitate spatial Bayesian modeling, we rewrite the model in Equation (\ref{eq:preprocessed_model}) to represent data across all locations and times at once in Equation (\ref{eq:BayesianGLM}). Here, the preprocessed response $\mathbf{y} = (\mathbf{y}_1',\ldots,\mathbf{y}_N')' \in \mathbb{R}^{TN}$ at times $t = 1,\ldots,T$ and locations $v = 1,\ldots,N$ is explained by linear effects $\boldsymbol\beta_{k} \in \mathbb{R}^N$ for each of $K$ tasks ($\mathbf{X}_k = \text{block-diagonal}(\mathbf{x}_{1,k},\ldots,\mathbf{x}_{N,k}) \in \mathbb{R}^{TN \times N}$) and an error term $\mathbf{e} \in \mathbb{R}^{TN}$. 
\begin{align}
    \mathbf{y}  = \sum_{k = 1}^K \mathbf{X}_k \boldsymbol{\beta}_k + \mathbf{e}, \quad \mathbf{e} \sim \text{Normal}(\mathbf{0},\sigma^2\mathbf{I}) \label{eq:BayesianGLM}
\end{align}
This linear model facilitates incorporating spatial dependence, in contrast to the $N$ separate linear models constituting the classical GLM, which does not include spatial dependence at the model-fitting stage. We fit this model across an entire cortical hemisphere, and the two cortical surface hemispheres are analyzed separately for computational reasons, which does not affect the results because they are represented by separate surface meshes that are physically distinct.

The SBSB GLM incorporates spatial dependence in the activation amplitudes $\boldsymbol\beta_k$ using a special class of Gaussian Markov random field (GMRF) processes known stochastic partial differential equation (SPDE) priors \citep{mejia2020bayesian,spencer2022spatial,lindgren2011explicit,bolin2013comparison}. The specific construction of the prior is  
\begin{align}
	\boldsymbol{\beta}_k & = \boldsymbol{\Psi}_k \mathbf{w}_k, &
	\mathbf{w}_k & \sim \text{Normal}(\mathbf{0},\mathbf{Q}_k^{-1}), &
	\mathbf{Q}_k & = \tau_k^2(\kappa_k^4 \mathbf{C} + 2\kappa_k^2 \mathbf{G} + \mathbf{G}\mathbf{C}^{-1} \mathbf{G}) \label{eq:SPDEprior} \\
	&&\kappa_k & \sim \text{log-Normal}(\mu_\kappa,\sigma_\kappa^2), &
	\tau_k & \sim \text{log-Normal}(\mu_\tau,\sigma_\tau^2), \nonumber
\end{align}
where $\boldsymbol{\Psi}_k \in \mathbb{R}^{N \times n}$ maps the original $N$ data locations to a triangular mesh consisting of $n$ vertices. $\mathbf{C}$ is a fixed diagonal matrix describing the relative mesh vertex precisions. $\mathbf{G}$ is a fixed sparse matrix describing the neighborhood structure of the mesh vertices, in which entries corresponding to neighboring locations take nonzero values. The mesh is usually created to maximize the minimum interior angle across the triangles, and extra vertices may be added along the boundaries to avoid adverse boundary effects. An example of this mesh structure can be seen in \textbf{Figure \ref{fig:cs_mesh}}. The default setting for the INLA implementation used in \cite{mejia2020bayesian} sets the priors for $\tau_k$ and $\kappa_k$ to be log-Normal with mean 2 and variance 2. Under the SPDE representation in $\mathbb{R}^d$, the variance of the random field can be represented as $\phi_k = \Gamma(\nu)\left(\Gamma(\alpha) (4\pi)^{d/2}\kappa_k^{2\nu}\tau_k^2 \right)^{-1}$ where $\nu = \alpha - d / 2$. Spectral theory shows that an integer must be picked for $\alpha$ to obtain a Markov field. The form of the prior in Equation (\ref{eq:SPDEprior}) assumes $\alpha = 2$, which makes $\nu = 1$ for a surface, which has dimension $d = 2$. Therefore, the variance of the random field simplifies to
\begin{align}
    \phi_k = (4\pi \kappa_k^2 \tau_k^2 )^{-1}. \label{eq:phi}
\end{align}
The product $\kappa_k^2\tau_k^2$ in the variance remains the same if $\kappa_k^{2'} = c\kappa_k^2$ and $\tau_k^{2'} = \tau_k^2 / c$ for any constant $c$, and early experiments showed that this can cause the EM algorithm to produce values for $\kappa_k^2$ and $\tau_k^2$ in which one approaches infinity while the other approaches zero. Therefore, we rewrite the precision in the SPDE prior in terms of the variance of the random field as
\begin{align}
	\mathbf{Q}_k & = \frac{4\pi}{\phi_k}\left( \kappa_k^2 \mathbf{C} + 2\mathbf{G} + \kappa_k^{-2} \mathbf{GC}^{-1}\mathbf{G} \right) = \frac{4\pi}{\phi_k}\tilde{\mathbf{Q}}_k. 
\end{align}
In this form, $\phi_k$ controls the variance and $\kappa_k^2$ controls the spatial correlation of the random field.

To avoid blurring across structural boundaries, cortical surface data are already represented on a triangular mesh.  

For mathematical convenience, the model in Equation (\ref{eq:BayesianGLM}) can be rewritten using the notation in Equation (\ref{eq:SPDEprior}) to include all $k$ task covariates in matrix form:
\begin{align}
    \mathbf{y} = \mathbf{X}\boldsymbol{\Psi}\mathbf{w} + \mathbf{e}, \quad \mathbf{e} \sim \text{Normal}(\mathbf{0},\sigma^2\mathbf{I}),
\end{align}
where $\mathbf{X} \in \mathbb{R}^{TN \times NK}$ is a column-bound matrix of the covariates, $\boldsymbol{\Psi} \in \mathbb{R}^{NK \times nK}$ is a block diagonal matrix projecting the covariate values onto the triangular mesh, and $\mathbf{w} \in \mathbb{R}^{nK \times 1}$ is the vector of coefficients corresponding to the covariates on the mesh. 

\subsection{Expectation-Maximization Procedure}

In order to provide a computational alternative to INLA for the SBSB GLM, here we derive an EM algorithm \citep{dempster1977maximum, gelman2013bayesian} to find the mode of the posterior distribution of $\mathbf{w}$, and by extension $\boldsymbol{\beta}$, along with maximum likelihood parameter estimates for $\boldsymbol\theta = (\kappa_1^2,\ldots,\kappa_K^2,\phi_i,\ldots,\phi_K,\sigma^2)'$. This algorithm could be easily extended to incorporate priors on these hyperparameters to account for uncertainty in their values and avoid underestimation of the posterior variance of the latent fields. In addition, we obtain an estimate of the posterior precision of $\boldsymbol\beta_k$, which can be used in conjunction with the excursions method developed by \cite{bolin2015excursion,bolin2017quantifying,bolin2018calculating} to determine areas of activation in the latent fields.

\subsubsection{Expectation of the Log-Likelihood Density (E-step)}

Beginning with the definition of the joint log likelihood $\mathcal{L}(\boldsymbol\Theta | \mathbf{y},\mathbf{w}) = p(\mathbf{y}|\mathbf{w},\sigma^2) p(\mathbf{w}| \kappa^2,\phi)$, the expectation of the log likelihood density with respect to $\mathbf{w}$ given $(\kappa_1^{2(s)},\ldots,\kappa_K^{2(s)},\phi_1^{(s)},\ldots,\phi_K^{(s)},\sigma^{2(s)})' = \hat{\boldsymbol{\theta}}^{(s)}$ at algorithm step $s$ is found to be:
\begin{align}
    R(\boldsymbol\theta|\hat{\boldsymbol\theta}^{(s)}) = E_{\mathbf{w} | \mathbf{y},\hat{\boldsymbol{\theta}}^{(s)}}(\log p(\mathbf{y},\mathbf{w}|\boldsymbol\theta)) & = \int \log p(\mathbf{y}|\mathbf{w},\boldsymbol\theta)p\left(\mathbf{w}| \mathbf{y}, \hat{\boldsymbol\theta}^{(s)}\right) d\mathbf{w}  +  \int \log p(\mathbf{w}|\boldsymbol\theta)p\left(\mathbf{w}| \mathbf{y}, \hat{\boldsymbol\theta}^{(s)}\right) d\mathbf{w}, \label{eq:EoldFormula} \\ 
    & \propto R_1\left(\boldsymbol\theta|\hat{\boldsymbol\theta}^{(s)}\right) + R_2\left(\boldsymbol\theta|\hat{\boldsymbol\theta}^{(s)}\right) \nonumber
\end{align}
where $\mathbf{w} | \mathbf{y},\boldsymbol\theta \sim \text{Normal}\left(\boldsymbol{\mu}, \boldsymbol{\Sigma}\right)$, such that 
\begin{align*}
    \mathbf{Q} & = \text{block-diagonal}(\mathbf{Q}_1,\ldots,\mathbf{Q}_K), & 
    \boldsymbol{\Sigma} & = \left( \mathbf{Q} + \frac{1}{\sigma^2}\boldsymbol{\Psi}' \mathbf{X}' \mathbf{X} \boldsymbol{\Psi} \right)^{-1},  &
    \boldsymbol{\mu} & = \frac{1}{\sigma^2}\boldsymbol{\Sigma} \boldsymbol{\Psi}'\mathbf{X}'\mathbf{y}.
\end{align*}
In the interest of computational efficiency, we use several identities to avoid expensive matrix inversions (see Section \ref{sec:computation}). This avoids the calculation of $\boldsymbol{\Sigma}$ which is computationally prohibitive, as it requires the inversion of an $nK \times nK$ dense matrix, with $n$ between 1,000 and 30,000 in most practical applications. In the next section, we show the MLE of $\sigma^2$, and we outline the optimization strategy for $\kappa_k^2$ and $\phi_k$.

\subsubsection{Maximization of the Log Likelihood Density (M-step)}\label{sec:Mstep}

Next, we expand the expected joint log-likelihood in Equation (\ref{eq:EoldFormula}), which we maximize in order to find the MLEs of the elements in $\boldsymbol{\theta}$.
\begin{align*}
    R(\boldsymbol\theta|\hat{\boldsymbol\theta}^{(s)}) =& E_{\mathbf{w} | \mathbf{y},\hat{\boldsymbol{\theta}}^{(s)}}(\log p(\mathbf{y},\mathbf{w}|\boldsymbol\theta)) =  \int \log p(\mathbf{y}|\mathbf{w},\boldsymbol\theta)p\left(\mathbf{w}| \mathbf{y}, \hat{\boldsymbol\theta}^{(s)}\right) d\mathbf{w}  +  \int \log p(\mathbf{w}|\boldsymbol\theta)p\left(\mathbf{w}| \mathbf{y}, \hat{\boldsymbol\theta}^{(s)}\right) d\mathbf{w}, \\
    \propto & \int \log \left[ \sigma^{-TN} \exp \left\{ -\frac{1}{2\sigma^2}(\mathbf{y} - \mathbf{X}\boldsymbol{\Psi}\mathbf{w})'(\mathbf{y} - \mathbf{X}\boldsymbol{\Psi}\mathbf{w}) \right\} \right] d\mathbf{w} 
     + \int \log \left[ |\mathbf{Q}|^{1/2} \exp\left\{ -\frac{1}{2}\mathbf{w}'\mathbf{Q}\mathbf{w} \right\} \right] d\mathbf{w} \\
    \propto & -\frac{TN}{2} \log(\sigma^2) - \frac{1}{2\sigma^2} \mathbf{y}'\mathbf{y}  + \int \left[ \frac{1}{2\sigma^2} \mathbf{y}'\mathbf{X}\boldsymbol{\Psi}\mathbf{w} + \frac{1}{2\sigma^2} \mathbf{w}'\boldsymbol{\Psi}'\mathbf{X}'\mathbf{y} - \frac{1}{2\sigma^2} \mathbf{w}'\boldsymbol{\Psi}'\mathbf{X'X}\boldsymbol{\Psi}\mathbf{w} \right] d\mathbf{w} \\
    & + \frac{1}{2} \log |\mathbf{Q}| - \frac{1}{2} \int \mathbf{w'Qw} d\mathbf{w} \\
    \propto & -\frac{TN}{2} \log(\sigma^2) - \frac{1}{2\sigma^2} \mathbf{y}'\mathbf{y}  + \frac{1}{\sigma^2} \mathbf{y'X}\boldsymbol{\Psi}\text{E}\left(\mathbf{w}|\mathbf{y},\hat{\boldsymbol{\theta}}^{(s)}\right) - \frac{1}{2\sigma^2} \text{E}\left(\mathbf{w}'\boldsymbol{\Psi}'\mathbf{X'X}\boldsymbol{\Psi}\mathbf{w}|\mathbf{y},\hat{\boldsymbol{\theta}}^{(s)}\right) \\
    & + \frac{1}{2} \log |\mathbf{Q}| - \frac{1}{2} \text{E}\left(\mathbf{w'Qw}|\mathbf{y},\hat{\boldsymbol{\theta}}^{(s)}\right)
\end{align*}
The two quantities $\text{E}\left(\mathbf{w}'\boldsymbol{\Psi}'\mathbf{X'X}\boldsymbol{\Psi}\mathbf{w}|\mathbf{y},\hat{\boldsymbol{\theta}}^{(s)}\right)$ and $\text{E}\left(\mathbf{w'Qw}|\mathbf{y},\hat{\boldsymbol{\theta}}^{(s)}\right)$ are scalars, and are therefore equivalent to their trace. The trace operation is invariant to circular permutations, so we can rewrite these two expectations as
\begin{align*}
    \text{E}\left(\mathbf{w}'\boldsymbol{\Psi}'\mathbf{X'X}\boldsymbol{\Psi}\mathbf{w}|\mathbf{y},\hat{\boldsymbol{\theta}}^{(s)}\right) & = \text{Tr}\left(\text{E}\left(\boldsymbol{\Psi}'\mathbf{X'X}\boldsymbol{\Psi}\mathbf{ww'}|\mathbf{y},\hat{\boldsymbol{\theta}}^{(s)}\right)\right) 
    = \text{Tr}\left(\boldsymbol{\Psi}'\mathbf{X'X}\boldsymbol{\Psi}\text{E}\left(\mathbf{ww'}|\mathbf{y},\hat{\boldsymbol{\theta}}^{(s)}\right) \right) \\
    \text{E}\left(\mathbf{w'Qw}|\mathbf{y},\hat{\boldsymbol{\theta}}^{(s)}\right) & = \text{Tr}\left( \text{E}\left(\mathbf{Qww'}|\mathbf{y},\hat{\boldsymbol{\theta}}^{(s)}\right)\right) 
    = \text{Tr}\left(\mathbf{Q} \text{E}\left(\mathbf{ww'}|\mathbf{y},\hat{\boldsymbol{\theta}}^{(s)}\right)\right)
\end{align*}
We can now write the expected log-likelihood as 
\begin{align}
     R(\boldsymbol\theta|\hat{\boldsymbol\theta}^{(s)}) = & E_{\mathbf{w} | \mathbf{y},\hat{\boldsymbol{\theta}}^{(s)}}(\log p(\mathbf{y},\mathbf{w}|\boldsymbol\theta)) \nonumber \\
     \propto & \underbrace{-\frac{TN}{2} \log(\sigma^2) - \frac{1}{2\sigma^2} \mathbf{y}'\mathbf{y} + \frac{1}{\sigma^2} \mathbf{y'X}\boldsymbol{\Psi}\text{E}\left(\mathbf{w}|\mathbf{y},\hat{\boldsymbol{\theta}}^{(s)}\right) - \frac{1}{2\sigma^2} \text{Tr}\left(\boldsymbol{\Psi}'\mathbf{X'X}\boldsymbol{\Psi}\text{E}\left(\mathbf{ww'}|\mathbf{y},\hat{\boldsymbol{\theta}}^{(s)}\right)\right)}_{R_1(\boldsymbol{\theta}|\hat{\boldsymbol{\theta}}^{(s)}} \label{eq:R1R2} \\
    & + \underbrace{\frac{1}{2} \log |\mathbf{Q}| - \frac{1}{2} \text{Tr}\left(\mathbf{Q}\text{E}\left(\mathbf{ww'}|\mathbf{y},\hat{\boldsymbol{\theta}}^{(s)}\right) \right)}_{R_2\left(\boldsymbol{\theta}|\hat{\boldsymbol{\theta}}^{(s)}\right)} \nonumber
\end{align}
The MLE for $\sigma^2$ can be found through the maximization of $R_1\left(\boldsymbol\theta|\hat{\boldsymbol\theta}^{(s)}\right)$ with respect to $\sigma^2$:
\begin{align}
    \frac{\partial R_1}{\partial \sigma^2} & = -\frac{TN}{2\sigma^2} + \frac{1}{2(\sigma^2)^2}\mathbf{y'y} - \frac{1}{(\sigma^2)^2} \mathbf{y}'\mathbf{X}\boldsymbol{\Psi}\text{E}(\mathbf{w}|\mathbf{y},\hat{\boldsymbol\theta}^{(s)}) + \frac{1}{2(\sigma^2)^2}\text{Tr}(\boldsymbol{\Psi}' \mathbf{X}' \mathbf{X} \boldsymbol{\Psi} \text{E}(\mathbf{ww}'|,\mathbf{y},\hat{\boldsymbol\theta}^{(s)})), \nonumber \\
    \widehat{\sigma^2} & = \frac{1}{TN} \left[ \mathbf{y'y} - 2\mathbf{y'X}\boldsymbol{\Psi}E(\mathbf{w}|\mathbf{y},\hat{\boldsymbol\theta}^{(s)}) + \text{Tr}(\boldsymbol{\Psi}'\mathbf{X'X}\boldsymbol{\Psi} E(\mathbf{ww'}|,\mathbf{y},\hat{\boldsymbol\theta}^{(s)}) \right] \label{eq:sigma2_hat}
\end{align}
Next, values for $\phi_k$ and $\kappa_k^2$ must be found that maximize the log-likelihood. This is equivalent to optimizing $R_2(\boldsymbol\theta|\hat{\boldsymbol\theta}^{(s)})$ with respect to $\kappa_k^2$ and $\phi_k$. Setting $\tilde{\mathbf{Q}}_k = \left( \kappa_k^2 \mathbf{C} + 2\mathbf{G} + \kappa_k^{-2} \mathbf{GC}^{-1}\mathbf{G} \right)$, we rewrite $R_2\left(\boldsymbol\theta|\hat{\boldsymbol\theta}^{(s)} \right)$ as 
\begin{align}
    R_2\left(\boldsymbol\theta|\hat{\boldsymbol\theta}^{(s)} \right) & =  \frac{1}{2} \log \left| \mathbf{Q}\right| - \frac{1}{2} \text{Tr} \left(  \mathbf{Q} \text{E}\left(\mathbf{ww}'|\mathbf{y},\hat{\boldsymbol\theta}^{(s)}\right) \right) \nonumber \\
    & = \frac{1}{2} \log\left( \prod_{k=1}^K \left|\frac{1}{4\pi\phi_k}\tilde{\mathbf{Q}}_k\right| \right) -\frac{1}{2}  \text{Tr}\left( \mathbf{Q} \text{E}\left(\mathbf{ww}'|\mathbf{y},\hat{\boldsymbol\theta}^{(s)}\right) \right) \nonumber \\
    & = \frac{n}{2} \sum_{k=1}^K \log \left( \frac{1}{4\pi\phi_k} \right) +\frac{1}{2} \sum_{k=1}^K \log\left( |\tilde{\mathbf{Q}}_k| \right) -\frac{1}{2} \text{Tr}\left( \mathbf{Q} \text{E}\left(\mathbf{ww}'|\mathbf{y},\hat{\boldsymbol\theta}^{(s)}\right) \right) \nonumber \\ 
    & = - \frac{nK}{2} \log (4\pi) - \frac{n}{2} \sum_{k=1}^K \log (\phi_k) +\frac{1}{2} \sum_{k=1}^K\log\left( |\tilde{\mathbf{Q}}_k| \right) - \frac{1}{2} \sum_{k=1}^K \text{Tr}\left(\frac{1}{4\pi\phi_k} \tilde{\mathbf{Q}}_k \text{E}\left(\mathbf{w}_k\mathbf{w}_k'|\mathbf{y},\hat{\boldsymbol\theta}^{(s)}\right) \right) \nonumber \\
    \mathbf{Q} & = \text{block-diagonal}\left(\frac{1}{4\pi\phi_1}\tilde{\mathbf{Q}}_1,\ldots,\frac{1}{4\pi\phi_K}\tilde{\mathbf{Q}}_K\right) \label{eq:blockQtilde}
\end{align}
so that 
\begin{align*}
    \frac{\partial R_2}{\partial \phi_k} \left(\boldsymbol\theta|\hat{\boldsymbol\theta}^{(s)}\right) & = -\frac{n}{2\phi_k} + \frac{1}{8\pi\phi_k^2} \text{Tr} \left(\tilde{\mathbf{Q}}_k \text{E}\left(\mathbf{w}_k\mathbf{w}_k'|\mathbf{y},\hat{\boldsymbol\theta}^{(s)}\right)\right), \\
    \hat{\phi}_k & = \frac{1}{4\pi n} \text{Tr}\left(\tilde{\mathbf{Q}}_k\text{E}\left(\mathbf{w}_k\mathbf{w}_k'|\mathbf{y},\hat{\boldsymbol\theta}^{(s)}\right)\right).
\end{align*}
The optimal value for $\kappa_k^2$ is found by maximizing 
\begin{align}
	f(\kappa_k^2|\hat\phi_k) = & \frac{1}{2}\log\left( |\tilde{\mathbf{Q}}_k| \right) - \frac{1}{8\pi\hat{\phi}_k} \text{Tr}\left( \tilde{\mathbf{Q}}_k \text{E}\left(\mathbf{w}_k\mathbf{w}_k'|\mathbf{y},\hat{\boldsymbol\theta}^{(s)}\right) \right) \label{eq:optim_kappa}
\end{align}
with respect to $\kappa_k^2$. Convergence of the EM algorithm occurs when the average of the squared difference between $\boldsymbol{\theta^{(s)}}$ and $\boldsymbol{\theta}^{(s-1)}$ drops below a given stopping rule tolerance, $\epsilon$, which we set to 0.001 in our analyses. This value is based on a study of different stopping rule tolerances using simulated datasets (see \textbf{\ref{app:epsilon}}). 

\subsubsection{Identities for computational efficiency} \label{sec:computation}

In order to perform each step of the EM algorithm, computationally-efficient identities are used in order to reduce time and memory requirements. To begin, to calculate $\boldsymbol{\mu}_k = \frac{1}{\sigma^2}\boldsymbol{\Sigma}_k \boldsymbol{\Psi}_k'\mathbf{X}_k'\mathbf{y}$, we solve the linear equation
\begin{align*}
    \boldsymbol{\Sigma}_k^{-1}\boldsymbol{\mu}_k & = \frac{1}{\sigma^2} \boldsymbol{\Psi}_k'\mathbf{X}_k'\mathbf{y}.
\end{align*}
Here, $\boldsymbol{\Sigma}_k^{-1}$ is sparse and inexpensive to calculate, and the Eigen C++ library \citep*{guennebaud2010eigenweb} has efficient routines for solving linear equations of the form $\mathbf{Ax} = \mathbf{b}$ when $\mathbf{A}$ is sparse. 

Finding the MLE values for $\phi_k$ and $\kappa_k^2$ does not calculating the posterior precision in full, but approximating a trace for the matrix product involving the block-diagonal precision task component $\boldsymbol{\Sigma}_k^{-1}$. This allows for operations to be performed on smaller matrices and in parallel across tasks. The EM algorithm requires finding a trace of the form $\text{Tr}(\mathbf{A}\text{E}(\mathbf{w}_k\mathbf{w}_k'|\mathbf{y},\hat{\boldsymbol{\theta}}^{(s)}))$ for the hyperparameters $\phi_k$ and $\kappa_k^2$. This trace form can be simplified as
\begin{align*}
    \text{Tr}\left(\mathbf{A}\text{E}(\mathbf{w}_k\mathbf{w}_k'|\mathbf{y},\hat{\boldsymbol{\theta}}^{(s)})\right) & = \text{Tr}\left(\mathbf{A}\left(\boldsymbol{\Sigma}_k - \boldsymbol{\mu}_k\boldsymbol{\mu}_k'\right)\right) \\
    & = \text{Tr}\left( \mathbf{A}\boldsymbol{\Sigma}_k\right) - \text{Tr}\left(\mathbf{A}\boldsymbol{\mu}_k \boldsymbol{\mu}_k'\right) \\
    & = \text{Tr}\left( \mathbf{A}\boldsymbol{\Sigma}_k\right) - \text{Tr}\left(\boldsymbol{\mu}_k' \mathbf{A}\boldsymbol{\mu}_k \right) \\
    & = \text{Tr}\left( \mathbf{A}\boldsymbol{\Sigma}_k\right) - \boldsymbol{\mu}_k' \mathbf{A}\boldsymbol{\mu}_k
\end{align*}
The calculation of $\text{Tr}(\mathbf{A}\boldsymbol{\Sigma}_k)$ is done via approximation using the Hutchinson trace estimator \citep{hutchinson1989stochastic}. To begin, consider the following linear system $\boldsymbol{\Sigma}^{-1}\mathbf{P} = \mathbf{V}$, where $\mathbf{P}$ is a lower-dimensional approximation to the inverse of $\boldsymbol{\Sigma}$, $\mathbf{V} \in \mathbb{R}^{nT \times N_s}$ is a random matrix taking the values of 1 and -1 with equal probability, and $N_s$ is the size of the estimator. We choose to use $N_s = 50$ in our analyses because it limits the relative error of the approximation to about 10\% with 95\% probability \citep{skorski2021modern}. The symbolic representation of the Cholesky decomposition of $\boldsymbol{\Sigma}$ is calculated before beginning the EM algorithm, reducing the memory and time requirements at each iteration when solving for $\mathbf{P}$. We update $\mathbf{V}$ with new random values at each EM iteration to reduce sampling bias in the approximation. Combining these strategies results in the trace approximation
$$\text{Tr}(\mathbf{A}\boldsymbol{\Sigma}) \approx \frac{1}{N_s} \text{Tr}\left( \mathbf{P'AV} \right).$$
This method reduces the computation time of estimating the trace from $\mathcal{O}(n^3)$ to $\mathcal{O}(N_s^3)$ where $N_s \ll n$.

The speed of the EM algorithm is improved through the use of the squared extrapolation (SQUAREM) method developed by \cite{varadhan2004squared} in order to reduce the total number of evaluations of the fixed-point function $\boldsymbol{\theta}^{(s+1)} = f(\boldsymbol{\theta}^{(s)})$ at any EM algorithm iteration $s$. Code for the SQUAREM method in C++ is adapted from the \texttt{ashr} R package \citep{stephens2022ashr}.

All computations are implemented in the \texttt{BayesfMRI} package in R \citep{Mejia2022BayesfMRI}, which uses the \texttt{Rcpp} and \texttt{RcppEigen} packages \citep{eddelbuettel2011Rcpp,eddelbuettel2013seamless,bates2013fast} to implement C++ code. We rely upon the Eigen template library for linear algebra in C++ \citep{guennebaud2010eigenweb} to achieve significant speed improvements for linear algebra operations.

Time to convergence is substantially decreased by using the classical GLM estimates for $\mathbf{w}_k$ to inform the initial hyperparameter estimates $\hat{\boldsymbol\theta}^{(0)}$. Initial values $\hat{\phi}_k^{(0)}$ and $\hat{\kappa}_k^{2(0)}$ are found by iteratively solving the following two equations until a threshold of convergence is met.
\begin{align*}
    \hat{\phi}_k^{(0)} & = \frac{1}{4\pi n} \hat{\mathbf{w}}_k'\tilde{\mathbf{Q}}_k \hat{\mathbf{w}}_k , &
    \hat{\kappa}_k^{2(0)} & = \argmax_{\kappa_k^2} \frac{1}{2} \log |\tilde{\mathbf{Q}}_k| - \frac{1}{8\pi \hat{\phi}_k^{(0)}} \hat{\mathbf{w}}_k'\tilde{\mathbf{Q}}_k \hat{\mathbf{w}}_k
\end{align*}
Finally, the variance in the residuals from the classical GLM is used as the initial value for the observed variance ($\hat\sigma^{2(0)}$). This computationally-efficient algorithm allows for rapid convergence to Bayesian estimates for the hyperparameters at the subject level, which can then be combined to allow for group-level inference.

\subsection{Multi-subject analyses}

\cite{mejia2020bayesian} develops a joint multi-subject model that combines results across multiple single-subject models by taking a weighted average of their hyperparameters ($\boldsymbol{\theta}_G$), which results in the multi-subject posterior distribution for the SPDE hyperparameters. Our own multi-subject model differs because each single-subject result returns a point estimate of the maximum \textit{a posteriori} values for $\boldsymbol{\theta}_m$ for each subject $m$, rather than a distribution. For our multi-subject model, we assume that the values found in the single-subject analysis are noisy estimates of the fixed true values of $\boldsymbol{\theta}_G$. Therefore, in order to find the maximum \textit{a posteriori} values for $\boldsymbol{\theta}_G$ across all $M$ subjects, the multi-subject estimate of the SPDE hyperparameters is found as
\begin{align*}
    \hat{\boldsymbol{\theta}}_G & = (\phi_{1,G},\ldots,\phi_{K,G}, \kappa_{1,G}^2, \ldots, \kappa_{K,G}^2, \sigma_G^2)' = \sum_{m = 1}^M \lambda_m\hat{\boldsymbol{\theta}}_m,
\end{align*}
where $\hat{\boldsymbol{\theta}}_m$ is a vector of the maximum \textit{a posteriori} values found for each subject $m$ and $\lambda_m$ is a weight assigned to each subject's observation, which can be assigned based on the amount of data in each subject's scan and their respective noise levels. We then adjust the posterior distributions of $\mathbf{w}_m$ found for each of the $M$ subjects to reflect these updated hyperparameter values, calculating $\mathbf{Q}_G$ as in Equation (\ref{eq:blockQtilde}) using $\hat{\boldsymbol{\theta}}_G$:
\begin{align*}
    \mathbf{w}_{m}^* | \mathbf{y}, \boldsymbol{\theta}_{G} & \sim \text{Normal}\left(\boldsymbol{\mu}_{m}, \boldsymbol{\Sigma}_{m} \right), &
    \boldsymbol{\Sigma}_{m} & = \left( \mathbf{Q}_{G} + \frac{1}{\sigma_G^2}\boldsymbol{\Psi}' \mathbf{X}_m' \mathbf{X}_m \boldsymbol{\Psi} \right)^{-1},  &
    \boldsymbol{\mu}_{m} & = \frac{1}{\sigma_G^2}\boldsymbol{\Sigma}_{m} \boldsymbol{\Psi}'\mathbf{X}_m'\mathbf{y}.
\end{align*}
Next, $H$ samples are taken from these updated posterior distributions for the subjects, and a contrast vector $\mathbf{c} \in \mathbb{R}^{KM \times 1}$ is specified in order to find the multi-subject posterior distribution for a linear combination of tasks and subjects. For example, if an equally-weighted average of the task activation amplitudes across all subjects for the first task is desired, then the contrast vector would take the form
\begin{align*}
    \mathbf{c}  = (\mathbf{c}_1^{\prime},\mathbf{c}_2^{\prime},\ldots, \mathbf{c}_M^{\prime})', \quad
    \mathbf{c}_m  = (1/M,0,\ldots,0)' \in \mathbb{R}^K.
\end{align*}
A contrast matrix is then created using the Kronecker product such that $\mathbf{C} = \mathbf{I}_N \otimes \mathbf{c}' \in \mathbb{R}^{N \times NKM}$ and the multi-subject posterior draws for the average activation amplitude for task 1 is found to be $\boldsymbol{\beta}^* = \mathbf{C} \boldsymbol{\beta}^{(1:H)}$, where $\boldsymbol{\beta}^{(1:H)} = (\boldsymbol{\beta}_1^* \cdots  \boldsymbol{\beta}_H^*) \in \mathbb{R}^{NKM \times H}$, and $\boldsymbol{\beta}_h^* \in \mathbb{R}^{NKM \times 1} = (\boldsymbol{\beta}_{1,h}^{*\prime} \cdots \boldsymbol{\beta}_{M,h}^{*\prime})'$ is a column-bound matrix constructed where $\boldsymbol{\beta}_{m,h}^{*'} \in \mathbb{R}^{NK \times 1} = \boldsymbol{\Psi}\mathbf{w}_{m,h}^*$ and $\mathbf{w}_{m,h}^* \in \mathbb{R}^{nK \times 1}$ is the vector of the $h$th draw from the posterior distribution of $\mathbf{w}_m^*$. This principled combination of single-subject results provides the distribution of a multi-subject average activation amplitude effect, which can be used with the excursions method to determine population-level areas of activation, as discussed in the next section. 

\subsection{Determination of Activation} \label{sec:activations}

In addition to assessing an estimate of an effect size for brain activation, it is also desirable to determine which regions of the brain are significantly activated above a given threshold. The Bayesian GLM proposed by \cite{mejia2020bayesian} and validated by \cite{spencer2022spatial} takes advantage of the posterior distributions for the activation effects at all data locations on the cortical surface or subcortical volume through the use of the excursions set method proposed by \cite{bolin2015excursion}. The excursions method examines the joint distribution of the activation $\boldsymbol{\beta}_k = (\beta_{1,k},\ldots,\beta_{V,k})'$ for task $k$ and location $v = 1,\ldots,V$. Specifically, the excursions method finds the largest set $\mathcal{V} = (a_1,a_2,\ldots,a_p)'$ of data locations such that $P(\beta_{a_1,k} > \gamma, \beta_{a_2,k} > \gamma, \ldots, \beta_{a_p,k} > \gamma) \geq 1-\alpha$ for a given threshold $\gamma$ and credible level $1-\alpha$. This is done using the posterior distributions, either at the single- or multi-subject, for the activation amplitudes $\boldsymbol\beta_k$. Implementation of the excursions method has been done in the \texttt{excursions} package in R \citep{bolin2015excursion,bolin2017quantifying,bolin2018calculating}, which is used by the \texttt{BayesfMRI} package.

\section{Simulated Data Analysis}
\label{sec:simulated}

Data were simulated for cortical surface in R using our \texttt{brainSim}\footnote{https://github.com/danieladamspencer/brainSim/tree/1.0} package. This package generates functional MRI data on the cortical surface using design matrices with spatially-dependent coefficients and temporally autoregressive error terms. The software can simulate data for a variable number of subjects, scanning sessions, and scanning runs, which allows for nested individual, session, and run variances from a ``true" global coefficient. 

\subsection{Single-subject analysis}

We begin with simulations for a single subject, simulating data under nine different conditions. These conditions are the combination of resampling resolution at $n = \{2,000; 5,000; 10,000\}$ cortical surface vertices per hemisphere, and setting the number of tasks to be $K = \{2,5,8\}$. Under these conditions, the length of the scan is set to $T = 300$, and the maximum value for the spatial coefficient is set to 2. Under these conditions, the average value for the nonzero true coefficients is approximately 0.5 due to spatial smoothing in the data generation. The error term was set to have a variance of 1, and the error was assumed to be temporally independent.

\begin{figure}
    \centering
    \begin{subfigure}[b]{\textwidth}
        \centering
        \includegraphics[trim={0 3cm 0 0},clip,width=0.88\textwidth]{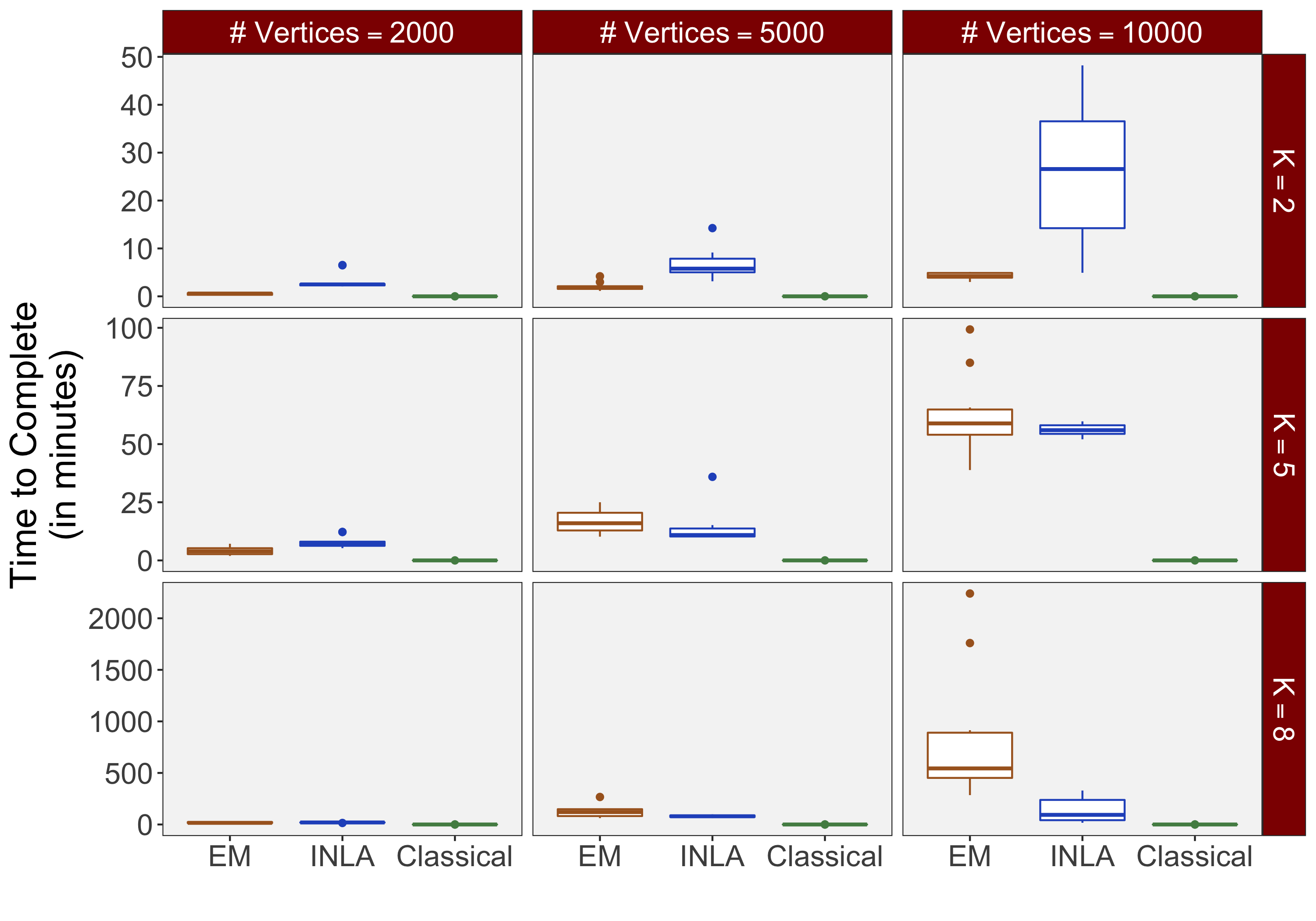}
        \caption{Time to run a single-subject analysis after preprocessing}
        \label{fig:time_sim}
    \end{subfigure}
    \begin{subfigure}[b]{\textwidth}
        \centering
        \includegraphics[trim={0 3cm 0 0},clip,width=0.88\textwidth]{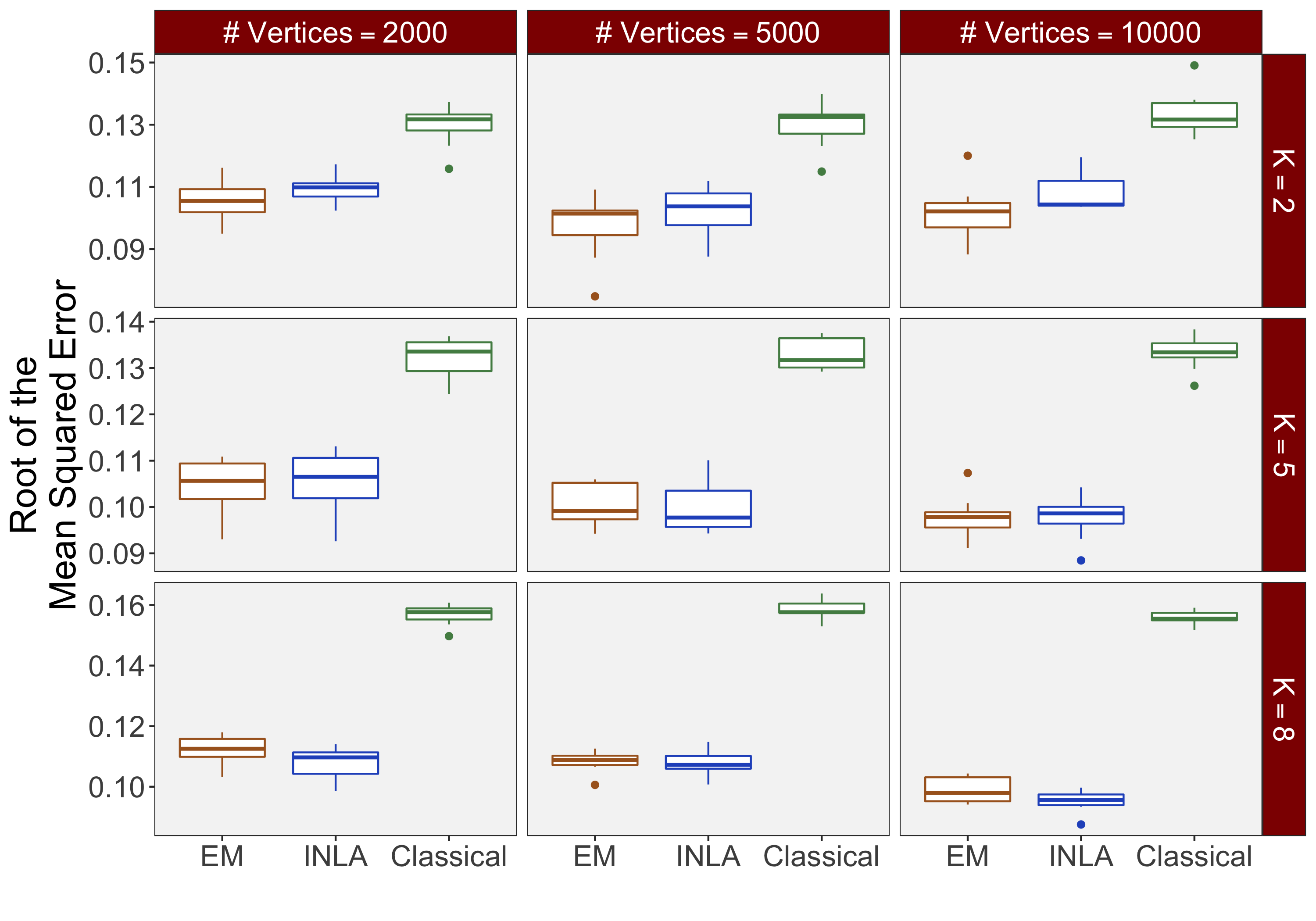}
        \caption{The square root of the mean squared error in single-subject analyses}
        \label{fig:rmse_sim}
    \end{subfigure}
    \caption{Performance comparison for the single-subject, simulated data under the four different simulation scenarios. Each boxplot summarizes the distribution of times and errors for ten simulated datasets under each of nine scenarios analyzed using the three methods. RMSE performance of the INLA method is approximate, as the algorithm did not converge for 13 of the 90 datasets. Twelve of these cases occurred when $K = 2$.}
\end{figure}

\begin{figure}
    \begin{subfigure}{0.49\textwidth}
        \begin{tabularx}{\textwidth}{c|c|c|}
    		\multicolumn{1}{c}{} & 
    		\multicolumn{1}{c}{\textbf{1 Run}} & 
    		\multicolumn{1}{c}{\textbf{2-Run Average}} \\ 
    		\cline{2-3} 
    		\rotatebox[origin=l]{90}{\quad \textbf{Truth} \qquad \,} &
    		\Includegraphics[width=.4\textwidth]{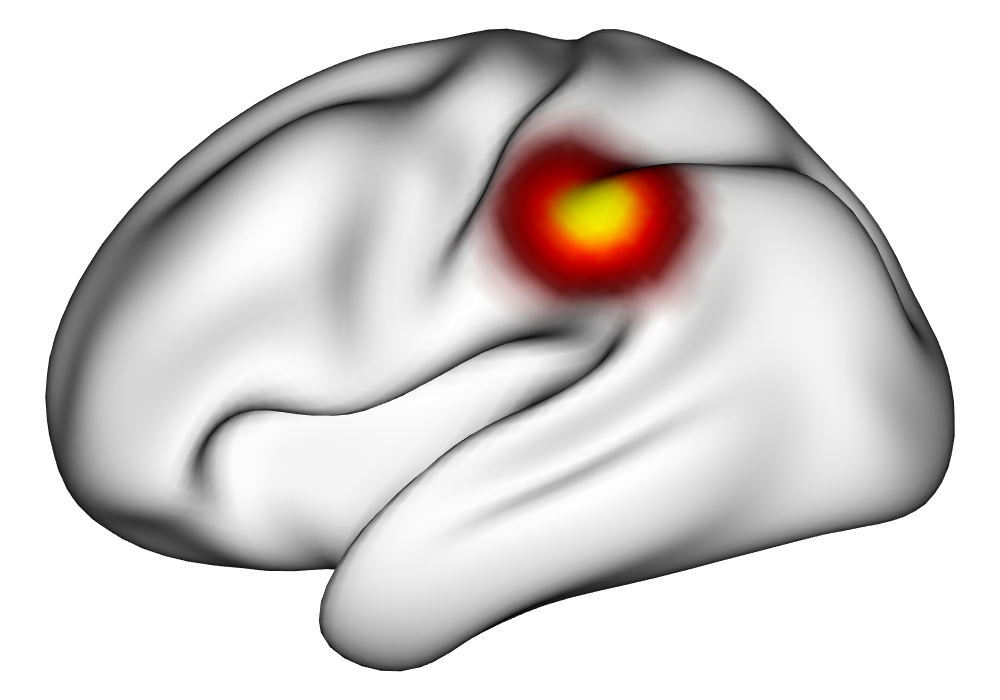} &
    		\Includegraphics[width=.4\textwidth]{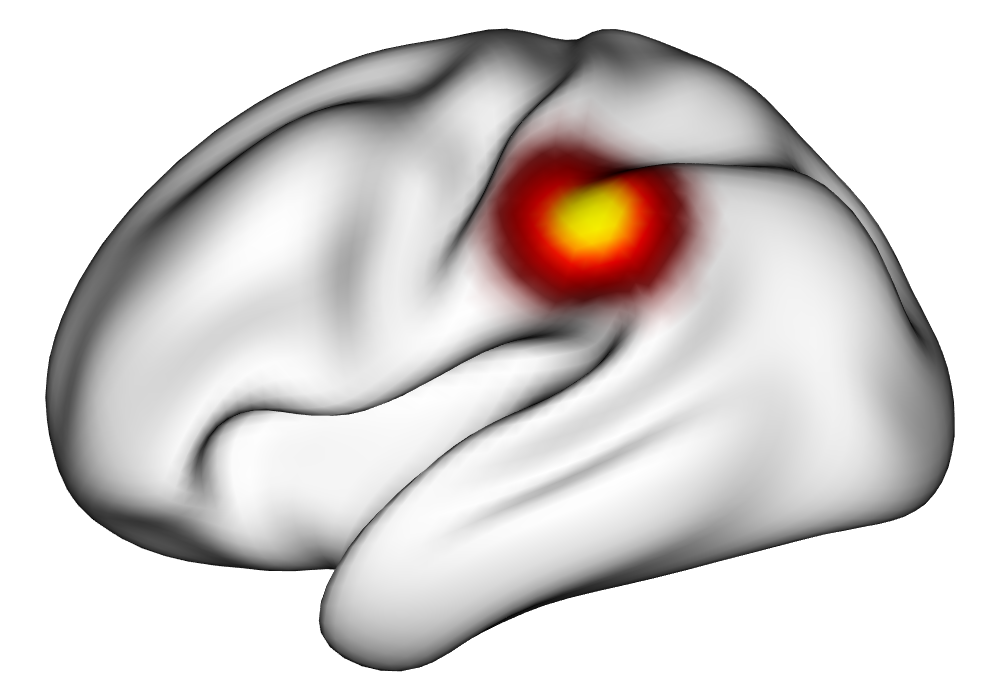} \\
    		\cline{2-3}
    		\multicolumn{1}{c}{\rotatebox[origin=l]{90}{\qquad}} & \multicolumn{2}{c}{\includegraphics[width=.5\textwidth]{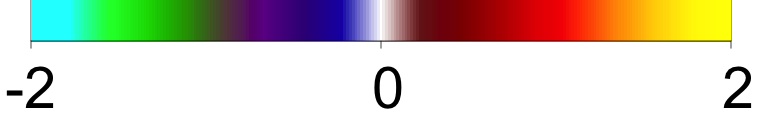} \vspace{2mm}} \\
    		\cline{2-3}
    		\rotatebox[origin=l]{90}{\quad \textbf{EM} \qquad \,} &
    		\Includegraphics[width=.4\textwidth]{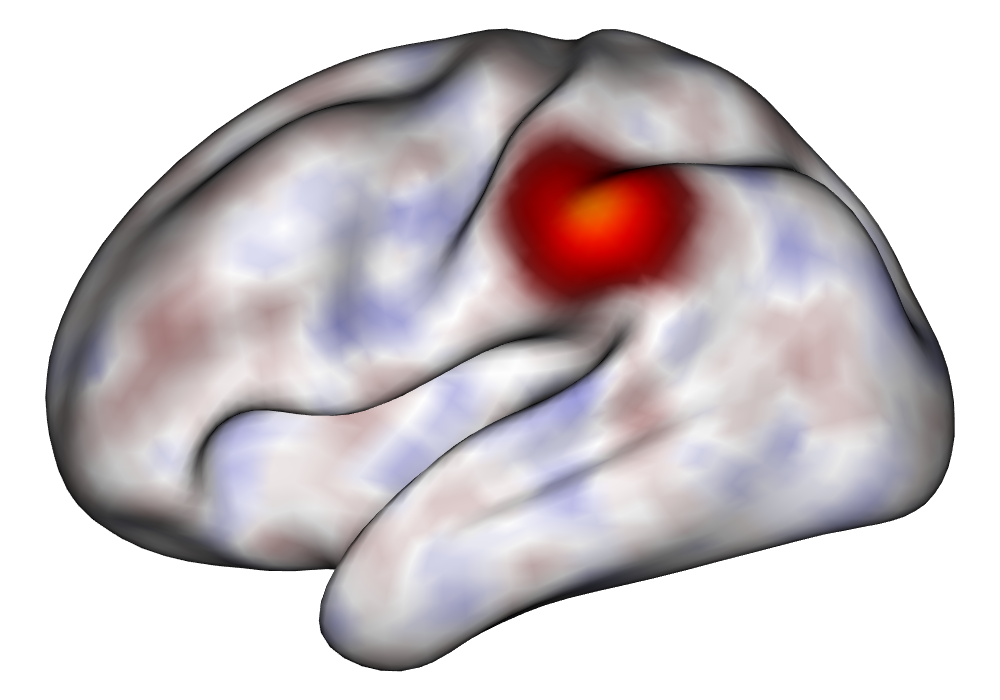} &
    		\Includegraphics[width=.4\textwidth]{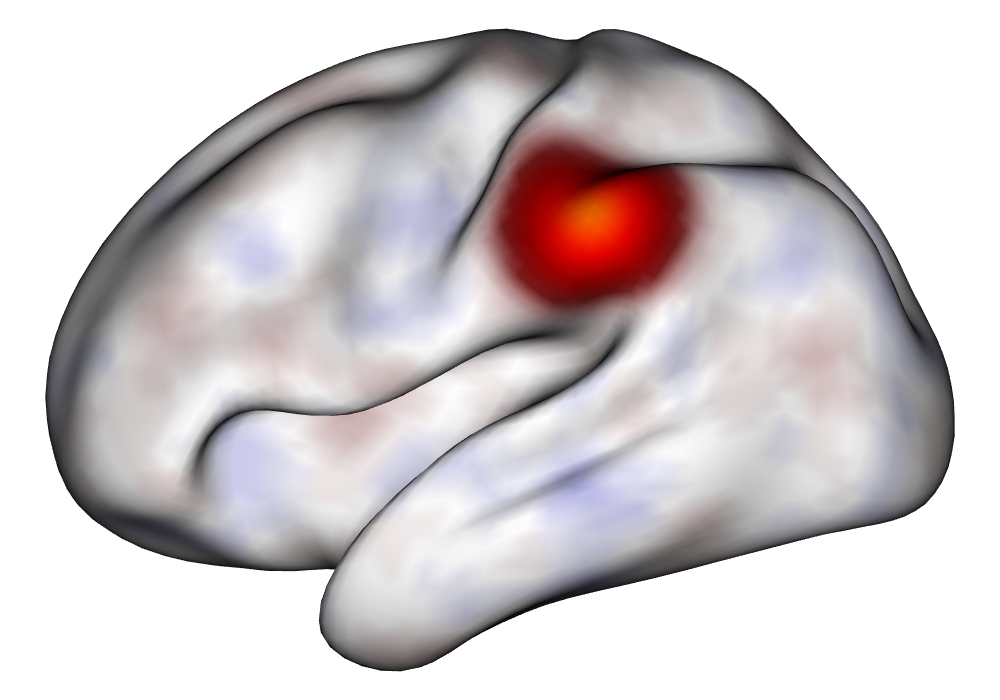} \\
    		\cline{2-3}
    		\rotatebox[origin=l]{90}{\quad \textbf{INLA} \qquad \,} &
    		\Includegraphics[width=.4\textwidth]{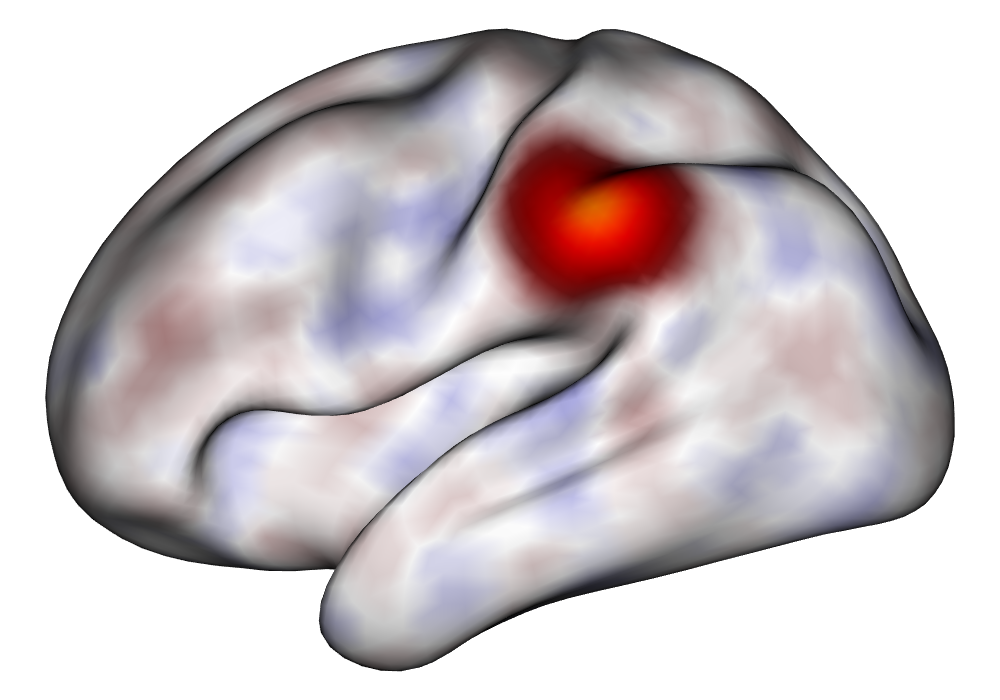} &
    		\Includegraphics[width=.4\textwidth]{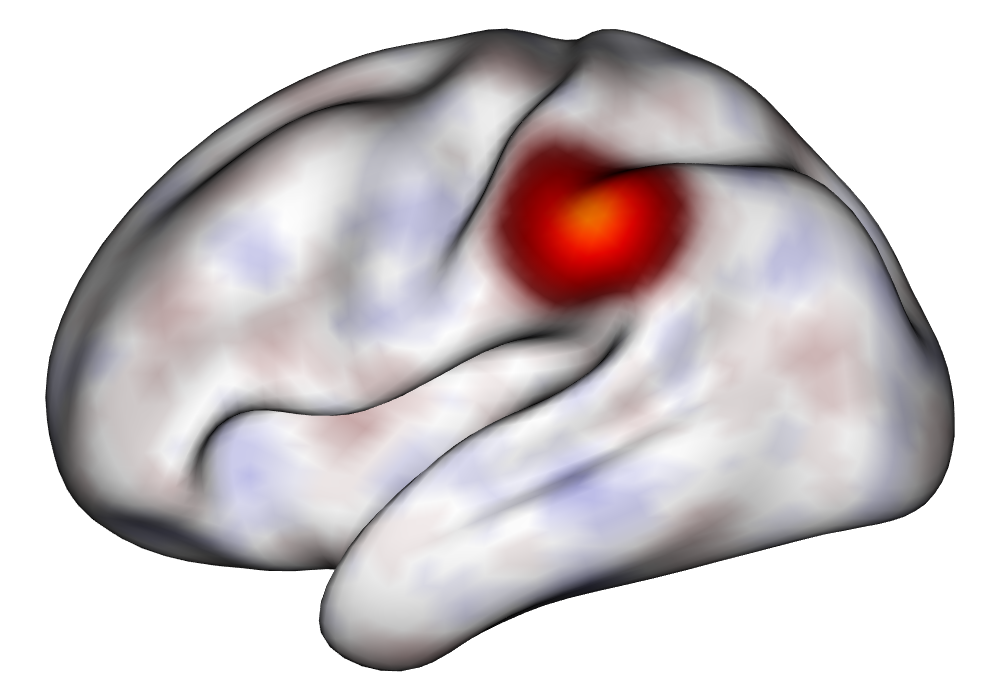} \\ 
    		\cline{2-3}
    		\rotatebox[origin=l]{90}{\, \textbf{Classical}} &
    		\Includegraphics[width=.4\textwidth]{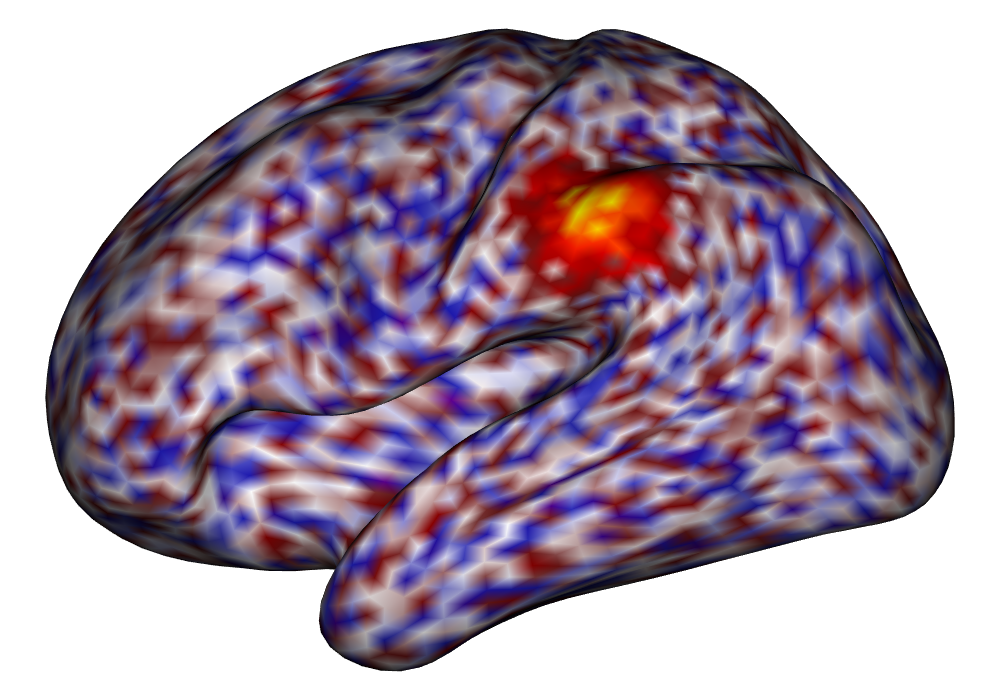} &
    		\Includegraphics[width=.4\textwidth]{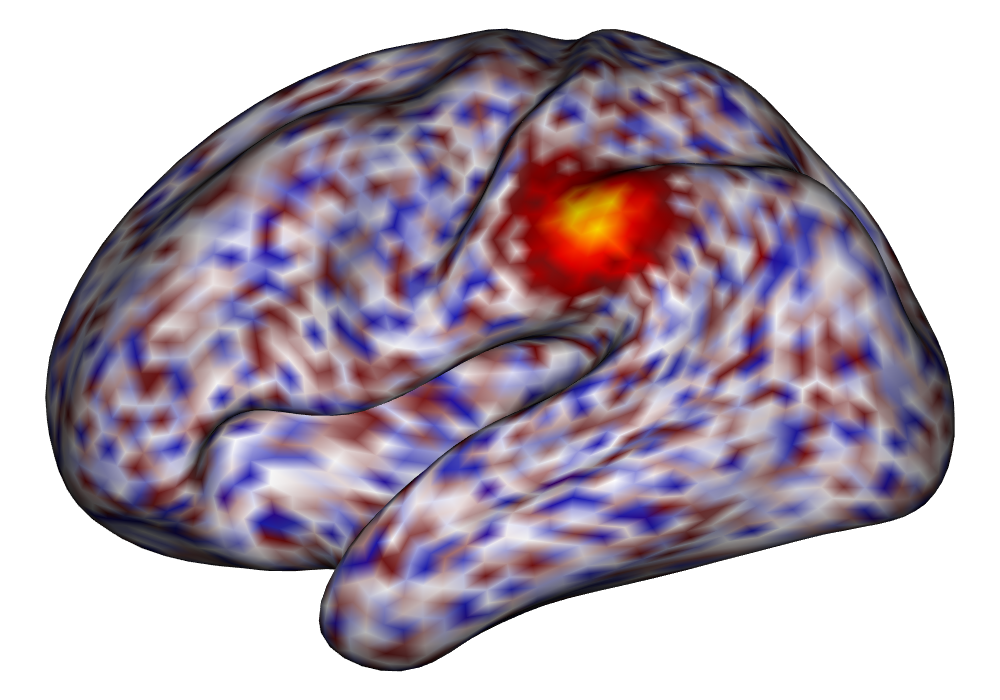} \\ 
    		\cline{2-3}
    		\multicolumn{1}{c}{\rotatebox[origin=l]{90}{\qquad}} & \multicolumn{2}{c}{\includegraphics[width=.5\textwidth]{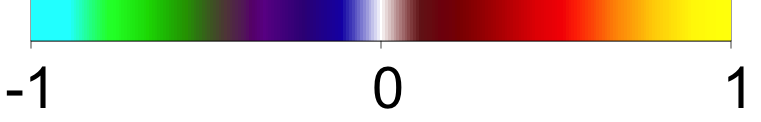}}
	    \end{tabularx}
	    \caption{Amplitude of activations}
	    \label{fig:sim_estimations}
    \end{subfigure}
    \begin{subfigure}{0.49\textwidth}
        \begin{tabularx}{\textwidth}{c|c|c|}
            \multicolumn{1}{c}{} & 
            \multicolumn{1}{c}{\textbf{1 Run}} & 
            \multicolumn{1}{c}{\textbf{2-Run Average}} \\ 
    		\cline{2-3} 
    		\rotatebox[origin=l]{90}{\quad \textbf{Truth} \qquad \,} &
    		\Includegraphics[width=.4\textwidth]{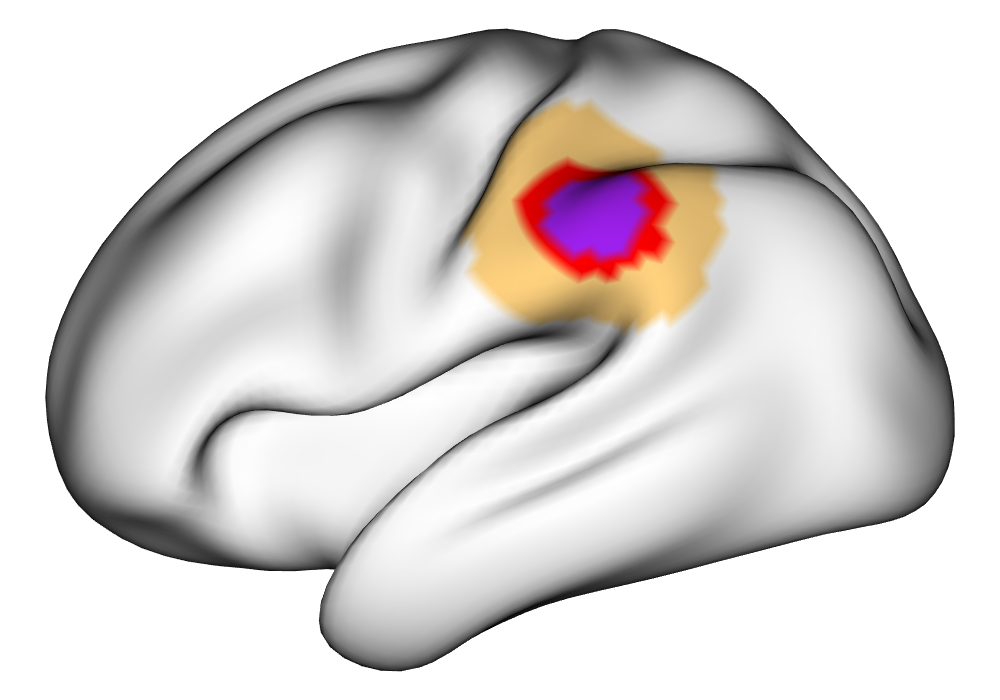} &
    		\Includegraphics[width=.4\textwidth]{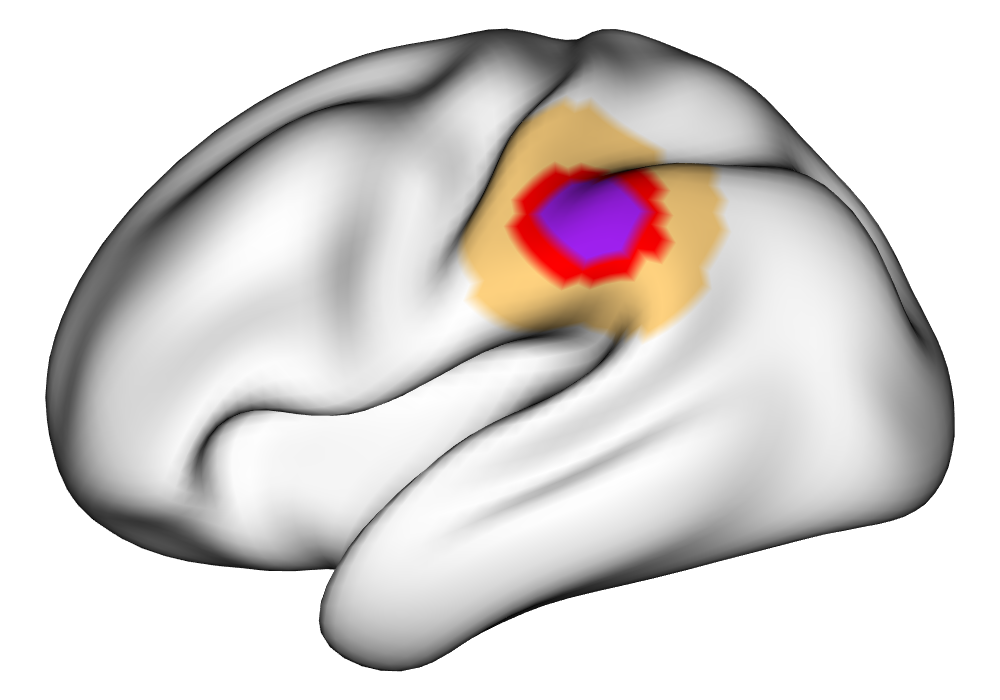} \\
    		\cline{2-3}
    		\multicolumn{1}{c}{\rotatebox[origin=l]{90}{\qquad}} & \multicolumn{2}{c}{$\gamma =$ \textcolor[HTML]{FFD27F}{$\blacksquare$} 0\% 
           \textcolor[HTML]{FF0000}{$\blacksquare$} 0.5\% 
           \textcolor[HTML]{A020F0}{$\blacksquare$} 1\%} \vspace{2mm} \\
    		\cline{2-3}
    		\rotatebox[origin=l]{90}{\quad \textbf{EM} \qquad \,} &
    		\Includegraphics[width=.4\textwidth]{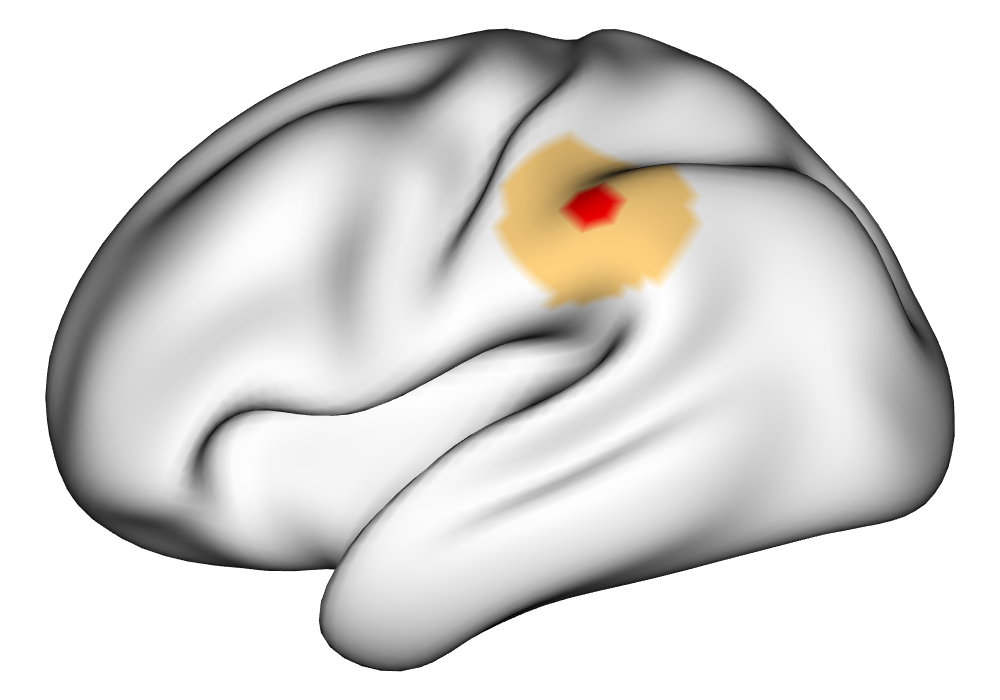} &
    		\Includegraphics[width=.4\textwidth]{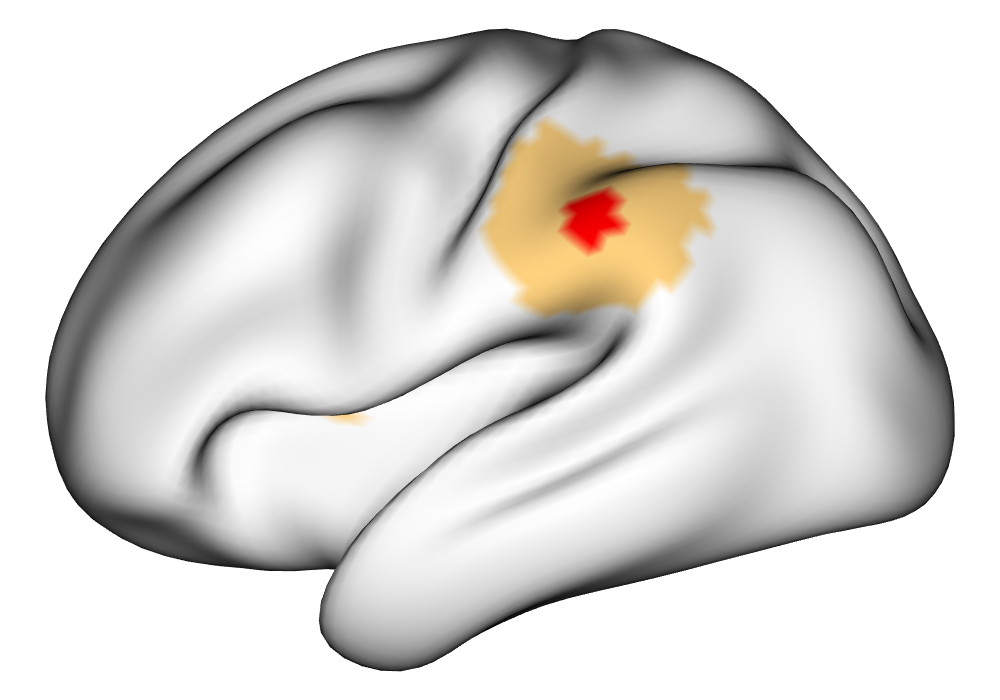} \\
    		\cline{2-3}
    		\rotatebox[origin=l]{90}{\quad \textbf{INLA} \qquad \,} &
    		\Includegraphics[width=.4\textwidth]{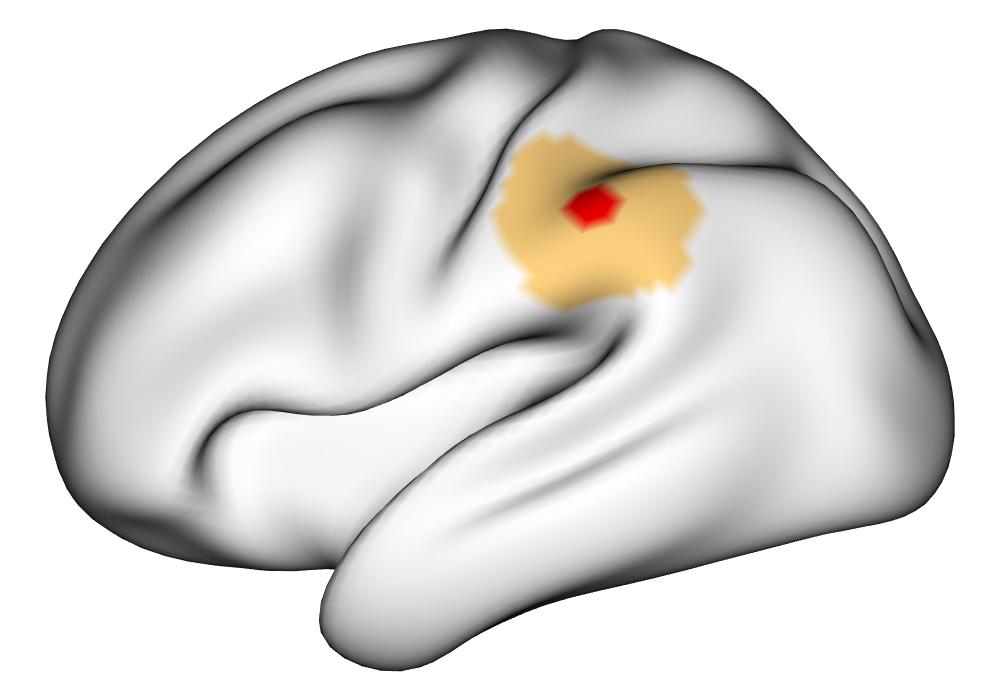} &
    		\Includegraphics[width=.4\textwidth]{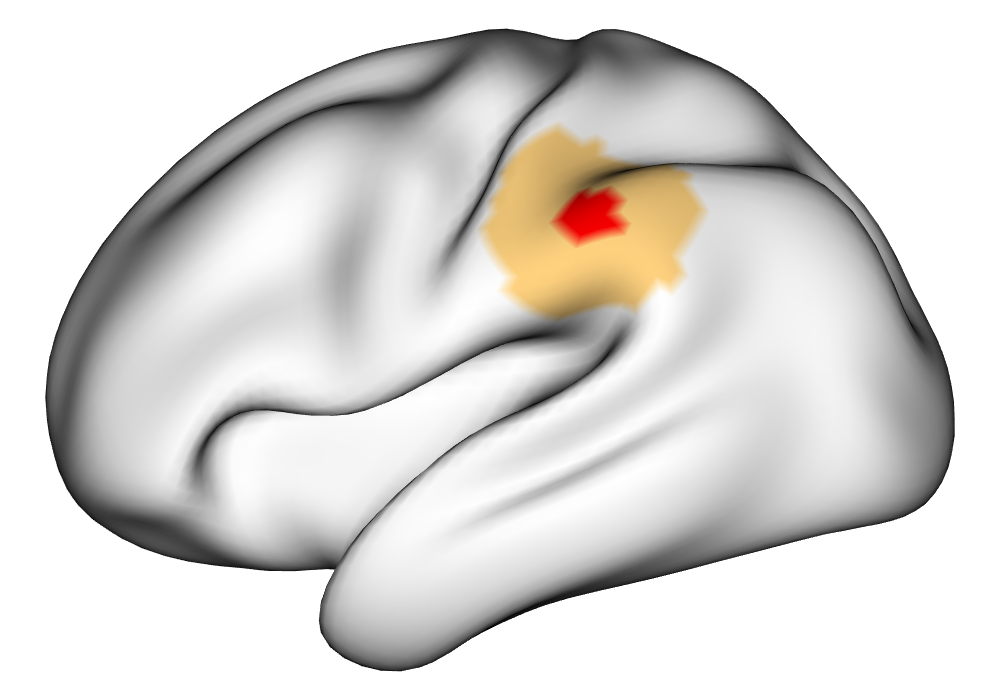} \\ 
    		\cline{2-3}
    		\rotatebox[origin=l]{90}{\, \textbf{Classical}} &
    		\Includegraphics[width=.4\textwidth]{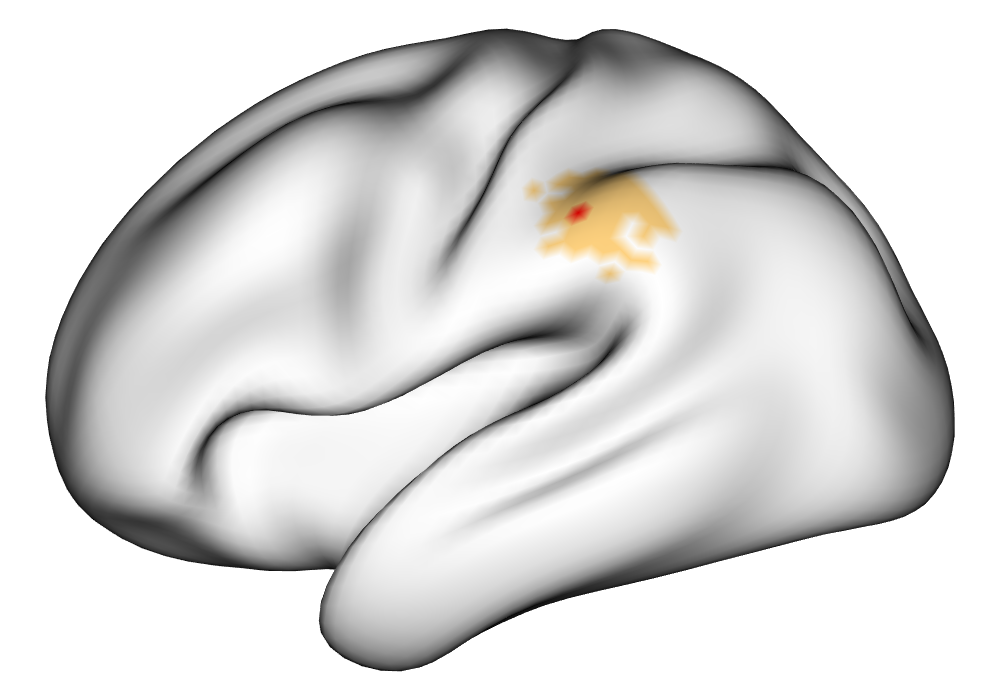} &
    		\Includegraphics[width=.4\textwidth]{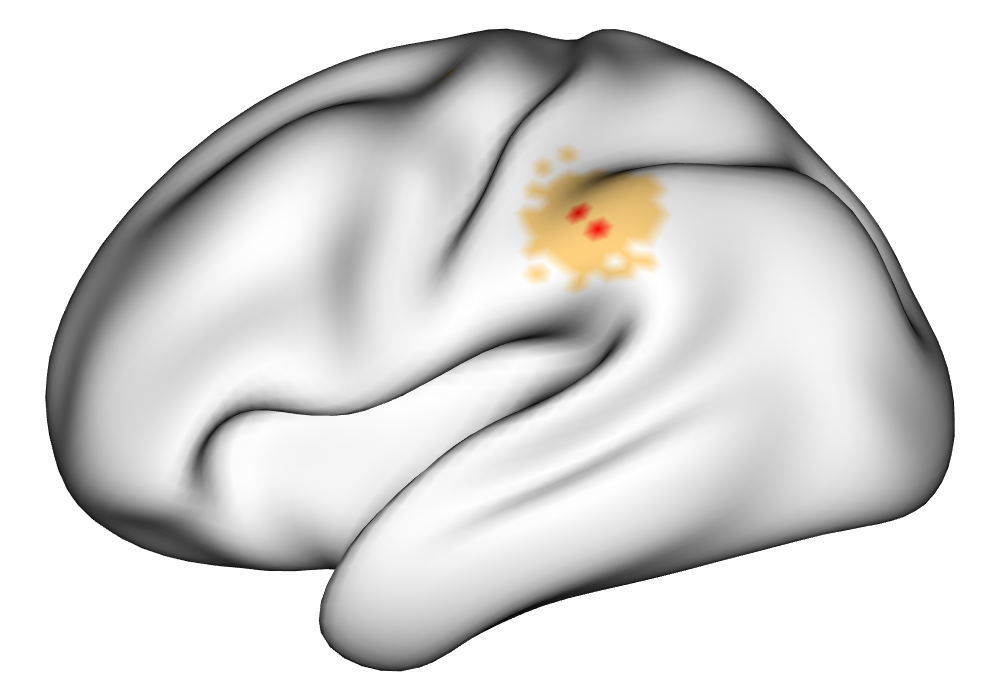} \\
    		\cline{2-3}
    		\multicolumn{1}{c}{\rotatebox[origin=l]{90}{\qquad}} & \multicolumn{2}{c}{$\gamma =$ \textcolor[HTML]{FFD27F}{$\blacksquare$} 0\% 
           \textcolor[HTML]{FF0000}{$\blacksquare$} 0.5\% 
           \textcolor[HTML]{A020F0}{$\blacksquare$} 1\%}
		\end{tabularx}
		\caption{Areas of Activation}
		\label{fig:sim_activations}
    \end{subfigure}
    \caption{Cortical surface estimates for the amplitude of activation and areas of activation in percent signal change for the INLA and EM implementations of the Bayesian GLM and the classical GLM. True amplitudes of activation and areas of activation are shown for comparison. Different color scales are shown for the true values and the estimates of the amplitudes of activation because the estimates are attenuated, as one would expect for models assuming sparsity. For the Bayesian GLM implementations, the areas of activation are found using the excursions method, and are found as areas deemed jointly greater than threshold $\gamma$ with probability 0.01. For the classical GLM, the areas of activation are found by fixing the false discovery rate at 0.01.}
    \label{fig:sim_est_and_act}
\end{figure}

Each of the nine conditions was used to generate 10 datasets, which were all fit to the classical GLM, and the SBSB GLM using the INLA and EM implementations. The INLA implementation allows for the use of the PARDISO linear algebra library \citep{pardiso-7.2a,pardiso-7.2b, pardiso-7.2c}, which allows for multiple processor threads to be used in expensive matrix solve computations. In these trials, the INLA implementation is allowed to use 6 threads as an advantageous setting for analysis. We compare the results in terms of the amount of time to perform the analysis after preprocessing (\textbf{Figure \ref{fig:time_sim}}) and in terms of their root mean squared error (RMSE) (\textbf{Figure \ref{fig:rmse_sim}}). These results show that the EM implementation is faster in terms of computation time than the INLA implementation when the number of tasks ($K$) is relatively small. The speeds are similar when $K=5$, but the INLA implementation is currently faster than the EM implementation due to faster solve times stemming from its use of the PARDISO library. Implementing the EM algorithm to take advantage of a solver that allows the use of multiple cores is likely to further increase its speed, potentially outstripping the INLA implementation in terms of computation time across all settings. However, no parallelization libraries are currently available across UNIX and Windows systems that are also compatible with the \texttt{RcppEigen} library. Using the RMSE measure, the EM implementation's accuracy is similar to that of the INLA implementation, though it is important to note that the INLA implementation failed to converge in the analysis of 13 of the 90 datasets, and thus did not produce estimates in these cases. Twelve of these datasets had the condition that $K = 2$, and of those, 7 had $n = 10000$. This suggests that the EM algorithm is more stable than the INLA implementation when evaluating experiments with few tasks as resolution increases. A visualization of an estimated and true coefficient surface for the sampling condition when $n = 5000$ and $K = 5$ can be seen in \textbf{Figure \ref{fig:sim_estimations}}.

Under the simulation condition with resolution $n = 5000$ and number of tasks $K = 5$, simulated data were generated for two runs in a single session to allow for sharing of the hyperparameters $\boldsymbol\theta = (\kappa_1,\ldots,\kappa_5,\phi_1,\ldots,\phi_5,\sigma^2)'$ across the runs. The INLA and EM implementations of the Bayesian GLM were then used to analyze the first run by itself and the two runs combined. The effect estimates and activations illustrating the difference in the 1- and 2-run analyses can be seen in \textbf{Figure \ref{fig:sim_est_and_act}}. As there is a small run effect simulated in the generation of the data, the true values of the activation amplitude are not identical between the two runs. Therefore, the true coefficient nonzero region in the single run is not centered exactly in the same location as the true nonzero coefficient region across both runs. In terms of both the coefficient estimation and activation, the EM implementation performs very similarly to the INLA implementation. The posterior standard deviations for the activation amplitudes for a single-run analysis are shown for the EM and INLA implementations in \textbf{Figure \ref{fig:sim_sd_comparison}}, which shows that the posterior standard deviations are slightly lower in the INLA implementation due to the prior imposed on the range and variance hyperparameters.

\subsection{Multi-subject analysis}

In order to compare the performance of the multi-subject analysis between the EM and INLA implementations of the SBSB GLM and the classical GLM, we generated a simulated population of 100 subjects with true amplitude maps with location translations from a population amplitude map by an average of 5mm. Each subject's data were simulated at a resolution of about 5,000 vertices per hemisphere, with a TR of 1 second for a total length of 300 seconds. Two tasks were included in the simulation, each with a maximum true signal-to-noise ratio of 2 in the simulation before the BOLD response and design matrices were scaled. Each subject's data was analyzed separately following the single subject analysis for the EM and INLA implementations of the SBSB GLM and the classical GLM as outlined in Section \ref{sec:single_subjectSBSBGLM}. Next, 100 different subsets of size 10 were drawn from the population, and the multi-subject analysis was performed using the single-subject analysis results from the subjects within the subset. For each multi-subject analysis, the square root of the mean squared error was found with respect to the true simulated population amplitudes. The Dice coefficient \citep{dice1945measures} is a measure of set overlap between sets $\mathcal{A}$ and $\mathcal{B}$ found as
\begin{align*}
    \text{Dice}(\mathcal{A},\mathcal{B}) =\frac{2 \times |\mathcal{A} \cap \mathcal{B}|}{|\mathcal{A}| + |\mathcal{B}|},
\end{align*}
where $|\mathcal{X}|$ is the number of elements in set $\mathcal{X}$ and $\mathcal{X} \cap \mathcal{Y}$ is the intersection of sets $\mathcal{X}$ and $\mathcal{Y}$. The Dice coefficient can take values between 0 and 1, and higher values indicate more overlap between sets. We calculate the Dice coefficient between the set of locations with true positive activation amplitudes in the population and the set of locations determined to have amplitudes significantly greater than zero ($\gamma = 0\%$) in each subset for each modeling method, as outlined in Section \ref{sec:activations}.

\begin{figure}
    \centering 
    \begin{subfigure}[b]{\textwidth}
        \begin{tabularx}{\textwidth}{c|c|c|c|c|}
    		\multicolumn{1}{c}{} & 
    		\multicolumn{1}{c}{\textbf{Truth}} & 
    		\multicolumn{1}{c}{\textbf{EM}} & 
    		\multicolumn{1}{c}{\textbf{INLA}} & 
    		\multicolumn{1}{c}{\textbf{Classical}} \\ 
    		\cline{2-5} 
    		\rotatebox[origin=l]{90}{\textbf{Amplitude}} &
    		\Includegraphics[width=.2\textwidth]{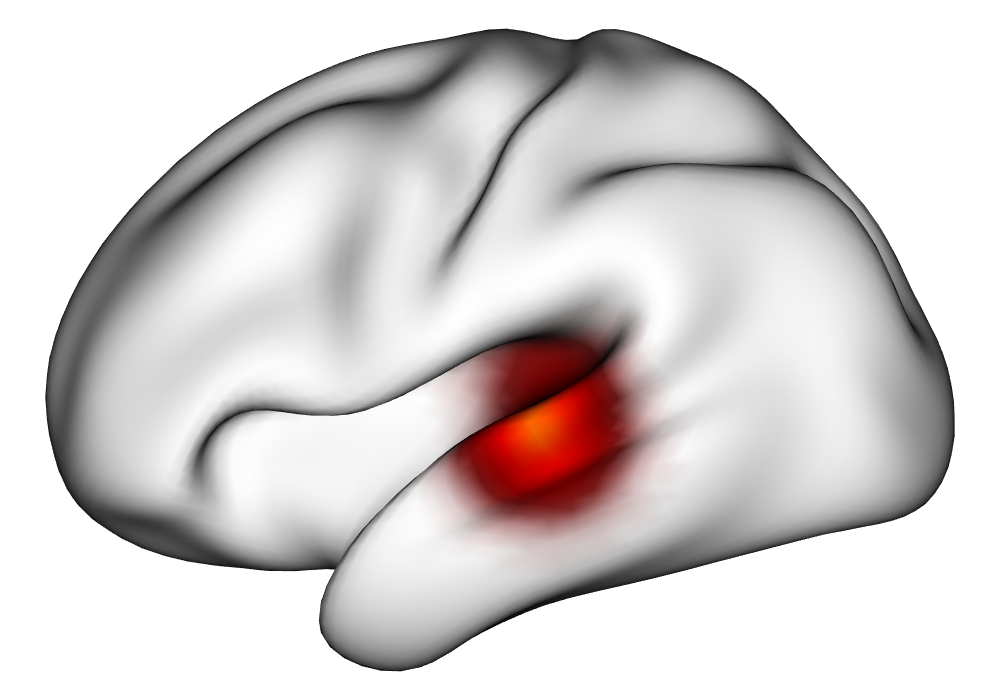} &
    		\Includegraphics[width=.2\textwidth]{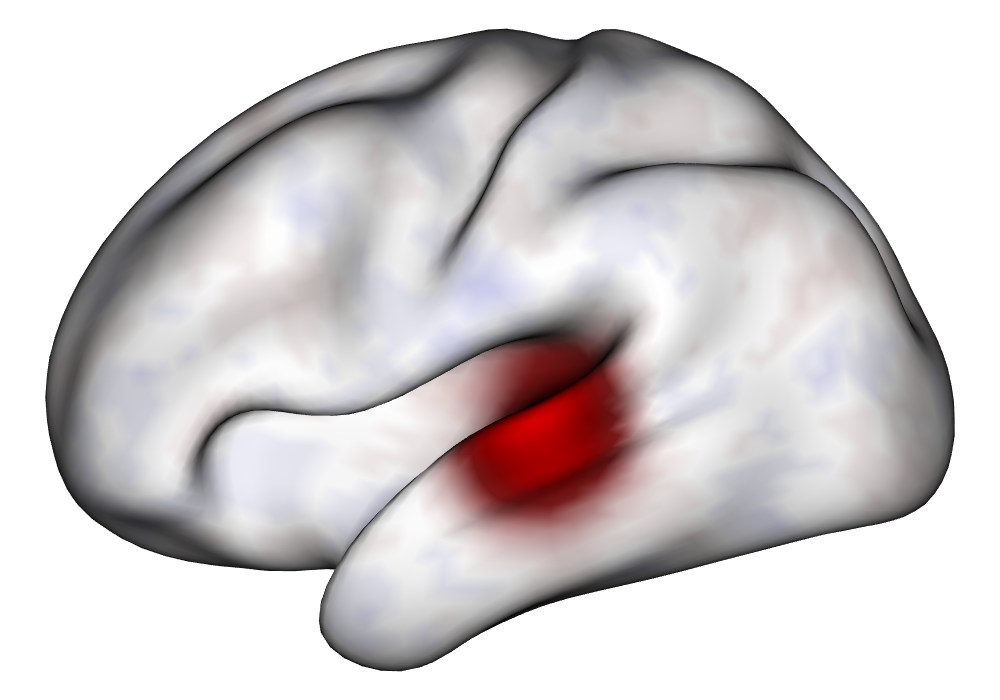} &
    		\Includegraphics[width=.2\textwidth]{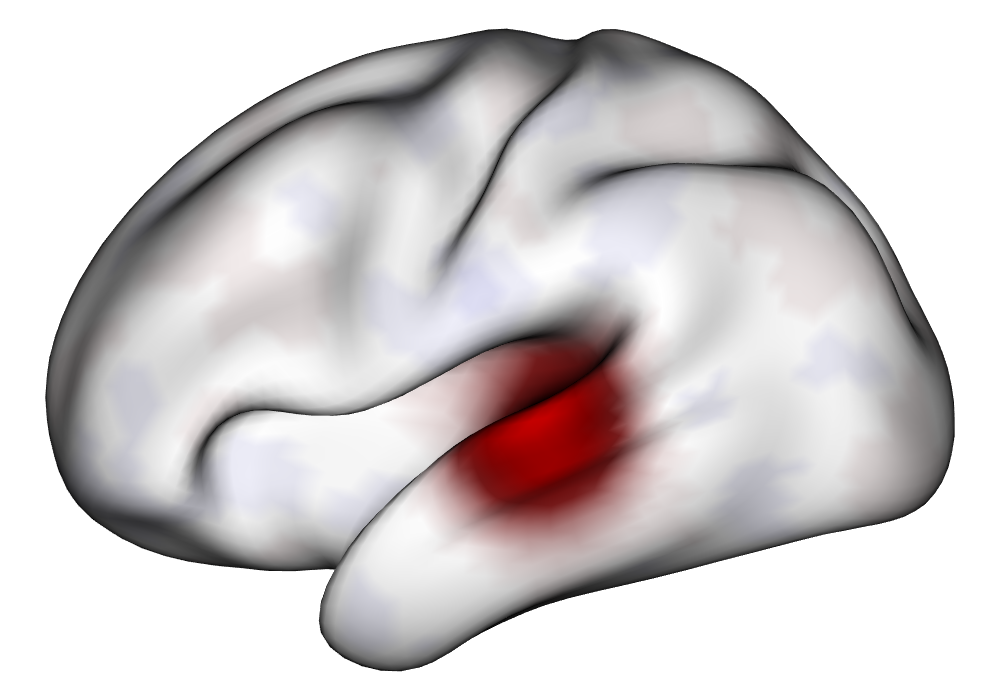} &
    		\Includegraphics[width=.2\textwidth]{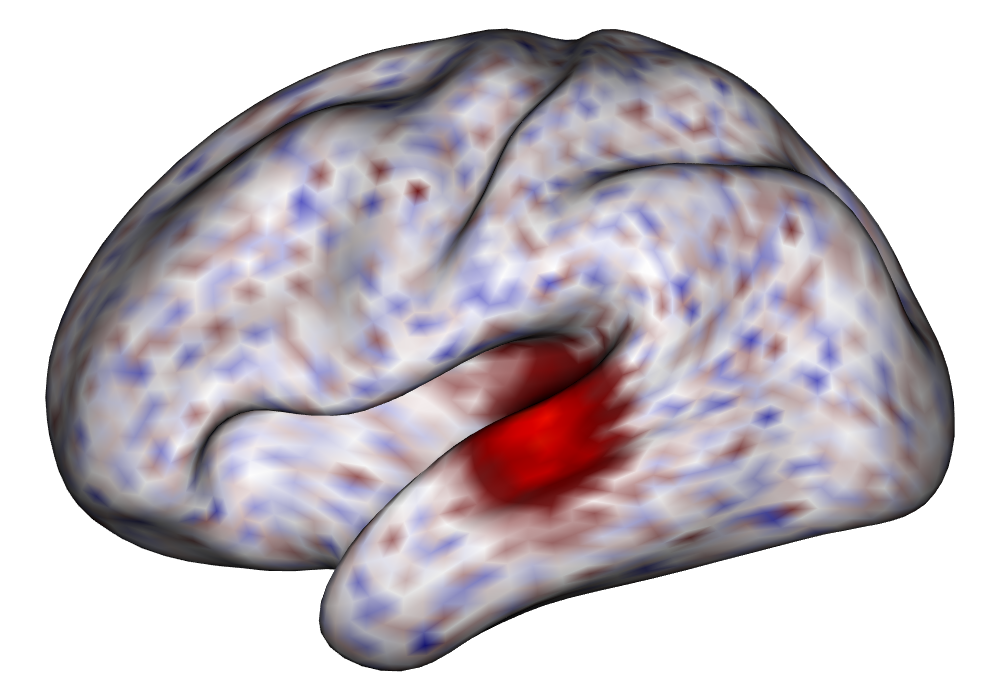} \\
    		\cline{2-5}
    		\multicolumn{1}{c|}{\rotatebox[origin=l]{90}{\qquad\,\,}} & \multicolumn{1}{c|}{\includegraphics[width=.2\textwidth]{3_legend_2.png}} &
    		\multicolumn{3}{c|}{\includegraphics[width=.2\textwidth]{3_legend_1.png}} \\
    		\cline{2-5}
    		\rotatebox[origin=l]{90}{\textbf{Activations}} &
    		\Includegraphics[width=.2\textwidth]{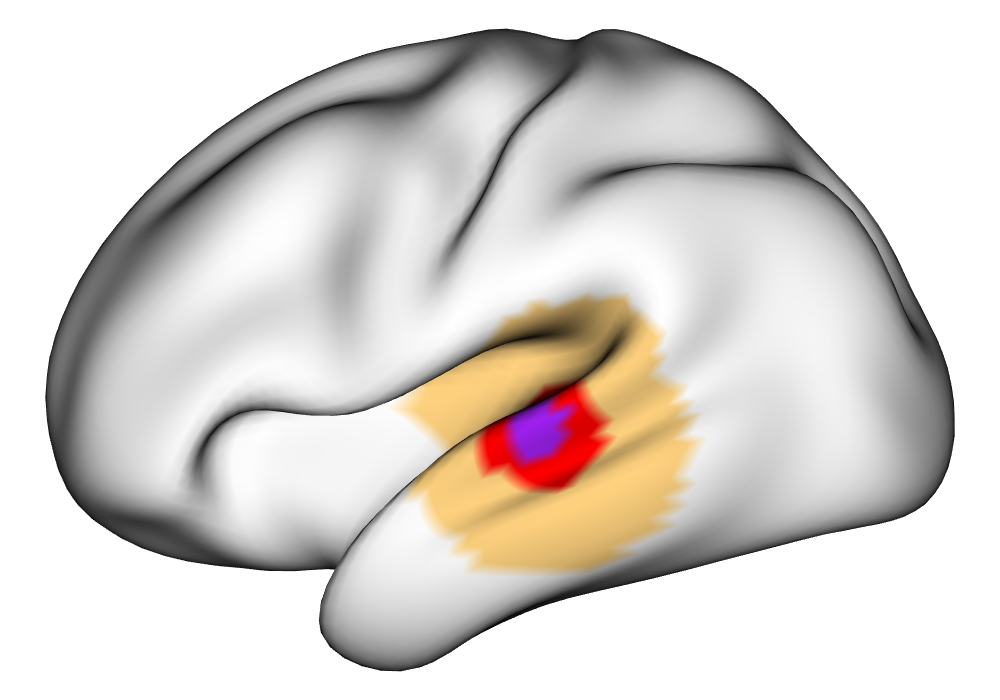} &
    		\Includegraphics[width=.2\textwidth]{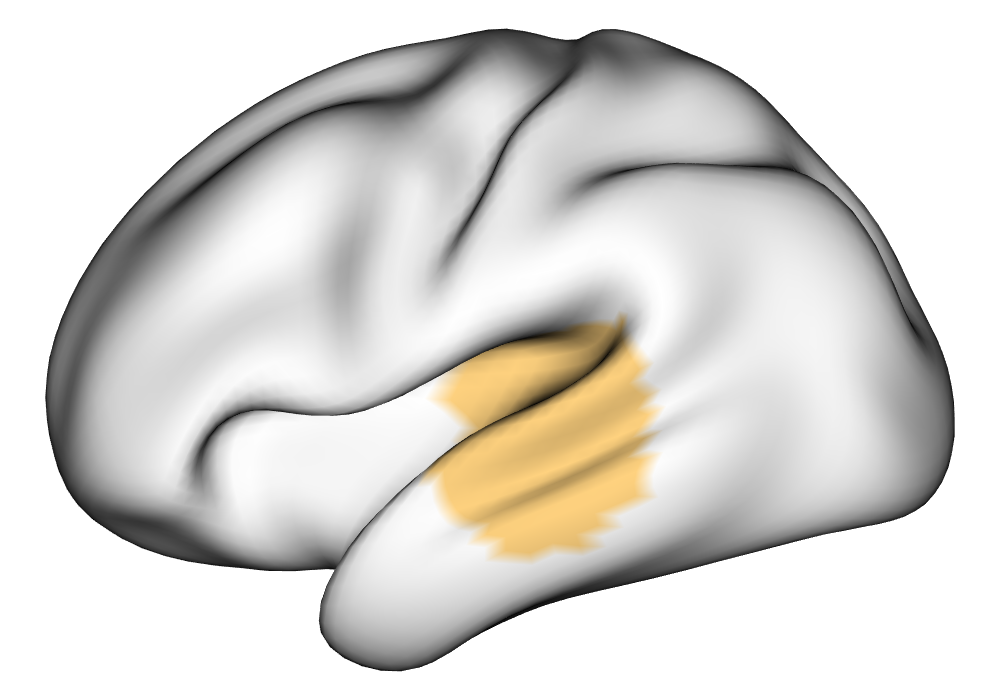} &
    		\Includegraphics[width=.2\textwidth]{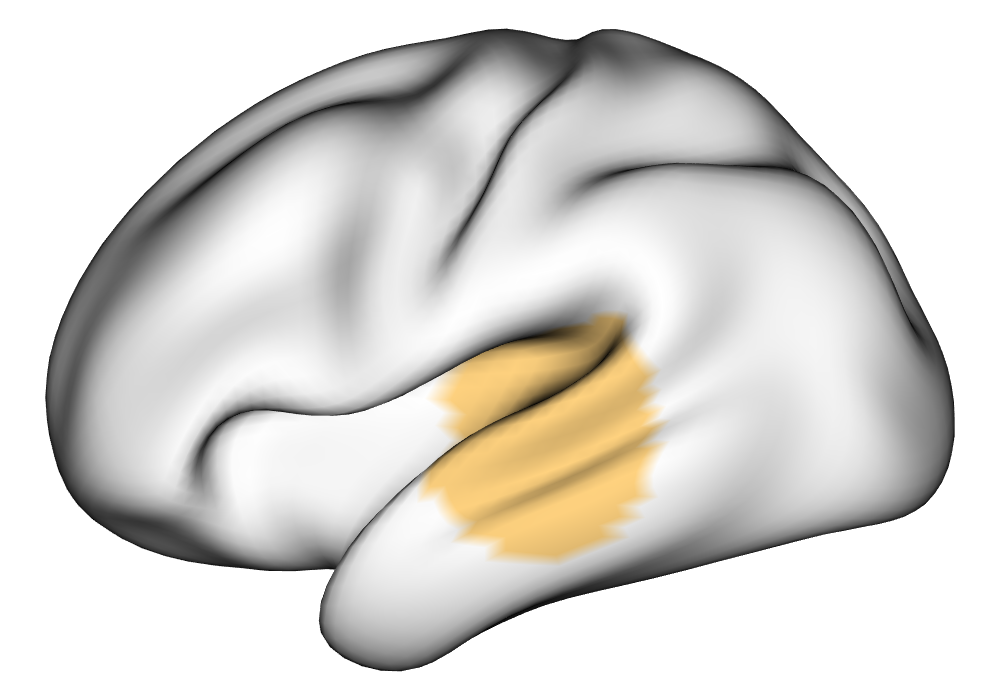} &
    		\Includegraphics[width=.2\textwidth]{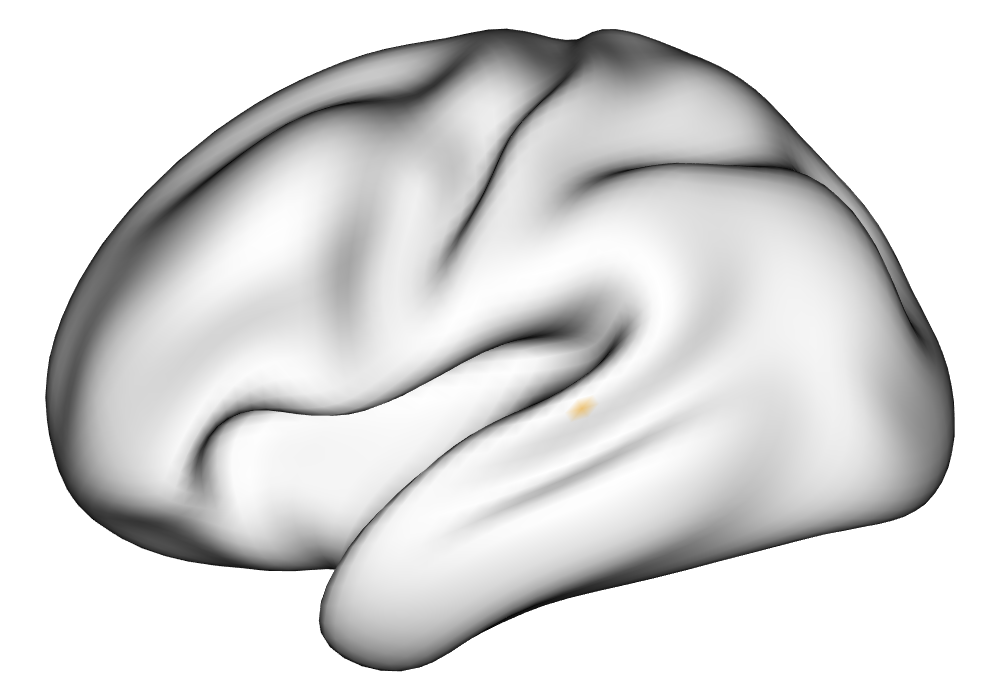} \\
    		\cline{2-5}
    		\multicolumn{1}{c}{\rotatebox[origin=l]{90}{\qquad}} & 
    		\multicolumn{4}{c}{$\gamma =$ \textcolor[HTML]{FFD27F}{$\blacksquare$} 0\% 
           \textcolor[HTML]{FF0000}{$\blacksquare$} 0.5\% 
           \textcolor[HTML]{A020F0}{$\blacksquare$} 1\%}
	    \end{tabularx}
	    \caption{Amplitude of activations and areas of activation from the multi-subject analysis of a single subset}
	    \label{fig:sim_group_estimations_activations}
    \end{subfigure}
    \begin{subfigure}[b]{\textwidth}
        \centering
        \includegraphics[width=0.8\textwidth]{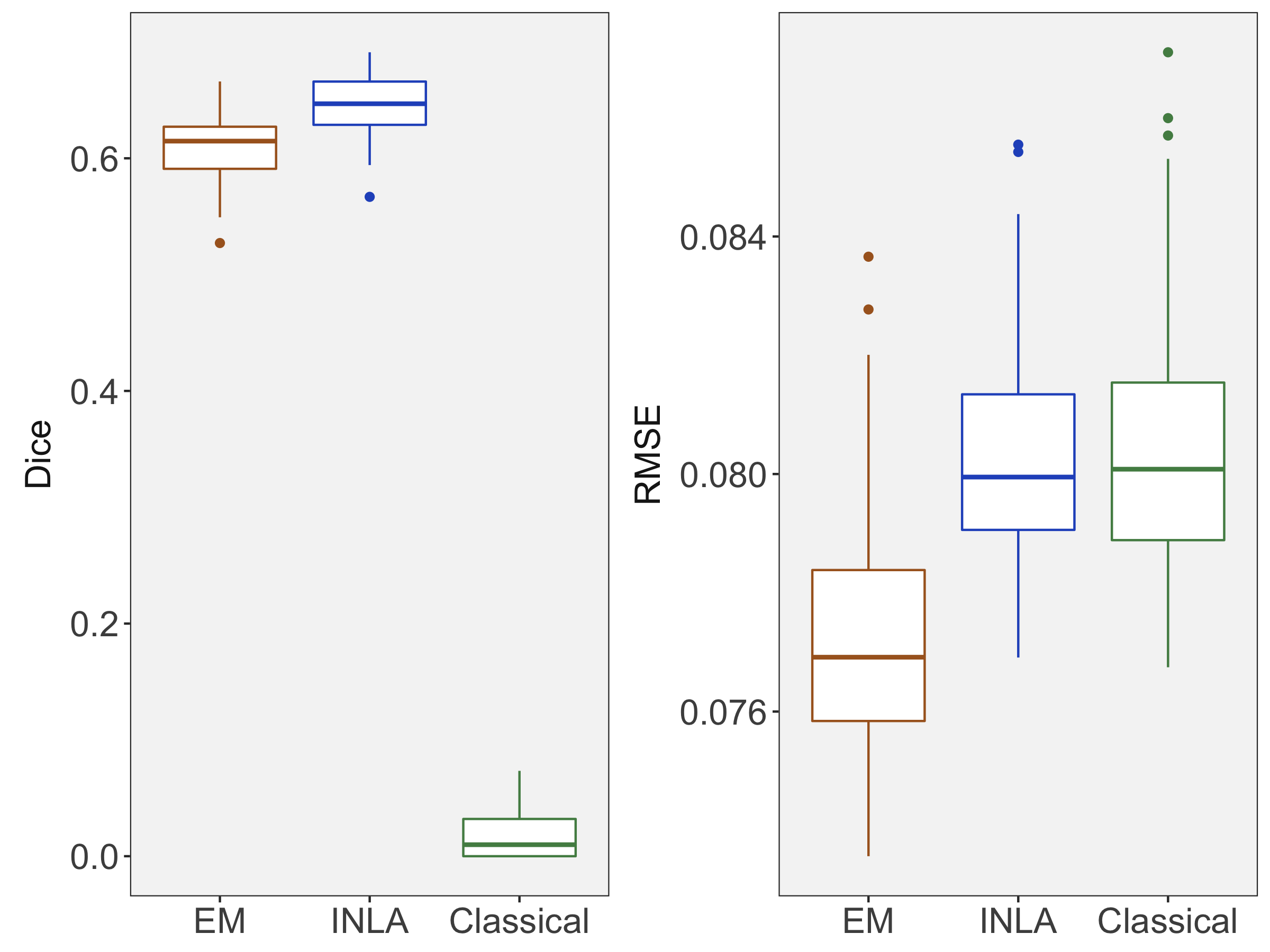}
        \caption{The RMSE values and the Dice coefficients for the 100 subsets}
	    \label{fig:sim_group_rmse_dice}
    \end{subfigure}
    \caption{Cortical surface estimates for the amplitude of activation and areas of activation in percent signal change for the INLA and EM implementations of the Bayesian GLM and the classical GLM. True amplitudes of activation and areas of activation are shown for comparison. Different color scales are shown for the true values and the estimates of the amplitudes of activation because the estimates are attenuated, as one would expect for models assuming sparsity. For the Bayesian GLM implementations, the areas of activation are found using the excursions method, and are found as areas deemed jointly greater than threshold $\gamma$ with probability 0.01. For the classical GLM, the areas of activation are found by fixing the false discovery rate at 0.01.}
    \label{fig:sim_group}
\end{figure}

\textbf{Figure \ref{fig:sim_group_estimations_activations}} shows the true value for the multi-subject activation amplitude for a single task, as well as the locations where the true amplitudes are greater than $\gamma = \{ 0\%, 0.5\%, 1\%\}$ signal change in the analysis of a single 10-subject subset. The EM and INLA implementations of the SBSB GLM produce similar amplitude estimates and areas of activation, showing attenuation between the true values and the estimates. In contrast, the point estimates from the classical GLM show many values that are appreciably distant from zero in the true zero-valued region, while only finding one location that is significantly different from zero within the true nonzero-valued region. None of the modeling methods found any locations in which the amplitude was found to be significantly greater than 0.5\% due to the attenuation expected in the Bayesian model and the low power of the classical GLM. For the Bayesian GLM implementations, the areas of activation are found using the excursions method, and are found as areas deemed jointly greater than threshold $\gamma$ with probability 0.01. For the classical GLM, the areas of activation are found by hypothesis test after fixing the false discovery rate at 0.01.

\textbf{Figure \ref{fig:sim_group_rmse_dice}} shows the calculated values for the square root of the mean squared error (RMSE) and the Dice coefficient for each of 100 subsets of ten subjects each in order to compare the accuracy between the different methods. The true amplitude values for the whole population of 100 simulated subjects was used to determine the values of the measures of accuracy, rather than the true values for each of the subsets, as population-level inference is the goal of the multi-subject analysis. The EM implementation of the SBSB GLM exhibits generally lower values for the RMSE than the INLA implementation and the classical GLM, while exhibiting only slightly lower values for the Dice coefficient. The EM estimates exhibit slightly less attenuation than in the INLA implementation (shown in \textbf{Figure \ref{fig:group_estimate_model_comparison}}), which is expected because no prior distributions are assumed on the hyperparameters in the EM implementation. The classical GLM performs poorly in terms of the Dice coefficient compared to both implementations of the SBSB GLM. These analyses suggest that the multi-subject analysis using the EM implementation of the SBSB GLM is likely to deliver powerful results in the analysis of real cs-fMRI data.

\section{Analysis of HCP Data}
\label{sec:real}

We selected 10 subjects from the Human Connectome Project (HCP) \citep{barch2013function} 1200-subject release in order to simulate small sample conditions common in many neuroimaging experiments as \cite{spencer2022spatial} shows the INLA implementation of the SBSB GLM to be powerful in such small studies. Motor task data were analyzed using the INLA and EM implementations of the SBSB GLM in order to compare the methods and their performance. These data were collected with two fMRI runs performed for each session, one with the phase acquisition from left-to-right, and the other from right-to-left. We analyze the average across these two runs because the acquisition-related effects are not of interest. The motor tasks in the experiment included tapping the fingers on the left and right hands, moving the toes on the left and right feet, and pressing the tongue to the inside of the cheek. A visual cue was used to indicate which task to perform.

These data were obtained from the HCP after being preprocessed with the minimal preprocessing pipeline \citep{van2013wu,barch2013function}. This pipeline includes the projection of the volumetric blood oxygen level-dependent (BOLD) values to the cerebral cortex and the subcortical regions and registering these surfaces to a common surface template. The preprocessing also creates high-resolution surface meshes with 164,000 vertices using structural T1-weighted and T2-weighted MRI scans, which are resampled to 32,000 vertices per hemisphere to match the resolution of the fMRI scans. In order to regularize the mapping process to the cortical surface, the fMRI timeseries were smoothed using a Gaussian smoothing kernel with a 2mm full-width half-maximum (FWHM). 

After these steps from the HCP minimal preprocessing pipeline, additional preprocessing of the fMRI time series as outlined in \ref{sec:preproc} are applied in order to remove nuisance effects and temporal autocorrelation in the data, and make the parameter estimates of $\boldsymbol{\beta}_k$ interpretable in terms of percent signal change, given the unitless nature of fMRI BOLD measures. The data are resampled to around 5,000 vertices per hemisphere using the HCP Workbench \citep{marcus2011informatics} via the R package \texttt{ciftiTools} \citep{pham2022ciftitools}, as this resolution is shown to be computationally efficient without sacrificing effective inferential resolution, as discussed in \cite{spencer2022spatial}. See Figure \ref{fig:surface_comparison} for a comparison of cortical surfaces between different subjects and resolutions. The preprocessed data for both runs of the first scanning session was analyzed separately for each hemisphere and each subject using the INLA and EM implementations of the SBSB GLM.

\begin{figure}
    \begin{subfigure}{0.49\textwidth}
        \begin{tabularx}{\textwidth}{c|c|c|}
            \multicolumn{1}{c}{} &
            \multicolumn{1}{c}{\textbf{INLA}} &
            \multicolumn{1}{c}{\textbf{EM}} \\
            \cline{2-3} 
    		\rotatebox[origin=l]{90}{\textbf{Subject A}} &
    		\Includegraphics[width=0.4\textwidth]{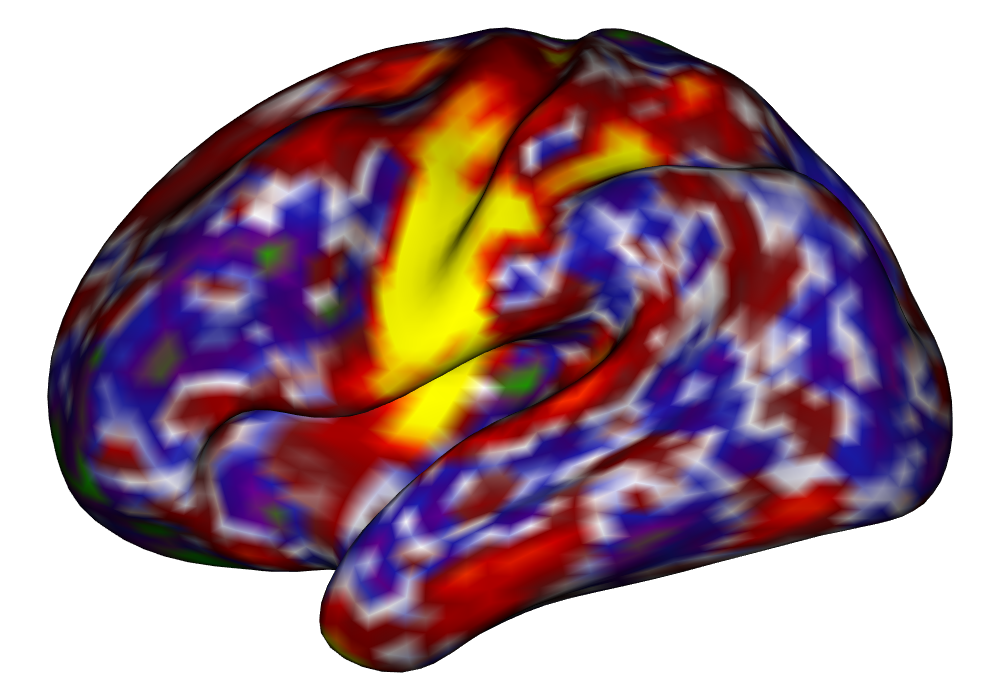} &
    		\Includegraphics[width=0.4\textwidth]{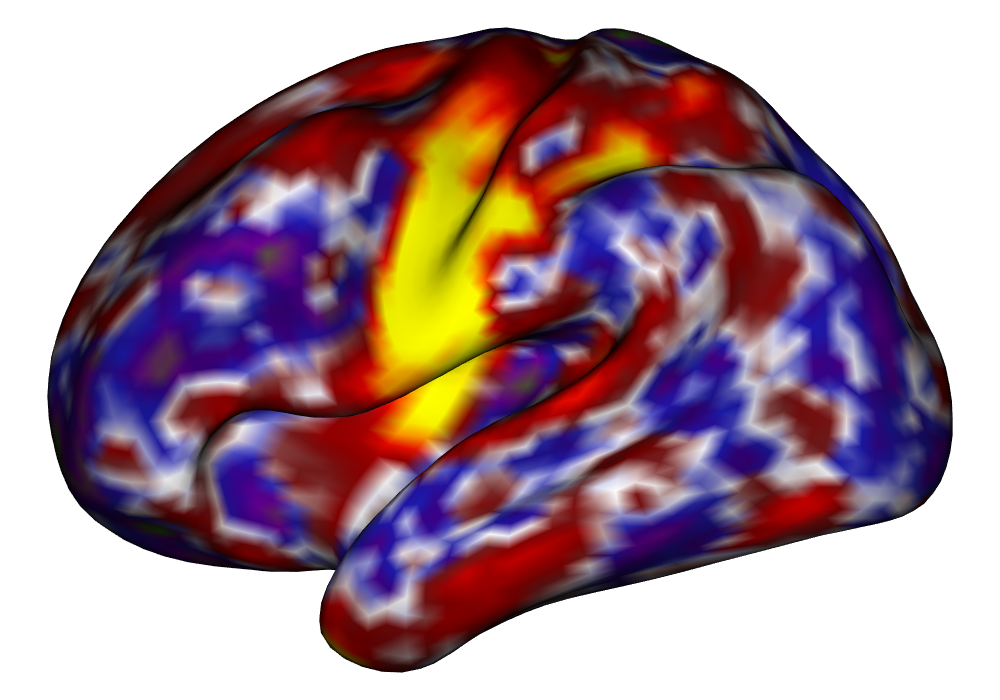} \\
    		\cline{2-3} 
    		\rotatebox[origin=l]{90}{\textbf{Subject B}} &
    		\Includegraphics[width=0.4\textwidth]{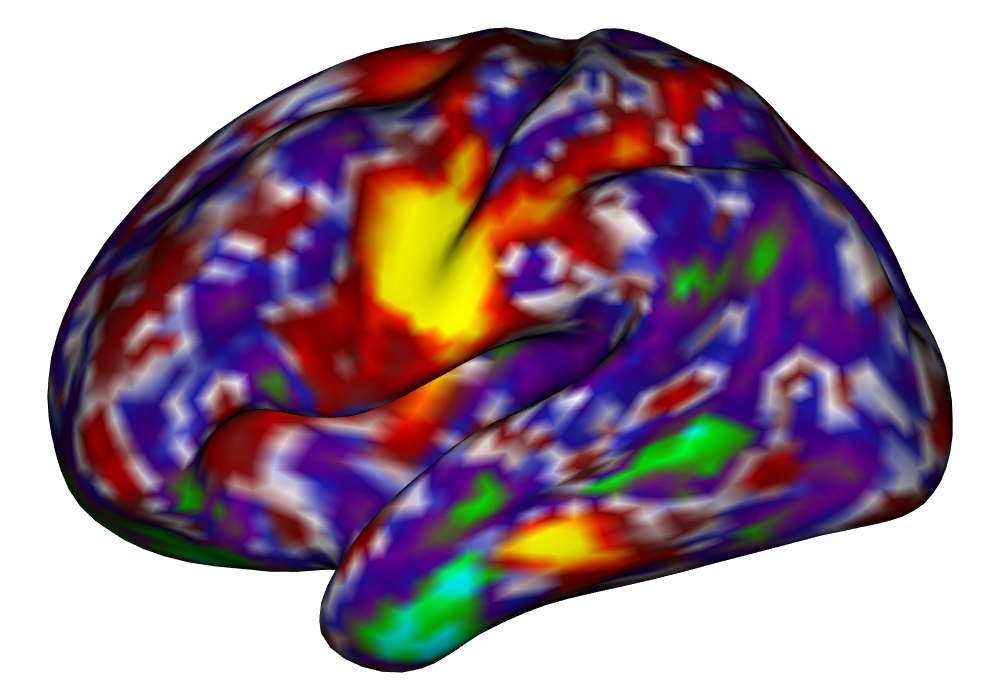} &
    		\Includegraphics[width=0.4\textwidth]{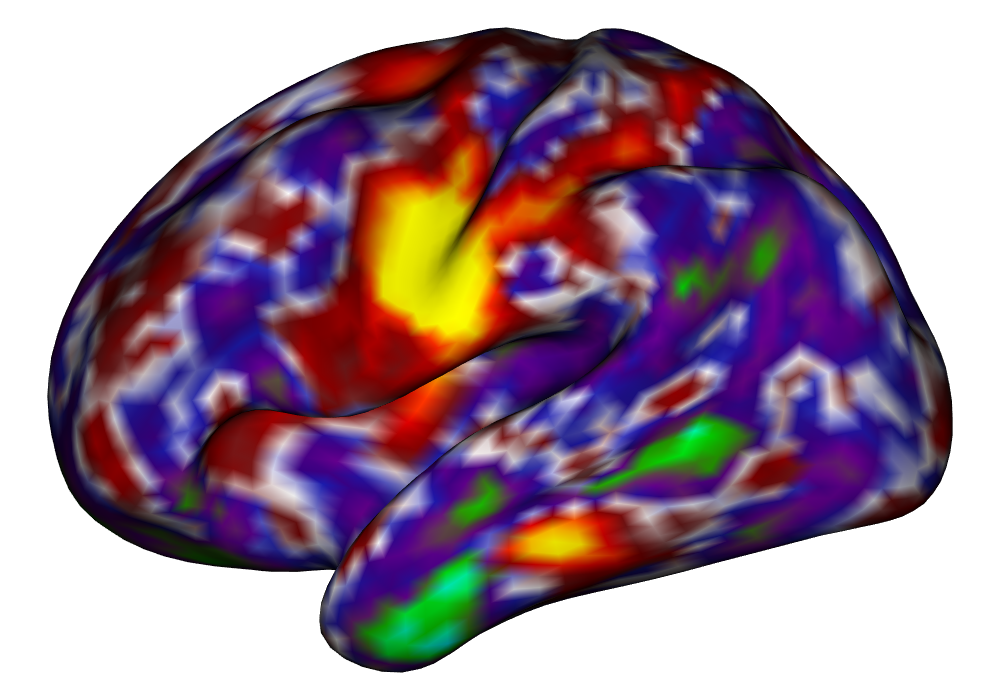} \\
    		\cline{2-3} 
    		\rotatebox[origin=l]{90}{\textbf{Subject C}} &
    		\Includegraphics[width=0.4\textwidth]{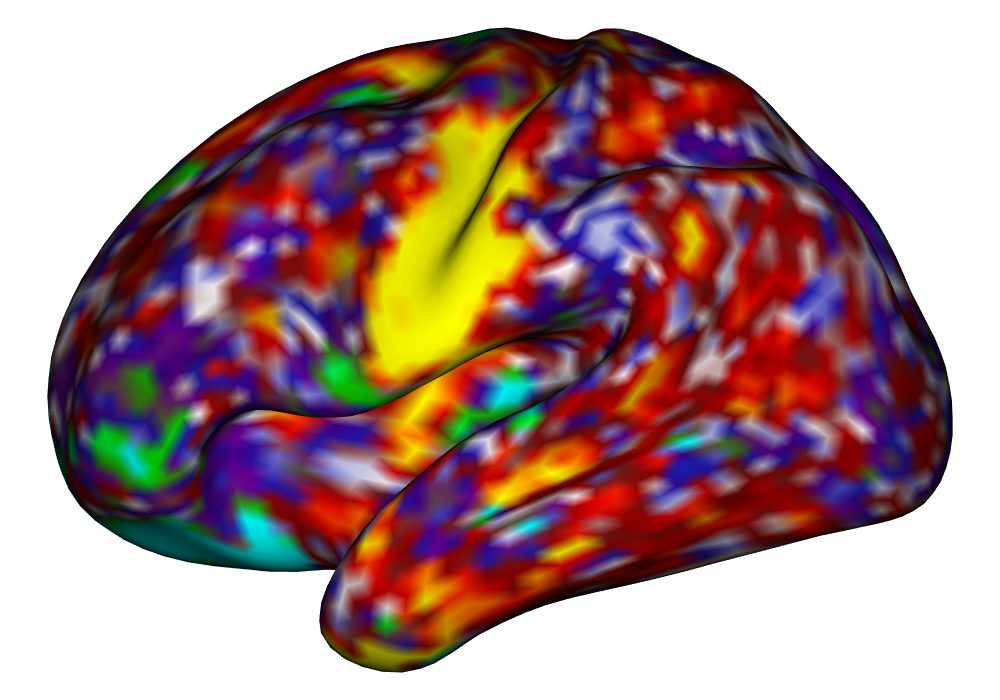} &
    		\Includegraphics[width=0.4\textwidth]{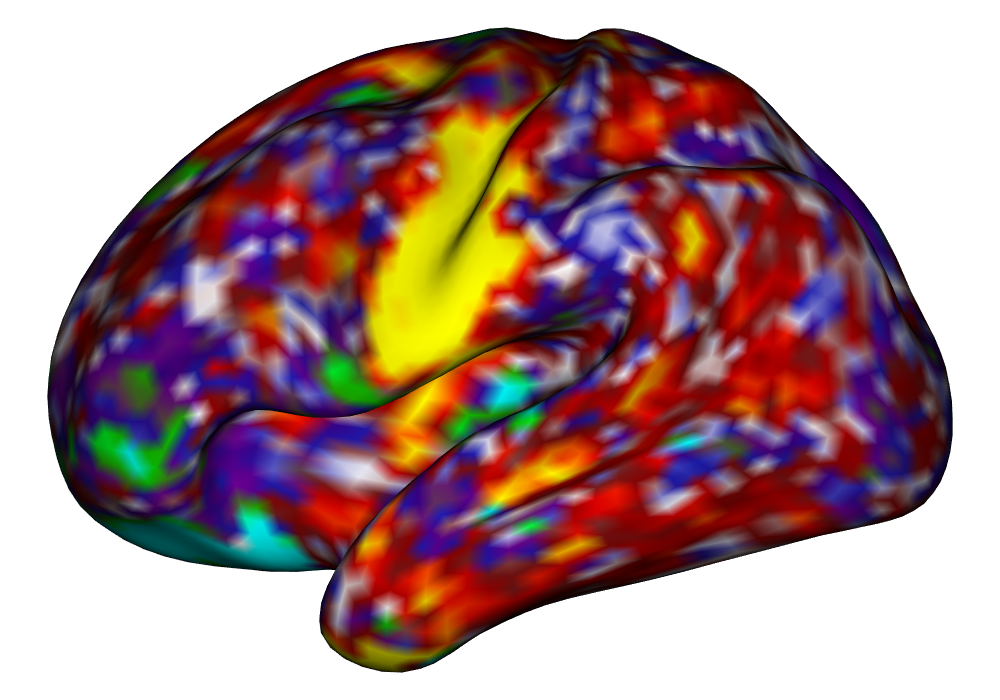} \\
    		\cline{2-3}
    		\multicolumn{1}{c}{\rotatebox[origin=l]{90}{\qquad \,}} &
    		\multicolumn{2}{c}{\includegraphics[width=0.6\textwidth]{3_legend_1.png}}
        \end{tabularx}
        \caption{Estimates}
    \end{subfigure}
    \begin{subfigure}{0.49\textwidth}
        \begin{tabularx}{\textwidth}{c|c|c|}
            \multicolumn{1}{c}{} &
            \multicolumn{1}{c}{\textbf{INLA}} &
            \multicolumn{1}{c}{\textbf{EM}} \\
            \cline{2-3} 
    		\rotatebox[origin=l]{90}{\textbf{Subject A}} &
    		\Includegraphics[width=0.4\textwidth]{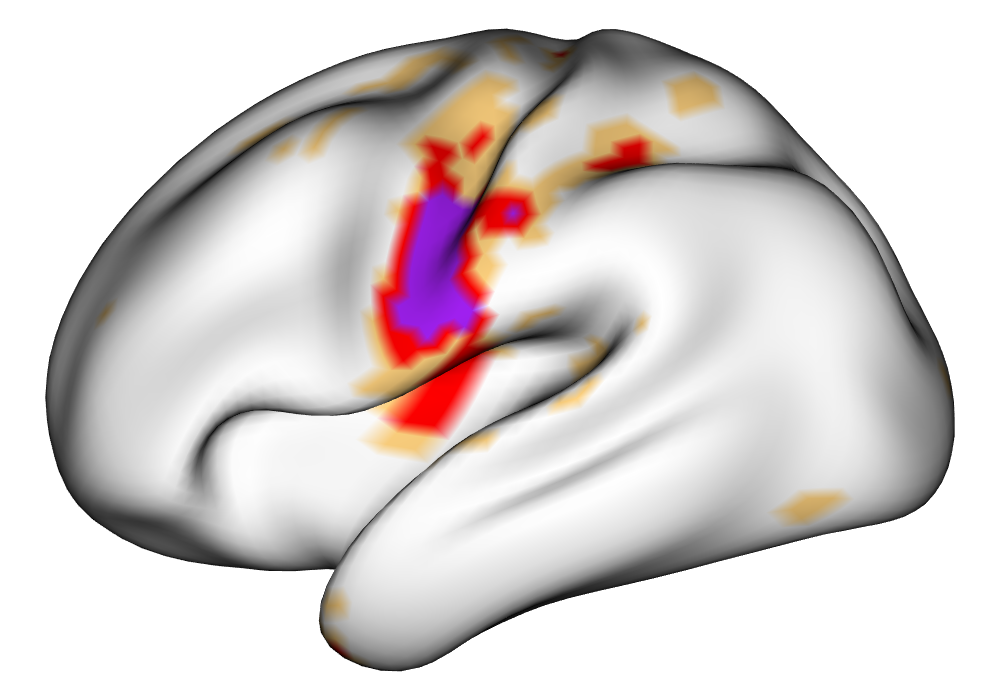} &
    		\Includegraphics[width=0.4\textwidth]{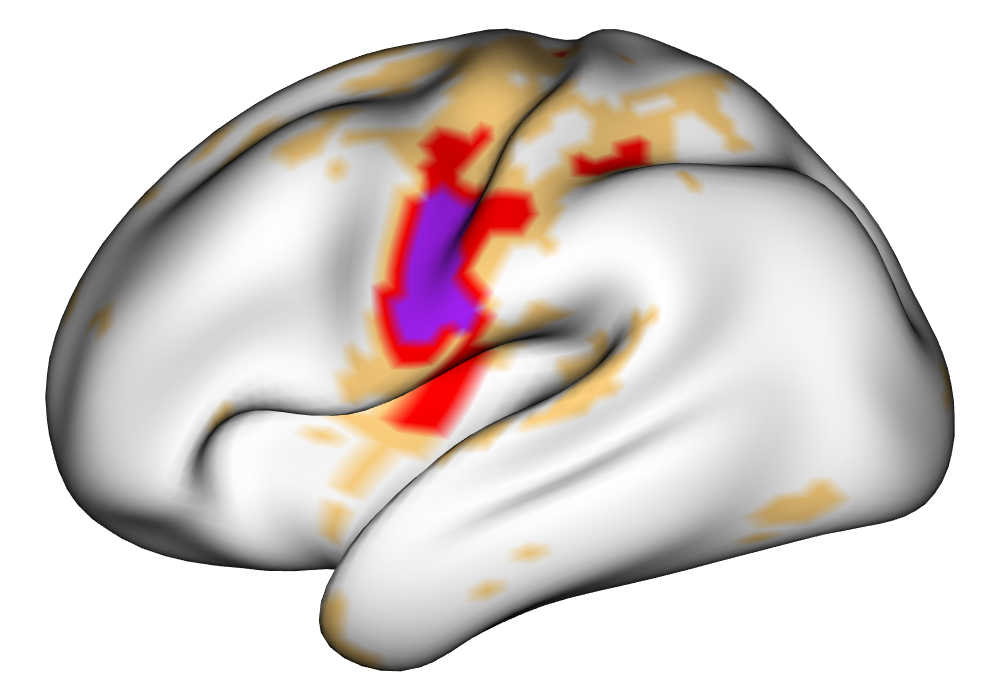} \\
    		\cline{2-3} 
    		\rotatebox[origin=l]{90}{\textbf{Subject B}} &
    		\Includegraphics[width=0.4\textwidth]{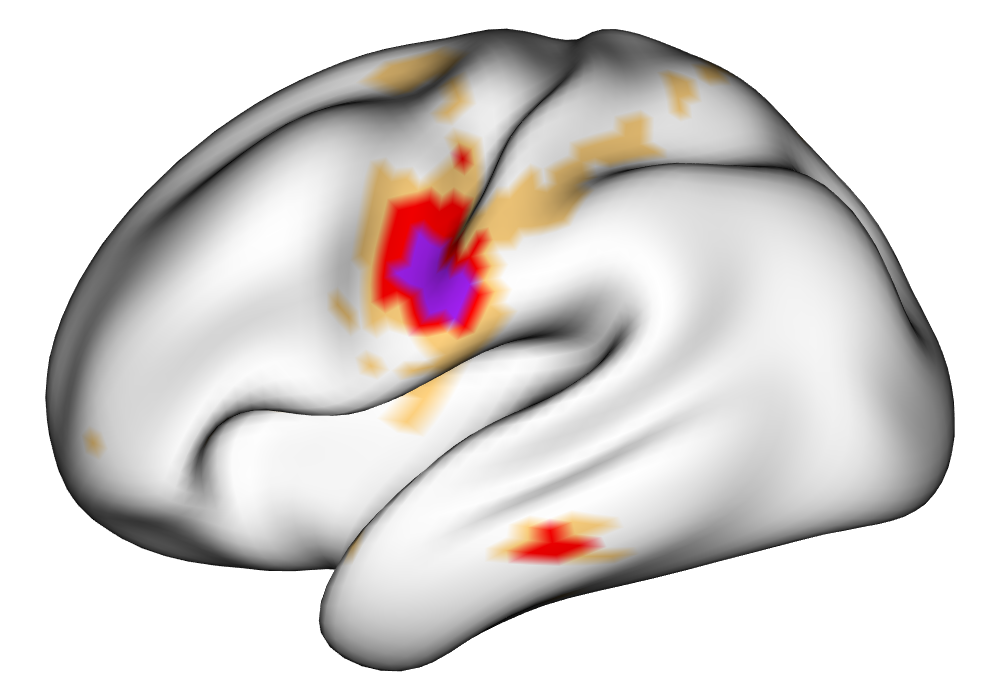} &
    		\Includegraphics[width=0.4\textwidth]{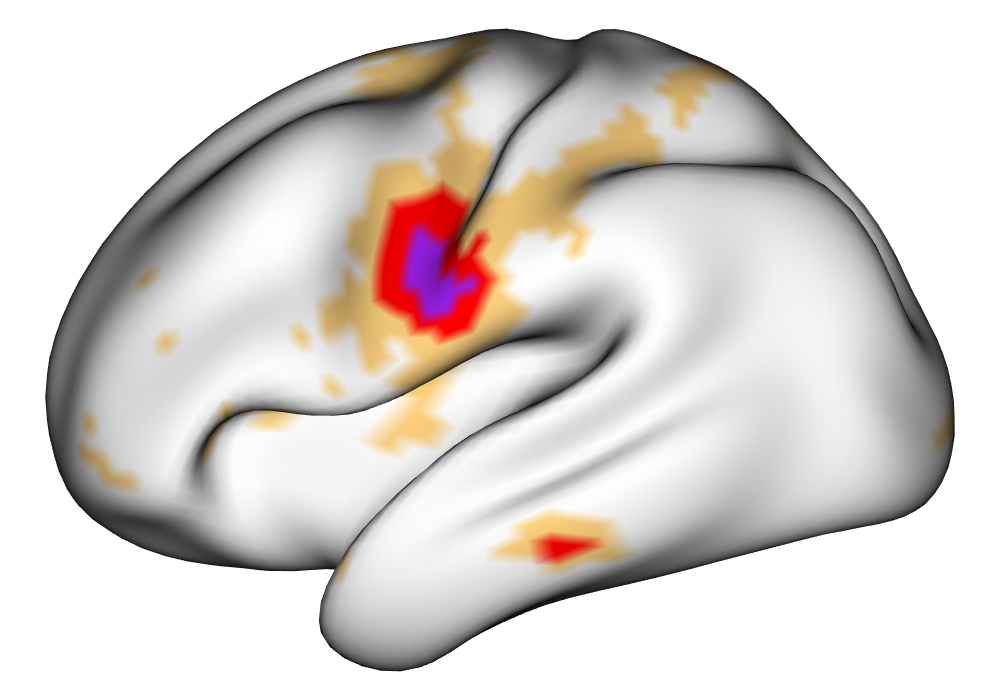} \\
    		\cline{2-3} 
    		\rotatebox[origin=l]{90}{\textbf{Subject C}} &
    		\Includegraphics[width=0.4\textwidth]{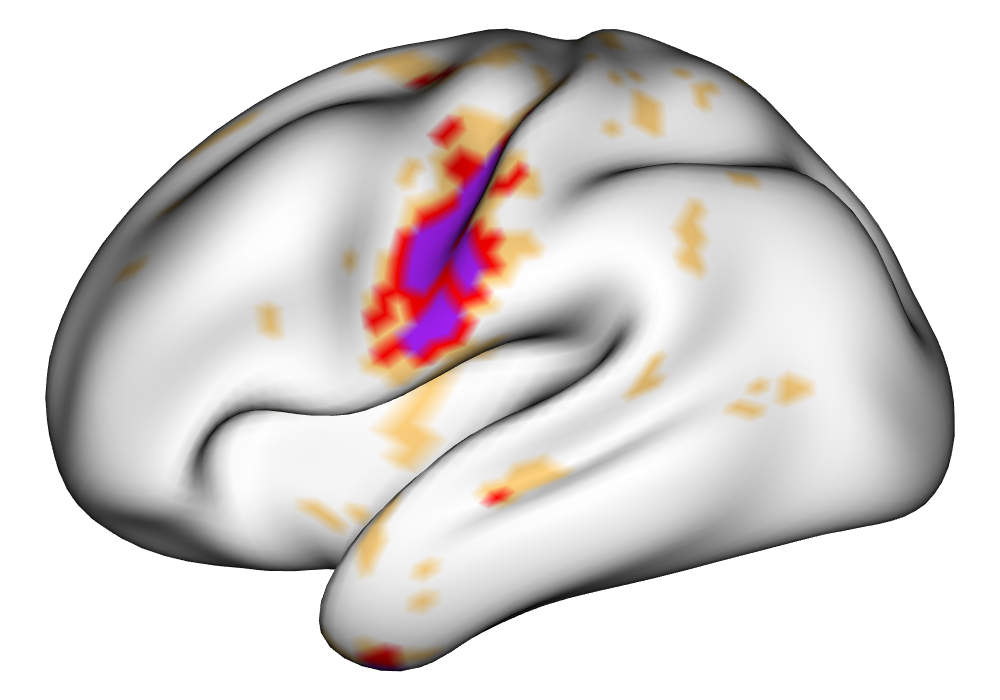} &
    		\Includegraphics[width=0.4\textwidth]{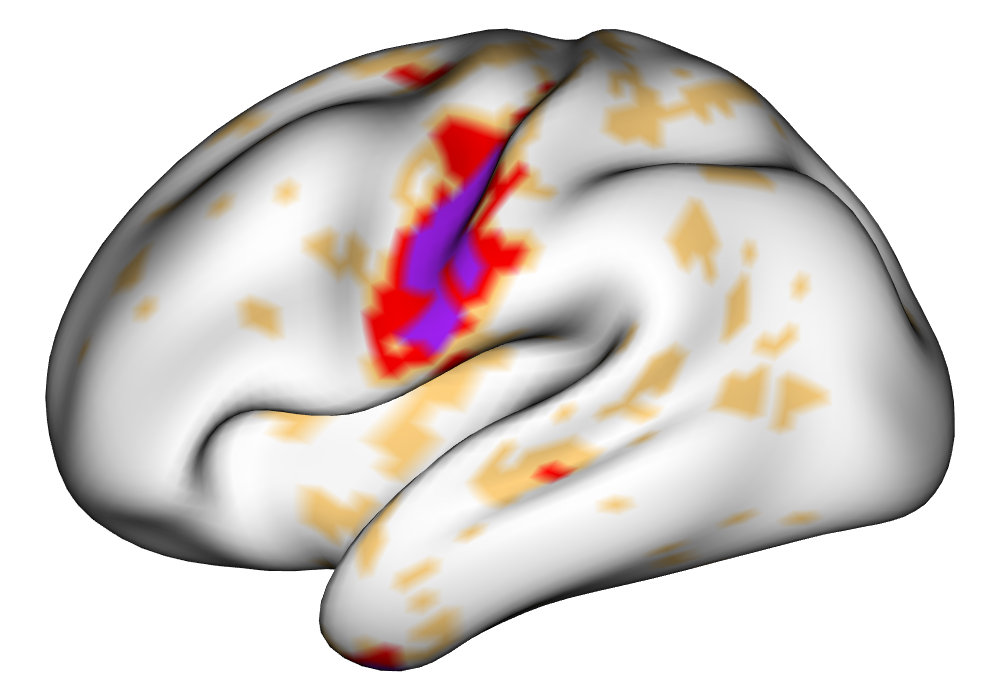} \\
    		\cline{2-3}
    		\multicolumn{1}{c}{\rotatebox[origin=l]{90}{\qquad}} &
    		\multicolumn{2}{c}{$\gamma =$ \textcolor[HTML]{FFD27F}{$\blacksquare$} 0\% 
           \textcolor[HTML]{FF0000}{$\blacksquare$} 0.5\% 
           \textcolor[HTML]{A020F0}{$\blacksquare$} 1\%}
        \end{tabularx}
        \caption{Activations}
    \end{subfigure}
    \caption{Coefficient estimates and activations from the INLA and EM implementations of the Bayesian GLM for the tongue task for three subjects. Estimates are shown with units in \% signal change. Activations are regions determined to be above the threshold $\gamma$ using the excursions method with joint probability of 0.99.}
    \label{fig:hcp_subject_est_and_act}
\end{figure}

\textbf{Figure \ref{fig:hcp_subject_est_and_act}} shows the estimates and areas of activation found in three different subjects for the tongue task. The tongue task is chosen for display due to the easily-visible pattern of activation in the sensorimotor cortex, which helps to highlight the individual subject differences found in the patterns of estimates and activations. It is clear here that the INLA and EM implementations perform very similarly in terms of the task coefficient estimates, and that the EM implementation shows slightly more activation, particularly at the lowest threshold, $\gamma = 0\%$. This is likely due to the reduced estimate of posterior variance in the EM method due to its treatment of the hyperparameters as fixed.

\begin{figure}
\centering
    \begin{subfigure}{0.6\textwidth}
        \begin{tabularx}{\textwidth}{c|c|c|}
            \multicolumn{1}{c}{} &
            \multicolumn{1}{c}{\textbf{INLA}} &
            \multicolumn{1}{c}{\textbf{EM}} \\
            \cline{2-3} 
    		\rotatebox[origin=l]{90}{\qquad\textbf{Estimates}} &
    		\Includegraphics[width=0.48\textwidth]{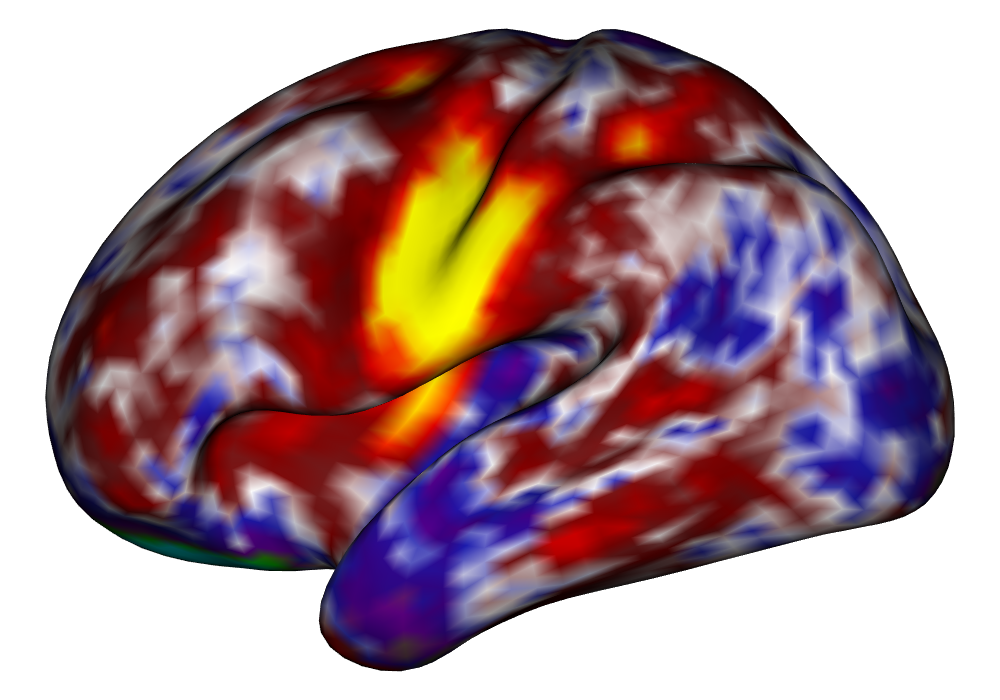} &
    		\Includegraphics[width=0.48\textwidth]{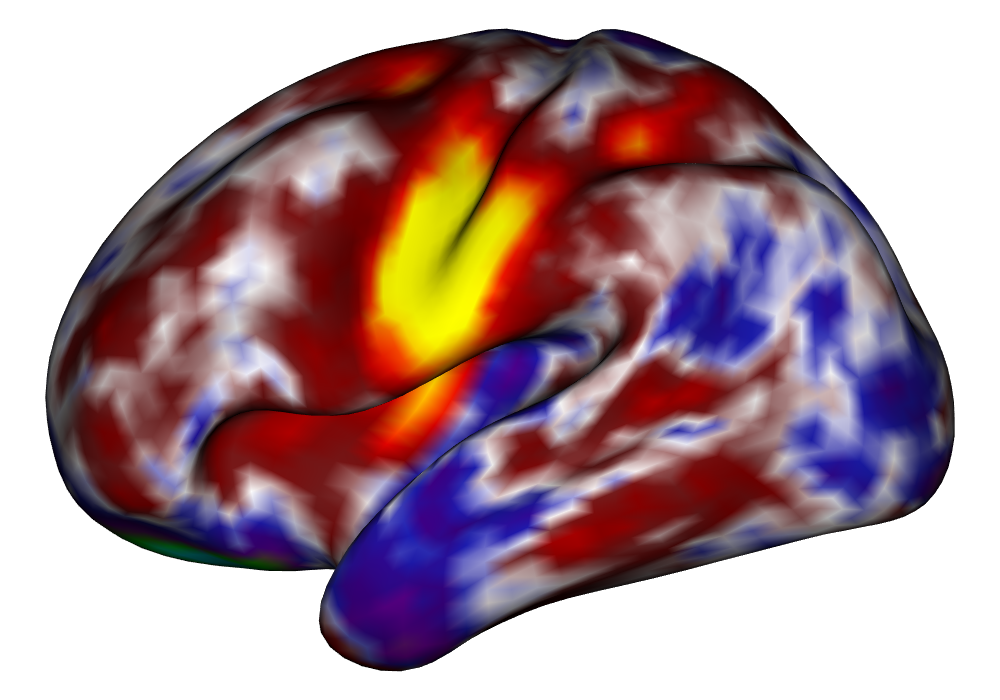} \\
    		\cline{2-3}
    		\multicolumn{1}{c}{\rotatebox[origin=l]{90}{\qquad \,}} &
    		\multicolumn{2}{c}{\includegraphics[width=0.48\textwidth]{3_legend_1.png}} \\
    		\cline{2-3} 
    		\rotatebox[origin=l]{90}{\quad \, \textbf{Activations}} &
    		\Includegraphics[width=0.48\textwidth]{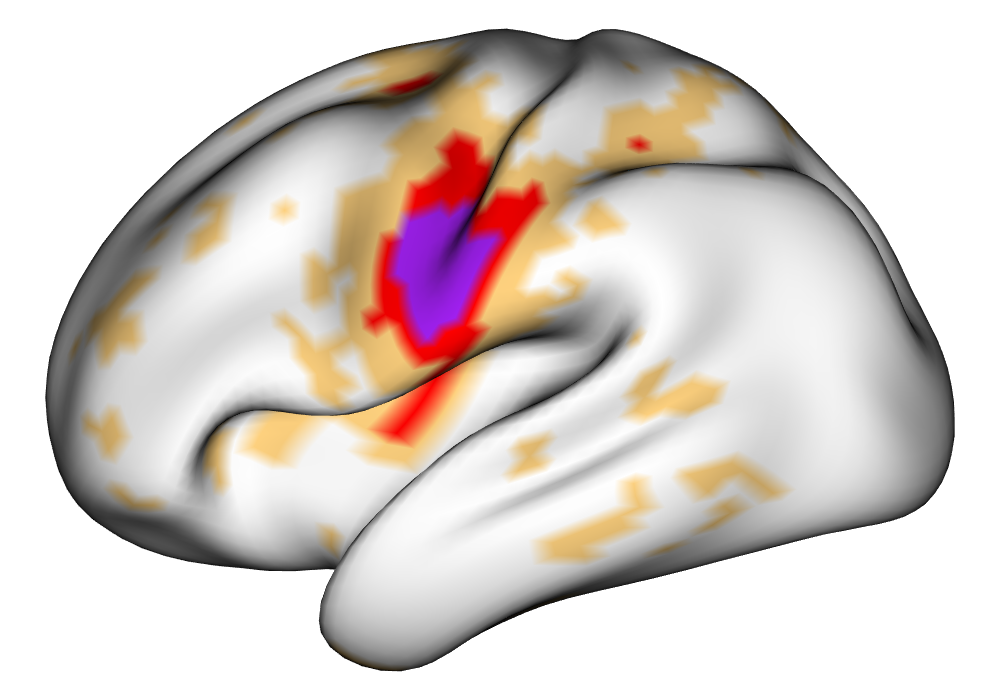} &
    		\Includegraphics[width=0.48\textwidth]{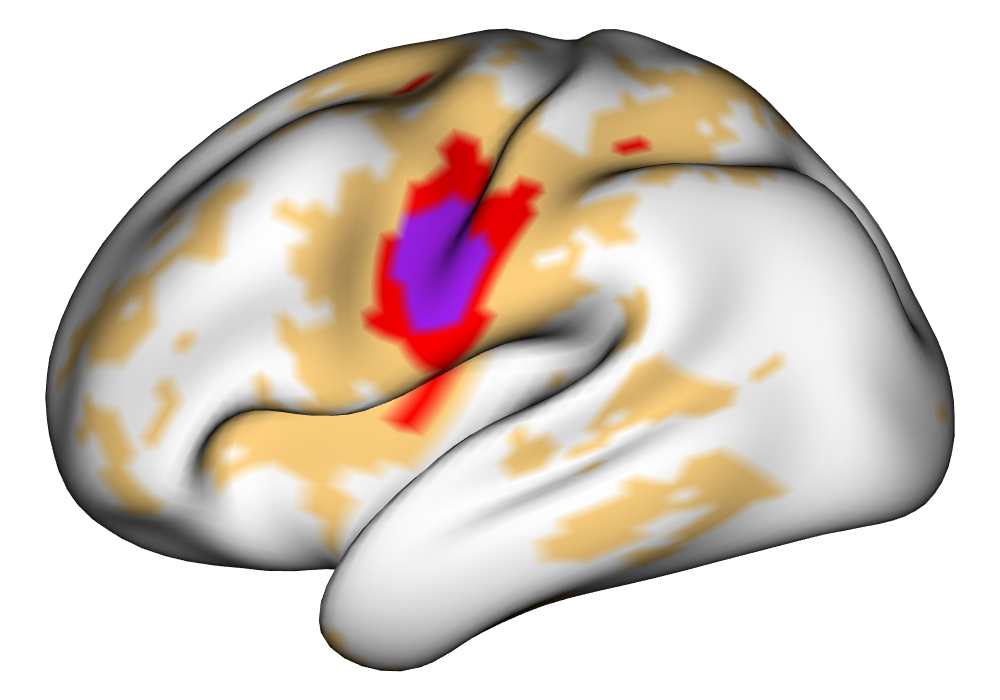} \\
    		\cline{2-3}
    		\multicolumn{1}{c}{\rotatebox[origin=l]{90}{\quad}} &
    		\multicolumn{2}{c}{$\gamma =$ \textcolor[HTML]{FFD27F}{$\blacksquare$} 0\% 
           \textcolor[HTML]{FF0000}{$\blacksquare$} 0.5\% 
           \textcolor[HTML]{A020F0}{$\blacksquare$} 1\%} 
        \end{tabularx}
    \end{subfigure}
    \caption{Group coefficient estimates and activations from the INLA and EM implementations of the Bayesian GLM for the tongue task for 10 subjects. Estimates are shown with units in \% signal change. Activations are regions determined to be above the threshold $\gamma$ using the excursions method with joint probability of 0.99.}
    \label{fig:hcp_group_est_and_act}
\end{figure}

\textbf{Figure \ref{fig:hcp_group_est_and_act}} shows the multi-subject estimates and areas of activation for the tongue task, based on all 10 subjects. The amplitude estimates from the EM implementation of the SBSB GLM are highly very similar to those found using the INLA implementation, and the areas of activation show that the EM algorithm detects more activations, especially at the $\gamma = 0\%$ threshold, again due to the underestimation of posterior variance. However, the activations found at the $\gamma = 0.5\%$ and $\gamma = 1\%$ thresholds are very similar, suggesting high levels of agreement at higher, more neurologically meaningful thresholds.

\section{Conclusions}
\label{sec:conclusion}

We develop and implement an EM algorithm to fit the surface-based spatial Bayesian GLM on cortical surface task fMRI data. This provides an exact method to fit the Bayesian GLM, reduce the memory consumption required by the INLA implementation of the Bayesian GLM, and also to reduce the dependence on the INLA package. The INLA package, while powerful, is not available on the Comprehensive R Archive Network (CRAN), and may be difficult or impossible to install on various computer systems. The increased memory requirement for the INLA package also presents an impediment to using the SBSB GLM to researchers without significant computing resources at their disposal.

Through analysis of simulated data, we determine that the EM algorithm performs well compared to the INLA implementation, both of which strongly outperform the surface-based classical GLM in terms of accuracy. The group-level results for the INLA and EM implementations are highly consistent, both in terms of effect size estimation and activation detection.

Analysis of data from the Human Connectome Project motor task data confirms that the EM implementation performs similarly to the INLA implementation, as both methods are able to find sparse, spatially-contiguous nonzero estimates of activation amplitude that display subject-specific characteristics. As in the simulated data, the EM finds more activated locations than INLA at a threshold of $\gamma = 0\%$, with similar numbers of activated locations found at higher thresholds.

Future work on this method includes improving computational efficiency through the use of parallelized linear algebra solvers and speed improvements through conjugate gradient methods. If such an optimized parallel solver were implemented, we expect the EM algorithm to take less time than INLA in all scenarios. Such speed improvements will likely make the spatial Bayesian GLM applicable to subcortical data, which have more dense neighborhood structures. Forthcoming work in a software paper for the \texttt{BayesfMRI} package in \texttt{R} will make the application of the functions used to perform these analyses accessible to anyone performing inference on relatively modest laptop computers.

\section{Acknowledgements}

Data were provided in part by the Human Connectome Project, WU- Minn Consortium (Principal Investigators: David Van Essen and Kamil Ugurbil; 1U54MH091657) funded by the 16 NIH Institutes and Centers that support the NIH Blueprint for Neuroscience Research; and by the McDonnell Center for Systems Neuroscience at Washington University.

The research of Daniel Spencer and Amanda Mejia is partially funded by the National Institute of Biomedical Imaging and Bioengineering (R01EB027119).

\bibliography{BayesGLMEM}

\begin{thebibliography}{49}
\providecommand{\natexlab}[1]{#1}
\providecommand{\url}[1]{\texttt{#1}}
\expandafter\ifx\csname urlstyle\endcsname\relax
  \providecommand{\doi}[1]{doi: #1}\else
  \providecommand{\doi}{doi: \begingroup \urlstyle{rm}\Url}\fi

\bibitem[Alappat et~al.(2020)Alappat, Basermann, Bishop, Fehske, Hager, Schenk,
  Thies, and Wellein]{pardiso-7.2a}
Christie Alappat, Achim Basermann, Alan~R. Bishop, Holger Fehske, Georg Hager,
  Olaf Schenk, Jonas Thies, and Gerhard Wellein.
\newblock A recursive algebraic coloring technique for hardware-efficient
  symmetric sparse matrix-vector multiplication.
\newblock \emph{ACM Trans. Parallel Comput.}, 7\penalty0 (3), June 2020.
\newblock ISSN 2329-4949.
\newblock \doi{10.1145/3399732}.
\newblock URL \url{https://doi.org/10.1145/3399732}.

\bibitem[Barch et~al.(2013)Barch, Burgess, Harms, Petersen, Schlaggar,
  Corbetta, Glasser, Curtiss, Dixit, Feldt, et~al.]{barch2013function}
Deanna~M Barch, Gregory~C Burgess, Michael~P Harms, Steven~E Petersen,
  Bradley~L Schlaggar, Maurizio Corbetta, Matthew~F Glasser, Sandra Curtiss,
  Sachin Dixit, Cindy Feldt, et~al.
\newblock Function in the human connectome: task-{fMRI} and individual
  differences in behavior.
\newblock \emph{Neuroimage}, 80:\penalty0 169--189, 2013.

\bibitem[Bates and Eddelbuettel(2013)]{bates2013fast}
Douglas Bates and Dirk Eddelbuettel.
\newblock Fast and elegant numerical linear algebra using the {RcppEigen}
  package.
\newblock \emph{Journal of Statistical Software}, 52\penalty0 (5):\penalty0
  1--24, 2013.
\newblock \doi{10.18637/jss.v052.i05}.

\bibitem[Bishop and Nasrabadi(2006)]{bishop2006pattern}
Christopher~M Bishop and Nasser~M Nasrabadi.
\newblock \emph{Pattern recognition and machine learning}, volume~4.
\newblock Springer, 2006.

\bibitem[Bolin and Lindgren(2013)]{bolin2013comparison}
David Bolin and Finn Lindgren.
\newblock A comparison between {Markov} approximations and other methods for
  large spatial data sets.
\newblock \emph{Computational Statistics \& Data Analysis}, 61:\penalty0 7--21,
  2013.

\bibitem[Bolin and Lindgren(2015)]{bolin2015excursion}
David Bolin and Finn Lindgren.
\newblock Excursion and contour uncertainty regions for latent {Gaussian}
  models.
\newblock \emph{Journal of the Royal Statistical Society, Series B (Statistical
  Methodology)}, 77\penalty0 (1):\penalty0 85--106, 2015.

\bibitem[Bolin and Lindgren(2017)]{bolin2017quantifying}
David Bolin and Finn Lindgren.
\newblock Quantifying the uncertainty of contour maps.
\newblock \emph{Journal of Computational and Graphical Statistics}, 26\penalty0
  (3):\penalty0 513--524, 2017.

\bibitem[Bolin and Lindgren(2018)]{bolin2018calculating}
David Bolin and Finn Lindgren.
\newblock Calculating probabilistic excursion sets and related quantities using
  {excursions}.
\newblock \emph{Journal of Statistical Software}, 86\penalty0 (5):\penalty0
  1--20, 2018.
\newblock \doi{10.18637/jss.v086.i05}.

\bibitem[Bollh{\"o}fer et~al.(2019)Bollh{\"o}fer, Eftekhari, Scheidegger, and
  Schenk]{pardiso-7.2c}
Matthias Bollh{\"o}fer, Aryan Eftekhari, Simon Scheidegger, and Olaf Schenk.
\newblock Large-scale sparse inverse covariance matrix estimation.
\newblock \emph{SIAM Journal on Scientific Computing}, 41\penalty0
  (1):\penalty0 A380--A401, 2019.
\newblock \doi{10.1137/17M1147615}.
\newblock URL \url{https://doi.org/10.1137/17M1147615}.

\bibitem[Bollh{\"o}fer et~al.(2020)Bollh{\"o}fer, Schenk, Janalik, Hamm, and
  Gullapalli]{pardiso-7.2b}
Matthias Bollh{\"o}fer, Olaf Schenk, Radim Janalik, Steve Hamm, and Kiran
  Gullapalli.
\newblock \emph{State-of-the-Art Sparse Direct Solvers}.
\newblock Springer International Publishing, Cham, 2020.
\newblock ISBN 978-3-030-43736-7.
\newblock \doi{10.1007/978-3-030-43736-7\_1}.
\newblock URL \url{https://doi.org/10.1007/978-3-030-43736-7\_1}.

\bibitem[Chatfield and Collins(2018)]{chatfield2018introduction}
Christopher Chatfield and Alexander~J Collins.
\newblock \emph{Introduction to multivariate analysis}.
\newblock Routledge, 2018.

\bibitem[Dempster et~al.(1977)Dempster, Laird, and Rubin]{dempster1977maximum}
Arthur~P Dempster, Nan~M Laird, and Donald~B Rubin.
\newblock Maximum likelihood from incomplete data via the {EM} algorithm.
\newblock \emph{Journal of the Royal Statistical Society: Series B
  (Methodological)}, 39\penalty0 (1):\penalty0 1--22, 1977.

\bibitem[Dice(1945)]{dice1945measures}
Lee~R Dice.
\newblock Measures of the amount of ecologic association between species.
\newblock \emph{Ecology}, 26\penalty0 (3):\penalty0 297--302, 1945.

\bibitem[Eddelbuettel(2013)]{eddelbuettel2013seamless}
Dirk Eddelbuettel.
\newblock \emph{Seamless {R} and {C++} Integration with {Rcpp}}.
\newblock Springer, New York, 2013.
\newblock \doi{10.1007/978-1-4614-6868-4}.
\newblock ISBN 978-1-4614-6867-7.

\bibitem[Eddelbuettel and Fran\c{c}ois(2011)]{eddelbuettel2011Rcpp}
Dirk Eddelbuettel and Romain Fran\c{c}ois.
\newblock {Rcpp}: Seamless {R} and {C++} integration.
\newblock \emph{Journal of Statistical Software}, 40\penalty0 (8):\penalty0
  1--18, 2011.
\newblock \doi{10.18637/jss.v040.i08}.

\bibitem[Eklund et~al.(2016)Eklund, Nichols, and Knutsson]{eklund2016cluster}
Anders Eklund, Thomas~E Nichols, and Hans Knutsson.
\newblock Cluster failure: Why {fMRI} inferences for spatial extent have
  inflated false-positive rates.
\newblock \emph{Proceedings of the national academy of sciences}, 113\penalty0
  (28):\penalty0 7900--7905, 2016.

\bibitem[Eklund et~al.(2019)Eklund, Knutsson, and Nichols]{eklund2019cluster}
Anders Eklund, Hans Knutsson, and Thomas~E Nichols.
\newblock Cluster failure revisited: Impact of first level design and
  physiological noise on cluster false positive rates.
\newblock \emph{Human brain mapping}, 40\penalty0 (7):\penalty0 2017--2032,
  2019.

\bibitem[Elliott et~al.(2020)Elliott, Knodt, Ireland, Morris, Poulton,
  Ramrakha, Sison, Moffitt, Caspi, and Hariri]{elliott2020test}
Maxwell~L Elliott, Annchen~R Knodt, David Ireland, Meriwether~L Morris, Richie
  Poulton, Sandhya Ramrakha, Maria~L Sison, Terrie~E Moffitt, Avshalom Caspi,
  and Ahmad~R Hariri.
\newblock What is the test-retest reliability of common task-functional {MRI}
  measures? {N}ew empirical evidence and a meta-analysis.
\newblock \emph{Psychological Science}, 31\penalty0 (7):\penalty0 792--806,
  2020.

\bibitem[Eshel(2003)]{eshel2003yule}
Gidon Eshel.
\newblock The {Y}ule {W}alker equations for the {AR} coefficients.
\newblock \emph{Internet resource}, 2:\penalty0 68--73, 2003.

\bibitem[Friston et~al.(1994)Friston, Holmes, Worsley, Poline, Frith, and
  Frackowiak]{friston1994statistical}
Karl~J Friston, Andrew~P Holmes, Keith~J Worsley, J-P Poline, Chris~D Frith,
  and Richard~SJ Frackowiak.
\newblock Statistical parametric maps in functional imaging: a general linear
  approach.
\newblock \emph{Human brain mapping}, 2\penalty0 (4):\penalty0 189--210, 1994.

\bibitem[Gelman et~al.(2013)Gelman, Carlin, Stern, Dunson, Vehtari, and
  Rubin]{gelman2013bayesian}
Andrew Gelman, John~B Carlin, Hal~S Stern, David~B Dunson, Aki Vehtari, and
  Donald~B Rubin.
\newblock \emph{Bayesian data analysis}.
\newblock CRC press, 2013.

\bibitem[Glasser et~al.(2013)Glasser, Sotiropoulos, Wilson, Coalson, Fischl,
  Andersson, Xu, Jbabdi, Webster, Polimeni, et~al.]{glasser2013minimal}
Matthew~F Glasser, Stamatios~N Sotiropoulos, J~Anthony Wilson, Timothy~S
  Coalson, Bruce Fischl, Jesper~L Andersson, Junqian Xu, Saad Jbabdi, Matthew
  Webster, Jonathan~R Polimeni, et~al.
\newblock The minimal preprocessing pipelines for the human connectome project.
\newblock \emph{Neuroimage}, 80:\penalty0 105--124, 2013.

\bibitem[Guennebaud et~al.(2010)Guennebaud, Jacob,
  et~al.]{guennebaud2010eigenweb}
Ga\"{e}l Guennebaud, Beno\^{i}t Jacob, et~al.
\newblock Eigen v3.
\newblock http://eigen.tuxfamily.org, 2010.

\bibitem[Guhaniyogi and Spencer(2021)]{guhaniyogi2021bayesian}
Rajarshi Guhaniyogi and Daniel Spencer.
\newblock Bayesian tensor response regression with an application to brain
  activation studies.
\newblock \emph{Bayesian Analysis}, 16\penalty0 (4):\penalty0 1221--1249, 2021.

\bibitem[Hutchinson(1989)]{hutchinson1989stochastic}
Michael~F Hutchinson.
\newblock A stochastic estimator of the trace of the influence matrix for
  {L}aplacian smoothing splines.
\newblock \emph{Communications in Statistics-Simulation and Computation},
  18\penalty0 (3):\penalty0 1059--1076, 1989.

\bibitem[Lindgren et~al.(2011)Lindgren, Rue, and
  Lindstr{\"o}m]{lindgren2011explicit}
Finn Lindgren, H{\aa}vard Rue, and Johan Lindstr{\"o}m.
\newblock An explicit link between {Gaussian} fields and {Gaussian} {Markov}
  random fields: The stochastic partial differential equation approach (with
  discussion).
\newblock \emph{Journal of the Royal Statistical Society B}, 73\penalty0
  (4):\penalty0 423--498, 2011.

\bibitem[Lindquist(2008)]{lindquist2008statistical}
Martin~A Lindquist.
\newblock The statistical analysis of {fMRI} data.
\newblock \emph{Statistical science}, 23\penalty0 (4):\penalty0 439--464, 2008.

\bibitem[Marcus et~al.(2011)Marcus, Harwell, Olsen, Hodge, Glasser, Prior,
  Jenkinson, Laumann, Curtiss, and Van~Essen]{marcus2011informatics}
Daniel~S Marcus, John Harwell, Timothy Olsen, Michael Hodge, Matthew~F Glasser,
  Fred Prior, Mark Jenkinson, Timothy Laumann, Sandra~W Curtiss, and David~C
  Van~Essen.
\newblock Informatics and data mining tools and strategies for the human
  connectome project.
\newblock \emph{Frontiers in neuroinformatics}, 5:\penalty0 4, 2011.

\bibitem[Mejia et~al.(2022)Mejia, Spencer, Pham, Bolin, Ryan, and
  Yue]{Mejia2022BayesfMRI}
Amanda Mejia, Daniel~A. Spencer, Damon Pham, David Bolin, Sarah Ryan, and
  Yu~(Ryan) Yue.
\newblock {BayesfMRI}, April 2022.
\newblock URL \url{https://github.com/mandymejia/BayesfMRI}.

\bibitem[Mejia et~al.(2020)Mejia, Yue, Bolin, Lindgren, and
  Lindquist]{mejia2020bayesian}
Amanda~F Mejia, Yu~Yue, David Bolin, Finn Lindgren, and Martin~A Lindquist.
\newblock A {Bayesian} general linear modeling approach to cortical surface
  {fMRI} data analysis.
\newblock \emph{Journal of the American Statistical Association}, 115\penalty0
  (530):\penalty0 501--520, 2020.

\bibitem[Opitz(2017)]{opitz2017latent}
Thomas Opitz.
\newblock Latent gaussian modeling and inla: A review with focus on space-time
  applications.
\newblock \emph{Journal de la soci{\'e}t{\'e} fran{\c{c}}aise de statistique},
  158\penalty0 (3):\penalty0 62--85, 2017.

\bibitem[Parlak et~al.(2022)Parlak, Pham, Spencer, Welsh, and
  Mejia]{parlak2022sources}
Fatma Parlak, Damon Pham, Daniel Spencer, Robert Welsh, and Amanda Mejia.
\newblock Sources of residual autocorrelation in multiband task {fMRI} and
  strategies for effective mitigation.
\newblock \emph{arXiv preprint arXiv:2209.06783}, 2022.

\bibitem[Penny et~al.(2005)Penny, Trujillo-Barreto, and
  Friston]{penny2005bayesian}
William~D Penny, Nelson~J Trujillo-Barreto, and Karl~J Friston.
\newblock Bayesian {fMRI} time series analysis with spatial priors.
\newblock \emph{NeuroImage}, 24\penalty0 (2):\penalty0 350--362, 2005.

\bibitem[Pham et~al.(2022)Pham, Muschelli, and Mejia]{pham2022ciftitools}
Damon~D Pham, John Muschelli, and Amanda~F Mejia.
\newblock ciftitools: A package for reading, writing, visualizing, and
  manipulating cifti files in r.
\newblock \emph{NeuroImage}, 250:\penalty0 118877, 2022.

\bibitem[Poldrack et~al.(2011)Poldrack, Mumford, and
  Nichols]{poldrack2011handbook}
Russell~A Poldrack, Jeanette~A Mumford, and Thomas~E Nichols.
\newblock \emph{Handbook of functional {MRI} data analysis}.
\newblock Cambridge University Press, 2011.

\bibitem[Poline and Mazoyer(1993)]{poline1993analysis}
Jean-Baptiste Poline and Bernard~M Mazoyer.
\newblock Analysis of individual positron emission tomography activation maps
  by detection of high signal-to-noise-ratio pixel clusters.
\newblock \emph{Journal of Cerebral Blood Flow \& Metabolism}, 13\penalty0
  (3):\penalty0 425--437, 1993.

\bibitem[Poline et~al.(1997)Poline, Worsley, Evans, and
  Friston]{poline1997combining}
Jean-Baptiste Poline, Keith~J Worsley, Alan~C Evans, and Karl~J Friston.
\newblock Combining spatial extent and peak intensity to test for activations
  in functional imaging.
\newblock \emph{Neuroimage}, 5\penalty0 (2):\penalty0 83--96, 1997.

\bibitem[Rue et~al.(2009)Rue, Martino, and Chopin]{rue2009approximate}
H{\aa}vard Rue, Sara Martino, and Nicholas Chopin.
\newblock Approximate {Bayesian} inference for latent {Gaussian} models using
  integrated nested {Laplace} approximations (with discussion).
\newblock \emph{Journal of the Royal Statistical Society B}, 71:\penalty0
  319--392, 2009.

\bibitem[Sid{\'e}n et~al.(2017)Sid{\'e}n, Eklund, Bolin, and
  Villani]{siden2017fast}
Per Sid{\'e}n, Anders Eklund, David Bolin, and Mattias Villani.
\newblock Fast {B}ayesian whole-brain {fMRI} analysis with spatial 3{D} priors.
\newblock \emph{NeuroImage}, 146:\penalty0 211--225, 2017.

\bibitem[Skorski(2021)]{skorski2021modern}
Maciej Skorski.
\newblock Modern analysis of {H}utchinson's trace estimator.
\newblock In \emph{2021 55th Annual Conference on Information Sciences and
  Systems (CISS)}, pages 1--5. IEEE, 2021.

\bibitem[Smith and Nichols(2009)]{smith2009threshold}
Stephen~M Smith and Thomas~E Nichols.
\newblock Threshold-free cluster enhancement: addressing problems of smoothing,
  threshold dependence and localisation in cluster inference.
\newblock \emph{Neuroimage}, 44\penalty0 (1):\penalty0 83--98, 2009.

\bibitem[Spencer et~al.(2020)Spencer, Guhaniyogi, and Prado]{spencer2020joint}
Daniel Spencer, Rajarshi Guhaniyogi, and Raquel Prado.
\newblock Joint {B}ayesian estimation of voxel activation and inter-regional
  connectivity in {fMRI} experiments.
\newblock \emph{Psychometrika}, 85\penalty0 (4):\penalty0 845--869, 2020.

\bibitem[Spencer et~al.(2022)Spencer, Yue, Bolin, Ryan, and
  Mejia]{spencer2022spatial}
Daniel Spencer, Yu~Ryan Yue, David Bolin, Sarah Ryan, and Amanda~F Mejia.
\newblock Spatial {B}ayesian {GLM} on the cortical surface produces reliable
  task activations in individuals and groups.
\newblock \emph{NeuroImage}, 249:\penalty0 118908, 2022.

\bibitem[Stephens et~al.(2022)Stephens, Carbonetto, Willwerscheid, Dai,
  et~al.]{stephens2022ashr}
Matthew Stephens, Peter Carbonetto, Jason Willwerscheid, Chaoxing Dai, et~al.
\newblock ashr: {An} {R} package for adaptive shrinkage.
\newblock \url{https://github.com/stephens999/ashr}, 2022.

\bibitem[Van~Essen et~al.(2013)Van~Essen, Smith, Barch, Behrens, Yacoub,
  Ugurbil, Consortium, et~al.]{van2013wu}
David~C Van~Essen, Stephen~M Smith, Deanna~M Barch, Timothy~EJ Behrens, Essa
  Yacoub, Kamil Ugurbil, Wu-Minn~HCP Consortium, et~al.
\newblock The {WU-Minn} human connectome project: an overview.
\newblock \emph{Neuroimage}, 80:\penalty0 62--79, 2013.

\bibitem[Varadhan and Roland(2004)]{varadhan2004squared}
Ravi Varadhan and Ch~Roland.
\newblock Squared extrapolation methods ({SQUAREM}): A new class of simple and
  efficient numerical schemes for accelerating the convergence of the {EM}
  algorithm.
\newblock \emph{Johns Hopkins University, Dept. of Biostatistics Working
  Papers}, 2004.

\bibitem[Wang and Titterington(2005)]{wang2005inadequacy}
Bo~Wang and D~Michael Titterington.
\newblock Inadequacy of interval estimates corresponding to variational
  {B}ayesian approximations.
\newblock In \emph{International Workshop on Artificial Intelligence and
  Statistics}, pages 373--380. PMLR, 2005.

\bibitem[Welvaert et~al.(2011)Welvaert, Durnez, Moerkerke, Verdoolaege, and
  Rosseel]{welvaert2011neurosim}
Marijke Welvaert, Joke Durnez, Beatrijs Moerkerke, Geert Verdoolaege, and Yves
  Rosseel.
\newblock {neuRosim}: An {R} package for generating {fMRI} data.
\newblock \emph{Journal of Statistical Software}, 44\penalty0 (10):\penalty0
  1--18, 2011.
\newblock URL \url{http://www.jstatsoft.org/v44/i10/}.

\bibitem[Zhang et~al.(2016)Zhang, Guindani, Versace, Engelmann, and
  Vannucci]{zhang2016spatiotemporal}
Linlin Zhang, Michele Guindani, Francesco Versace, Jeffrey~M Engelmann, and
  Marina Vannucci.
\newblock A spatiotemporal nonparametric {B}ayesian model of multi-subject
  {fMRI} data.
\newblock \emph{The Annals of Applied Statistics}, 10\penalty0 (2):\penalty0
  638--666, 2016.

\end{thebibliography}
\bibliographystyle{plainnat}

\appendix

\section{Preprocessing steps} \label{sec:preproc}

Consider cs-fMRI or subcortical fMRI data from a scan, represented as $\mathbf{y}_t \in \mathbb{R}^N$ for times $t = 1,\ldots,T$ at $N$ locations. These data are gathered while a subject completes $K$ different tasks as part of an experimental design. Data representing this design are represented as $\mathbf{x}_{t}^k \in \mathbb{R}^N$, which may not be identical across all locations after preprocessing (see section \ref{sec:preproc} for details). The $J$ nuisance regressors, accounting for unwanted effects such as motion and scanner drift, are represented through the notation $z_{t}^j$. Modeling brain activation for data in this form is done through the general linear model
\begin{align}
    \mathbf{y}_t = \boldsymbol\mu + \sum_{k=1}^K \mathbf{x}_t^k\boldsymbol\beta^k + \sum_{j = 1}^J z_t^jb^j + \mathbf{e}_t \label{eq:GLM}
\end{align}
where $\boldsymbol\mu$ is the mean value, $\boldsymbol\beta^k$ is the regression coefficient for task $k$, $\mathbf{b}^j$ is the nuisance regression coefficient for nuisance covariate $j$, and $\mathbf{e}_t$ is the error term. In our treatment of the model, the data are preprocessed to remove the mean effect $\boldsymbol\mu$. Please refer to section \ref{sec:preproc} for more details on data preprocessing. These data are then fitted to either the classical or Bayesian general linear model, with the main objective of performing inference about the parameters $\boldsymbol{\beta}^k$.

Data preprocessing is a common practice in the modeling of neuroimaging data, done with the intention of quickly removing variance stemming from known sources of error, such as movement and temporal autocorrelation. We preprocess the data using functions within the \texttt{BayesfMRI} R software package, following the steps outlined below.

The values from the design matrix are preprocessed to account for the delay in task stimulus and physiological response through convolution with a haemodynamic response function (HRF), $h(t)$,
\begin{align*}
    x_{t,k} = \int_0^t x_{t,k}(\tau) h(t - \tau) d\tau.
\end{align*}
We use the canonical (double-gamma) HRF characterized as 
\begin{align}
    h(t) = \left(\frac{t}{a_1b_1}\right)^{a_1} e^{-(t - a_1b_1) / b_1} - c\left( \frac{t}{a_2b_2} \right) ^{a_2} e^{-(t - a_2b_2) / b_2}.
\end{align}
with values set according to default values given in the \texttt{neuRosim} package in R \citep{welvaert2011neurosim}, specifically $a_1 = 6$, $a_2 = 12$, $b_1 = b_2 = 0.9$, and $c = 0.35$. The values of the HRF-convolved design matrix $\mathbf{X} \in \mathbb{R}^{T \times K}$ are then scaled by dividing each column by its maximum value, then centering these values around 0. This is done to keep estimates of $\boldsymbol\beta_k$ comparable across different tasks as measures of percent signal change. 

In order to facilitate spatial modeling given current computing memory capacity, the first step in preprocessing the response data is to resample the data to a lower resolution, which is done for the HCP data using an interpolation method outlined in \cite{glasser2013minimal}. This resampling presents a tradeoff between computational efficiency and spatial resolution in the inference, and needs to be considered thoroughly. This is examined in detail within \cite{spencer2022spatial}, showing that resampling to a resolution of around 5,000 vertices per hemisphere does not significantly alter inference on cortical surfaces. After resampling, the values from the response $\mathbf{y}_v$ at each data location $v$ are centered and scaled using the function 
\begin{align}
    f(\mathbf{y}_v) = 100 \times \frac{(\mathbf{y}_v - \bar y_v)}{\bar y_v}, \label{eq:scaleY}
\end{align}
where $\mathbf{y}_v$ is the cs-fMRI time series at location $v$, and $\bar y_v$ is the average value at that data location across time. This transformation makes the fMRI time series interpretable as percent signal change, while also removing the need for mean value parameters, as in Equation (\ref{eq:GLM}). 

Next, nuisance regression is performed to remove the effects of known confounding variables $z_t^j$ from the response data. This is done by regressing the centered and scaled response data against the nuisance variables, and then subtracting the estimated effects of the nuisance variables
$$ \tilde{\mathbf{y}}_v = f(\mathbf{y}_v) - \sum_{j=1}^J \mathbf{z}^j \hat{\mathbf{b}}^j, $$
where $f(\mathbf{y}_v)$ is as defined in Equation (\ref{eq:scaleY}), $\mathbf{z}^j \in \mathbb{R}^T$ is the value of the nuisance covariate across time, and $\hat{b}^j$ is the regression estimate of the nuisance parameter found by regressing $f(\mathbf{y}_v)$ against $\mathbf{Z} = (\mathbf{z}^1,\cdots,\mathbf{z}^J)$.

In order to remove temporal autocorrelation within the data, prewhitening is performed next. This process first finds the residual values from a regression of the response ($\tilde{\mathbf{y}}_v$) on the design matrix $\mathbf{X}_{v} \in \mathbb{R}^{T \times K}$, and then fits an AR(6) autoregressive model on these values using the method of solving the Yule-Walker equations \citep{eshel2003yule}. The coefficients and the residual variance of the AR(6) model are then spatially smoothed using a Gaussian kernel with a full-width half-maximum (FWHM) of 6 mm. These are used to create a temporal covariance matrix ($\mathbf{S}$) for the response at each location. The inverse of the square root of the covariance matrix ($\mathbf{D} = \mathbf{S}^{-1/2}$) is found using singular value decomposition. Finally, both the response data and the design matrix are premultiplied by $\mathbf{D}$ to produce the preprocessed response and task covariate data at each data location. 

After all of these preprocessing steps are applied, the data can be fit via a general linear model of the form shown in Equation (\ref{eq:preprocessed_model}).



\section{Choice of stopping rule tolerance}
\label{app:epsilon}

\begin{figure}
    \centering
    \includegraphics[width=.7\textwidth]{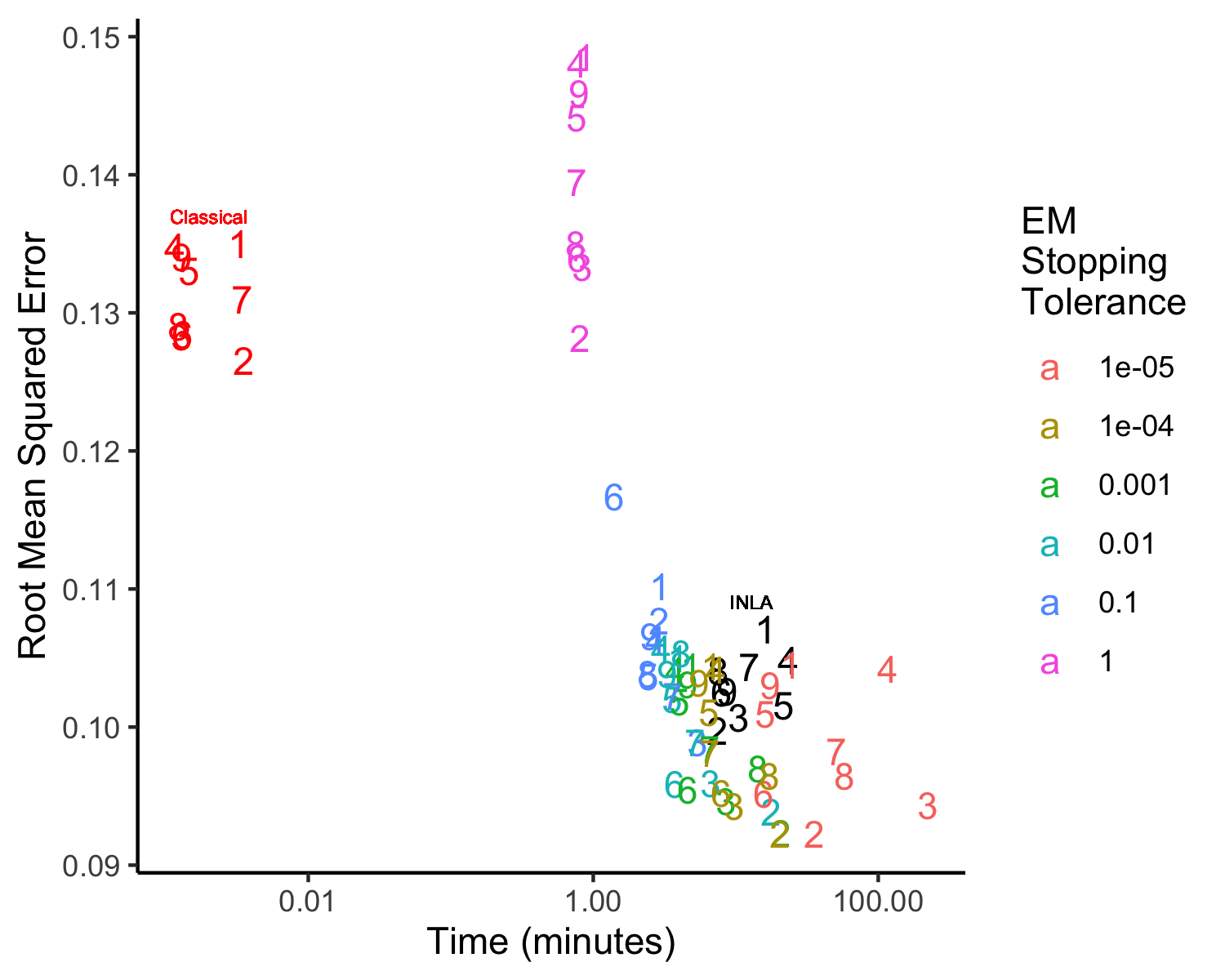}
    \caption{A comparison of computation time and inferential accuracy for different stopping rule tolerances in the EM algorithm.}
    \label{fig:time_rmse_tol}
\end{figure}

A simulation study was performed in which the stopping rule tolerance was allowed to vary from 1 down to 0.001 by powers of 10 to examine the effect of the choice of stopping rule on the speed and accuracy, measured through the square root of the mean squared error (RMSE), of the EM algorithm. In all cases, the model was fitted to simulated cortical surface data on the left hemisphere for four simulated tasks with a spatial resolution of 5,000 vertices per hemisphere. This data generation setting was used to generate 9 different datasets in order to give an idea of the variance in time and  \textbf{Figure \ref{fig:time_rmse_tol}} shows the differences in the accuracy and speed for the different tolerance levels. Based on this analysis, the stopping rule tolerance was set to $\epsilon = 0.001$ for all of the analyses in this paper.

\newpage 

\section{Additional figures}

\begin{figure}[h]
    \centering
    \includegraphics[width=0.4\textwidth]{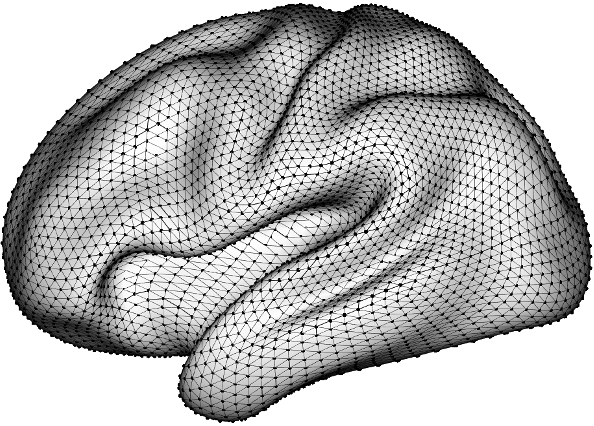}
    \caption{An example of the mesh structure on the cortical surface}
    \label{fig:cs_mesh}
\end{figure}

\begin{figure}[ht]
    \begin{tabularx}{\textwidth}{c|c|c|}
            \multicolumn{1}{c}{} &
            \multicolumn{1}{c}{\textbf{EM}} &
            \multicolumn{1}{c}{\textbf{INLA}} \\
            \cline{2-3} 
    		\rotatebox[origin=l]{90}{\qquad \, \textbf{Posterior SD}} &
    		\Includegraphics[width=0.45\textwidth]{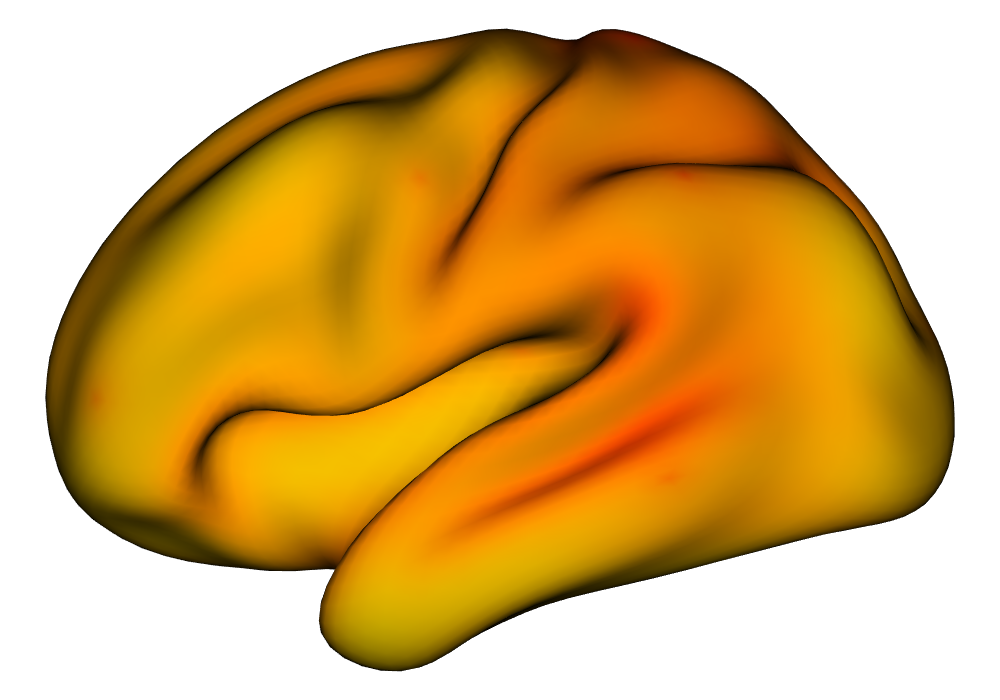} &
    		\Includegraphics[width=0.45\textwidth]{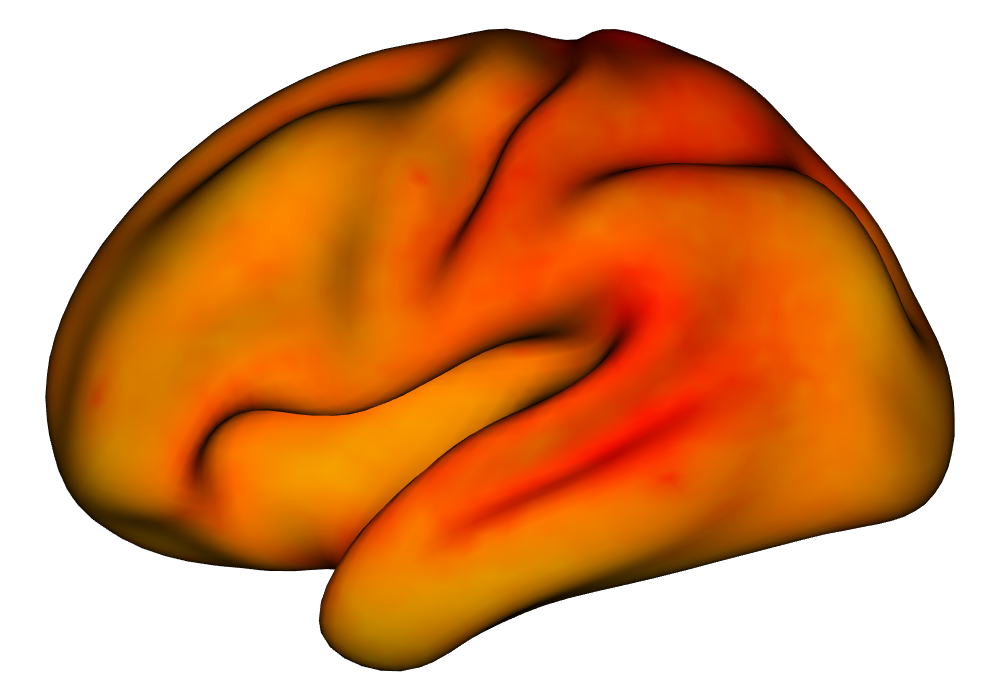} \\
    		\cline{2-3} 
    		\multicolumn{1}{c}{} &
            \multicolumn{2}{c}{\Includegraphics[width = 0.4\textwidth]{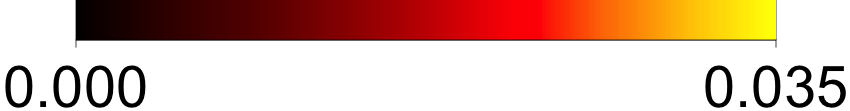}} 
    \end{tabularx}
    \caption{The posterior standard deviations for the activation amplitudes of a single task in a single run of simulated data found using the posterior mode estimates of $\boldsymbol{\theta}$ for the EM implementation and the posterior samples for the INLA implementation. The EM implementation shows a slightly higher standard deviation than the INLA method across much of the field due to the lack of a prior distribution on the hyperparameters, which influences the location of the posterior mode.}
    \label{fig:sim_sd_comparison}
\end{figure}

\begin{figure}[hb]
    \centering
    \includegraphics[width=\textwidth]{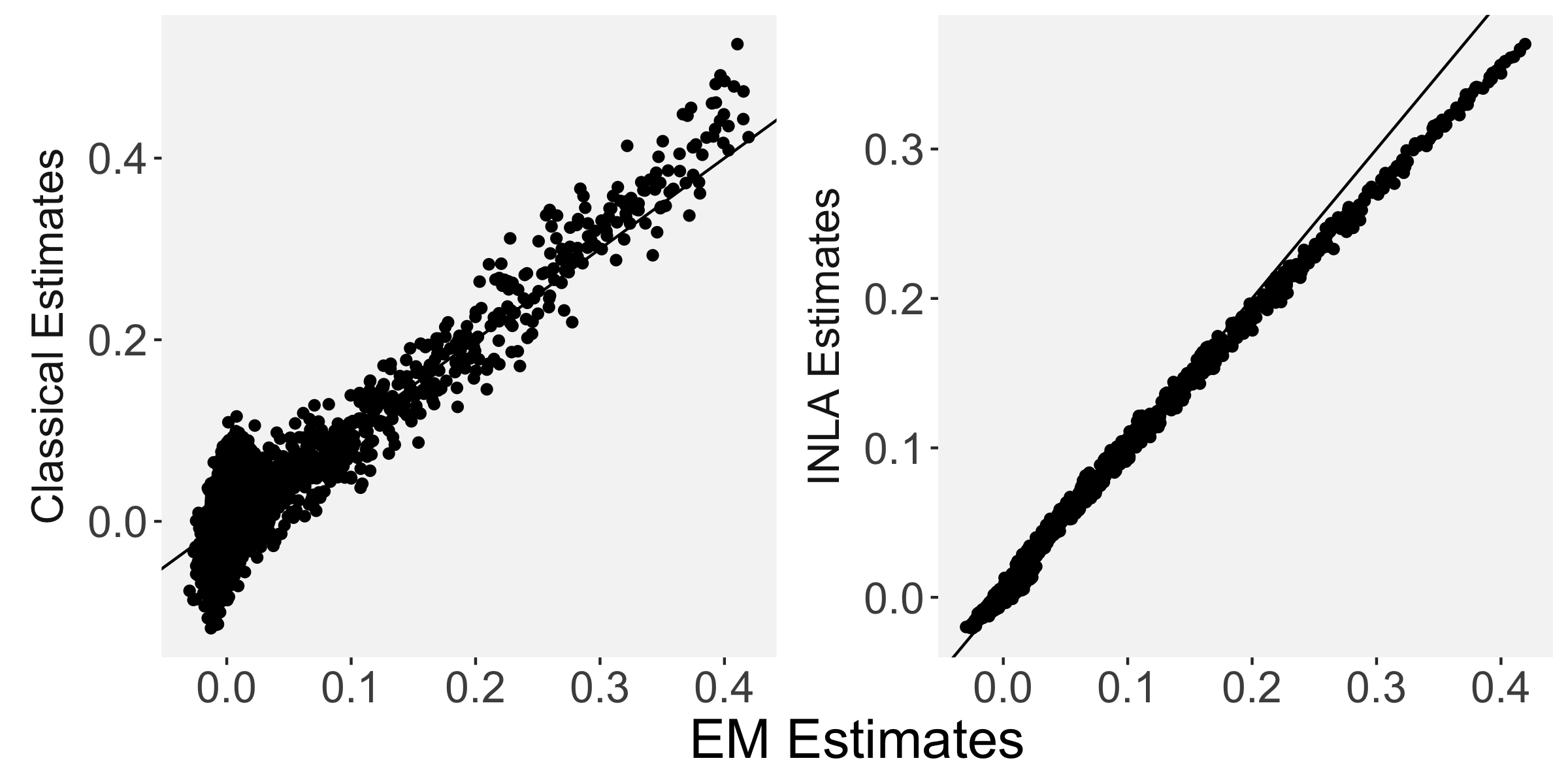}
    \caption{A comparison of the multi-subject point estimates between the EM and INLA implementations of the SBSB GLM and the classical GLM for a single subset of 10 subjects. The EM implementation exhibits less attenuation of estimates for higher values while still maintaining similar regularization for smaller values. For reference, the actual highest value in the population is about 1.4.}
    \label{fig:group_estimate_model_comparison}
\end{figure}

\begin{figure}
    \centering
    \begin{tabularx}{\textwidth}{c|c|c|}
            \multicolumn{1}{c}{} &
            \multicolumn{1}{c}{\textbf{32k Vertices}} &
            \multicolumn{1}{c}{\textbf{5k Vertices}} \\
            \cline{2-3} 
    		\rotatebox[origin=l]{90}{\qquad \quad\textbf{Subject A}} &
    		\Includegraphics[width=0.4\textwidth]{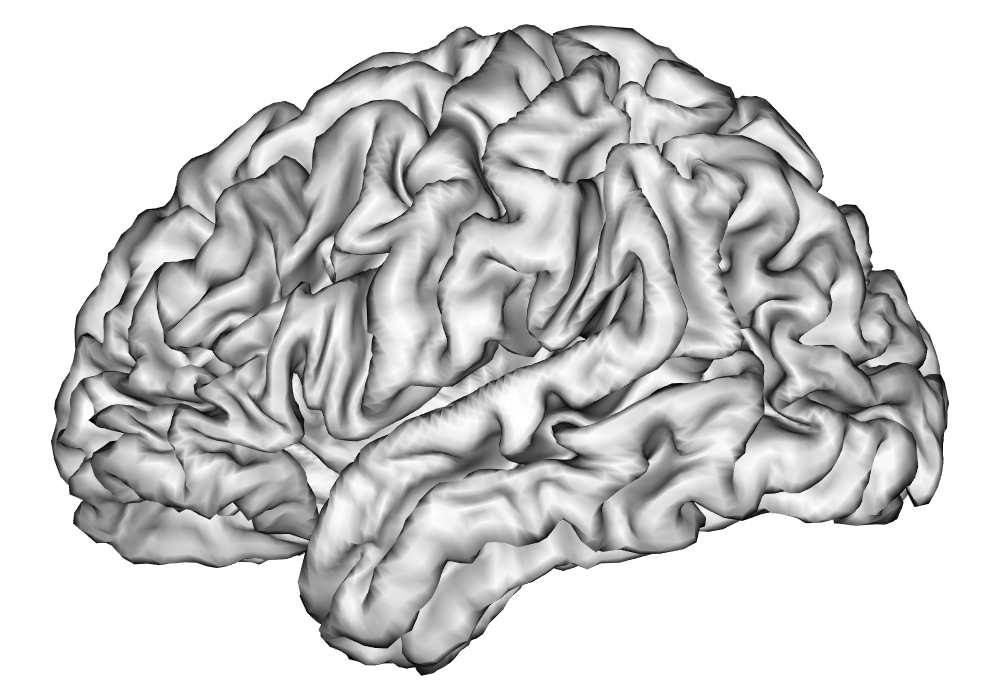} &
    		\Includegraphics[width=0.4\textwidth]{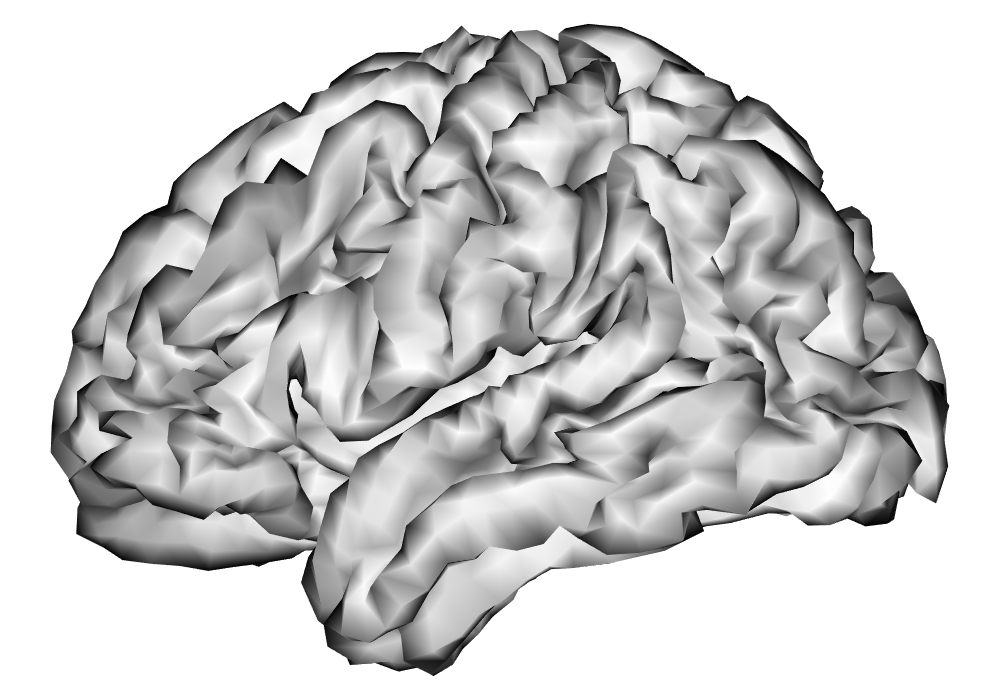} \\
    		\cline{2-3} 
    		\rotatebox[origin=l]{90}{\qquad \quad \textbf{Subject B}} &
    		\Includegraphics[width=0.4\textwidth]{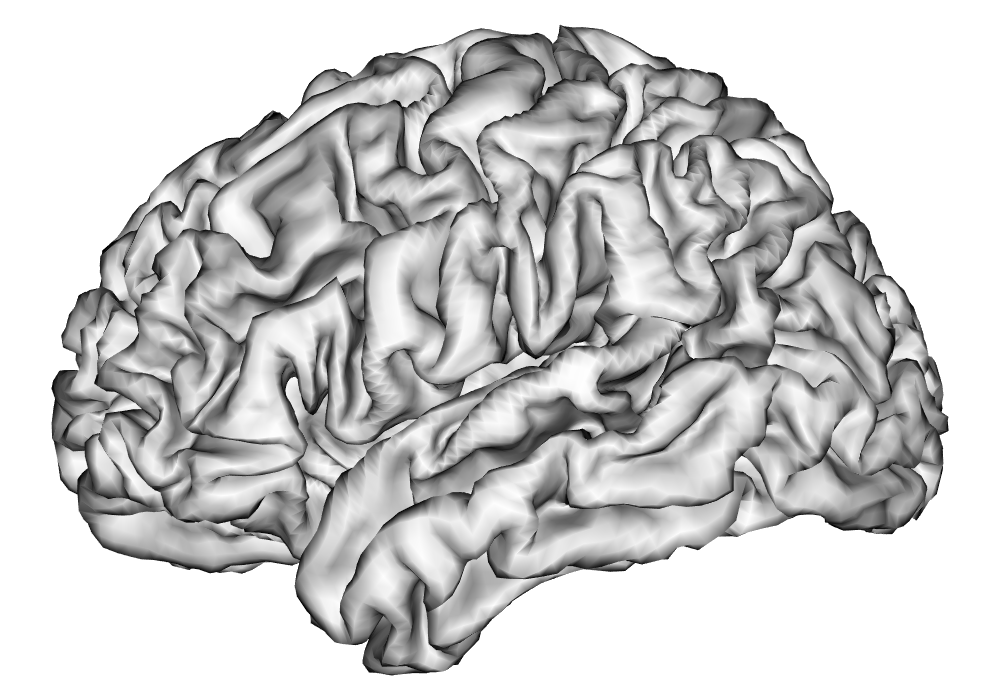} &
    		\Includegraphics[width=0.4\textwidth]{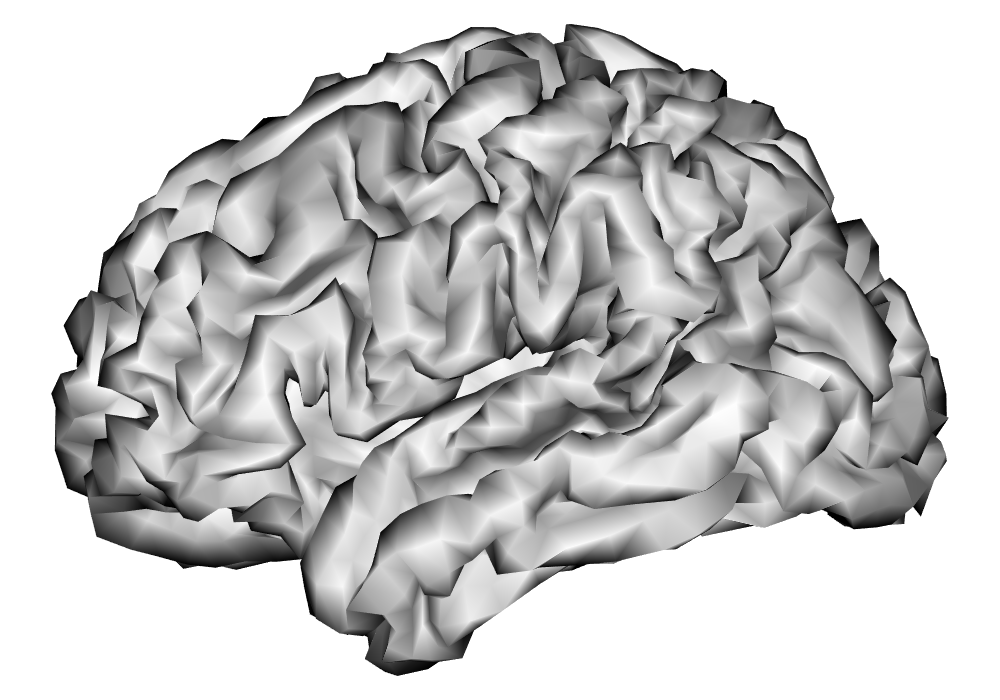} \\
    		\cline{2-3} 
        \end{tabularx}
    \caption{Full resolution (32k vertices) and resampled resolution (5k vertices) surfaces for two subjects showing the difference between subject-specific surfaces that are taken into account when using the SBSB GLM.}
    \label{fig:surface_comparison}
\end{figure}

\end{document}